\documentclass[a4paper,onecolumn,11pt]{quantumarticle}
\pdfoutput=1
\usepackage[utf8]{inputenc}
\usepackage[english]{babel}
\usepackage[T1]{fontenc}
\usepackage{amsmath}
\usepackage[numbers,sort&compress]{natbib}   
\usepackage{tikz}
\usepackage{lipsum}
\usepackage{amsfonts}
\usepackage{kotex}
\usepackage{physics}
\usepackage{amsthm}
\usepackage{enumitem}
\usepackage{mathtools}
\usepackage{array}
\usepackage{hyperref}

\allowdisplaybreaks[4]
\theoremstyle{definition}
\newtheorem{fact}{Fact}
\newtheorem{lemma}{Lemma}
\newtheorem*{fact*}{Fact}

\usepackage[colorinlistoftodos]{todonotes}

\begin{document}
\title{Shallow randomized measurement in noisy quantum devices}

\author{Gyungmin Cho}
\email{km950501@snu.ac.kr}
\affiliation{NextQuantum Center, Department of Physics and Astronomy, and Institute of Applied Physics, Seoul National University, Seoul 08826, Korea}

\author{Dohun Kim}
\email{dohunkim@snu.ac.kr}
\affiliation{NextQuantum Center, Department of Physics and Astronomy, and Institute of Applied Physics, Seoul National University, Seoul 08826, Korea}
\maketitle

\begin{abstract}
    Quantum hardware is steadily improving, but near-term quantum devices remain limited by noise and circuit depth. 
    This motivates measurement protocols that can use shallow-depth circuits while remaining robust to experimental errors. 
    In this work, we study the advantages of shallow randomized measurements over non-entangling single-qubit measurements for learning properties of quantum states. 
    Although shallow measurements have shown advantages in selected applications, their usefulness across different learning tasks has not been systematically understood. 
    Here, we develop a theoretical framework based on Clifford ensembles that incorporates shallow measurements into derandomized measurements, multi-shot protocols, common randomized measurements, error-mitigated estimators, and hybrid quantum-classical learning. 
    Finally, we validate these results on IBM quantum hardware in experiments with up to 40 qubits and 46 layers of two-qubit gates. 
    These results indicate that shallow-depth measurements can provide practical benefits on noisy quantum devices.
\end{abstract}

\section{Introduction}
Quantum computers are anticipated to demonstrate advantages over classical computers in tasks such as factorization~\cite{shor1999polynomial_1}, machine learning~\cite{huang2022quantum_2}, and optimization~\cite{jordan2025optimization_3}. Beyond these applications, quantum computers also serve as a new experimental tool in physics, facilitating the exploration of highly entangled quantum states~\cite{georgescu2014quantum_4}. Classically simulating such states is generally constrained by memory and computational costs that grow exponentially with system size; however, quantum computers enable the preparation of complex quantum states~\cite{semeghini2021probing_5, satzinger2021realizing_6} and measurement of various physical quantities~\cite{brydges2019probing_7, elben2020mixed_8}, creating new pathways to study unexplored physics. Nevertheless, in cases where theoretical frameworks are incomplete, it may be unclear which physical quantities to measure, potentially requiring quantum memory or additional experiments.

One solution is to store quantum information classically via quantum state tomography
(QST). However, without further assumptions, its memory and computational costs grow exponentially with the system size $n$~\cite{haah2016sample_10, o2016efficient_11}. To avoid full state reconstruction, shadow tomography~\cite{aaronson2018shadow_12} and, more recently, classical shadows~\cite{huang2020predicting_13} were proposed for predicting expectation values of many observables. These methods have been implemented on current noisy intermediate-scale quantum (NISQ) devices~\cite{struchalin2021experimental_14, zhang2021experimental_15, preskill2018quantum_16}.

Subsequent work has developed several variants of classical shadows. 
Derandomized classical shadows~\cite{huang2021efficient_17} improve the estimation of predetermined observables by selecting measurement bases deterministically. 
Robust shadows~\cite{chen2021robust_18} mitigate errors in randomized measurements, while multi-shot shadows~\cite{helsen2023thrifty_19, zhou2023performance_20} reduce the need to sample a new unitary for every shot. 
Other extensions include fermionic shadows~\cite{zhao2021fermionic_21}, tailored to fermionic systems, shallow shadows based on brickwork circuits~\cite{bertoni2024shallow_22, hu2023classical_23, hu2025demonstration_24}, and methods that incorporate prior information about the prepared state~\cite{vermersch2024enhanced_25}.

\subsection{Contributions and scope}

The developments reviewed above motivate a unified treatment of shallow
randomized measurements that can accommodate the main extensions of classical
shadows. In this work, we develop such a framework based on Clifford measurement ensembles. We first combine derandomization with shallow measurements, obtaining deterministic measurement bases whose confidence bound is no larger than the corresponding randomized average. While previous studies of shallow shadows primarily focused on brickwork circuits, we study block-shadow circuits, for which the shadow channel and its inverse can be analyzed explicitly. This structure enables extensions to multi-shot protocols, common randomized measurements with an approximate bias state, and error-mitigated estimators. We also show that classical machine-learning models trained on data from shallow measurements can outperform those trained on single-qubit measurement data in a many-body physics task. Finally, we validate these results on IBM quantum hardware and observe that the predicted advantages can persist in the presence of experimental noise.

\subsection{Related work}

Classical shadows~\cite{huang2020predicting_13} based on random Pauli
measurements and random Clifford measurements provide two standard reference
points. Random Pauli measurements are experimentally simple and are well suited
for local observables, whereas random Clifford measurements are effective for
low-rank observables but require global Clifford circuits with
\(\mathcal O(n)\) depth in standard implementations without ancilla
qubits~\cite{bravyi2021hadamard_26}. Shallow shadows
~\cite{bertoni2024shallow_22, hu2023classical_23,
hu2025demonstration_24} aim to interpolate between these two limits using
entangling measurement circuits of depth much smaller than \(\mathcal O(n)\);
in this work, we focus on the \(\mathcal O(\log n)\)-depth regime. Most
previous analyses of shallow shadows considered brickwork random circuits. Here,
we use Clifford measurement ensembles as a common framework for shallow
randomized measurements, and focus on block Clifford measurements when
closed-form expressions for the shadow channel, its inverse, and the moments
entering the variance analysis are needed. These explicit formulas allow us to
derive sample-complexity bounds and track post-processing costs without
introducing an additional approximation in the inverse channel.

Our setting is also related to recent approaches based on local random circuits
that form approximate unitary designs~\cite{schuster2025random_30}. These works
give worst-case guarantees by showing that the moment channel of the implemented
ensemble approximates the corresponding Haar moment channel in relative error.
In principle, one can define an unbiased estimator by using the exact inverse
channel of the implemented measurement ensemble. In practice, however, this
inverse channel may be difficult to compute or use directly. Moreover, since the
inverse shadow map is not a physical quantum channel, order-type inequalities
for moment channels do not directly translate into comparable inequalities
after applying the inverse map. For this reason, existing protocols often use
the inverse Haar shadow channel as a tractable substitute. This gives a
controllable weak bias for some observable-prediction tasks when the implemented
ensemble is sufficiently close to the target design, but it is not directly
applicable to tasks, such as single-shot purity estimation and quantum state
tomography, where a more refined inverse-channel treatment is needed. In the
block-shadow setting considered here, this issue is avoided because a
closed-form expression for the exact inverse channel is available.

Most of the favorable scaling results in this work should be understood as typical-case statements, except where worst-case bounds are stated explicitly. This viewpoint is close in spirit to previous analyses of brickwork shallow shadows~\cite{bertoni2024shallow_22, hu2023classical_23, hu2025demonstration_24}, which establish logarithmic-depth scaling for fixed, state-independent observables. The block Clifford structure allows us to extend this type of typical-case analysis to correlated state--observable settings, in which the observable is not fixed independently of the state being averaged over. Such settings are more difficult to treat for brickwork circuits. This situation arises, for example, in state verification, where the target observable may be the prepared state itself. In this correlated setting, we average the full state-dependent variance directly and show that shallow-depth block Clifford measurements achieve, in the typical-case regime, the same scaling as global random Clifford measurements. Similar moment calculations are also used in our analysis of purity estimation.

These typical-case results do not imply a worst-case guarantee. Indeed, we identify an explicit block-product family for which the relevant block-shadow norm scales as \(3^{n/k}\). These counterexamples are highly aligned with the chosen block partition and illustrate a structured failure mode of block shadows. Thus, logarithmic-depth block measurements should not be viewed as worst-case replacements for global random Clifford measurements. To probe the relevance of the typical-case analysis beyond random states, we perform numerical simulations for physically motivated many-body states, including ground states and states generated by quench dynamics.

Block shadows also have a practical advantage in post-processing. For brickwork shallow shadows, the inverse shadow channel is generally not available in closed form and is often approximated numerically~\cite{bertoni2024shallow_22, hu2023classical_23,
hu2025demonstration_24}. This requires an additional optimization or tensor-network approximation step. If an estimator is built from this approximate inverse, bounds derived for the exact inverse channel do not directly apply without controlling the additional approximation error. 
For block shadows, the inverse channel is exact and has a product structure. This makes the post-processing more efficient than the corresponding brickwork shallow-shadow calculation in the settings considered here; see Appendices~\ref{appx:fidel_comp} and~\ref{appx:purity_comp} for a detailed comparison.

Product-structured randomized measurements have also appeared in related contexts, including fidelity estimation~\cite{huggins2022unbiasing_28}, quantum state tomography~\cite{stilck2024efficient_59}, and measurements with analog quantum devices~\cite{PhysRevX.13.011049}. These works suggest that product structure can be useful in practice, but a general statistical explanation of the regimes in which such methods perform well has been less developed. Our analysis helps fill this gap by giving explicit variance and moment bounds for block Clifford measurements, and by identifying both typical-case advantages and structured worst-case counterexamples.

\section{Classical shadow}
The classical shadow is a method for predicting many properties of a quantum state without QST~\cite{huang2020predicting_13}. This method involves applying a unitary $U$, sampled from a random unitary ensemble $\mathbb{U}$, to the quantum state $\rho$, followed by a measurement that yields a bitstring $b\in\{0,1\}^n$ as the outcome. By repeating the process $T$ times, we can obtain the classical shadow ${S_T}(\rho ) = \{ {({U_i},{b_i})}\}_{i = 1}^T$. Using the shadow channel $\mathcal{M}_{\mathbb{U}}$ defined by 
\begin{equation}
    \mathcal{M}_{\mathbb{U}}(\rho)
    =
    \mathbb{E}_{U\sim\mathbb{U}}
    \sum_{b\in\{0,1\}^n}
    U^{\dagger}\ket{b}\bra{b}U\,
    \operatorname{Tr}\!
        (U\rho U^{\dagger}\ket{b}\bra{b}).
\end{equation}
Using this channel, we obtain an unbiased estimator of the quantum state $\rho$ as $\hat{\rho}=\mathcal{M}_{\mathbb{U}}^{-1}(U^{\dagger} \ket{b} \bra{b} U)$. To estimate the expectation value of $O$, we can use $\hat{o}=\text{Tr}(O\hat{\rho})$. The number of experiments required to predict with an additive error $\epsilon$ is determined by the state-dependent shadow norm 
\begin{equation}
    \lVert O \rVert_{\mathrm{sh,\rho,\mathbb{U}}}^2=\mathbb{E}_{U\sim\mathbb{U},b}\text{Tr}(O\hat{\rho})^2
\end{equation}
Then, state-independent shadow norm is $\lVert O \rVert_{\mathrm{sh,\mathbb{U}}}= \max_{\rho}\lVert O \rVert_{\mathrm{sh,\rho,\mathbb{U}}}$ and according to Chebyshev’s inequality, the required number of experiments is $T =  \mathcal{O}(\text{Var}[\hat{o}]/\epsilon^2) =  \mathcal{O}(\lVert O \rVert_{\text{sh},\mathbb{U}}^2/\epsilon^2)$.
In the case of $\mathbb{U}=\text{Cl}(1)^{\otimes n}$, where $\text{Cl}(m)$ denotes a $m$-qubit Clifford group, this corresponds to random Pauli measurement (RPM) suitable for measuring local observables. RPM offers practical advantages as it requires only single-qubit gates, making it implementable on noisy quantum hardware. In contrast, $\mathbb{U}=\text{Cl}(n)$, termed random Clifford measurement (RCM), is suitable for low-rank observables. However, implementing a $n$-qubit random Clifford gate requires a circuit depth proportional to $\mathcal{O}(n)$ without ancilla qubits~\cite{bravyi2021hadamard_26}, posing challenges for current noisy hardware. Shallow shadow~\cite{bertoni2024shallow_22, hu2023classical_23} serves as an intermediate approach between these unitary ensembles, providing a smooth transition from RPM to RCM. 

\begin{figure}[t]
    \centering
    \includegraphics[
    width=0.7\linewidth,
    trim={1cm 24.5cm 12cm 0cm},
    clip
    ]{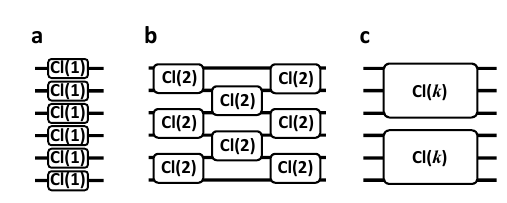}
    \caption{(a) Structure of single-qubit measurement. Each single-qubit gate is uniformly sampled from the set of single-qubit Clifford gates. (b) Structure of a brickwork circuit. Each two-qubit gate is uniformly sampled from the set of two-qubit Clifford gates. (c) Structure of a block circuit. Each block is uniformly sampled from the set of Clifford gates acting on $k$ qubits.}
    \label{fig1}
\end{figure}

Many previous studies~\cite{elben2020mixed_8, huang2020predicting_13, struchalin2021experimental_14, zhang2021experimental_15, huang2021efficient_17} have been conducted using RPM (Fig.~\ref{fig1}a). In shallow shadow, measurement circuits can be constructed using two different approaches: brickwork (shallow) shadow (Fig.~\ref{fig1}b) and block (shallow) shadow (Fig.~\ref{fig1}c). Brickwork measurements are composed of two-qubit gates acting on nearest-neighbor qubits. In the block measurements, we apply gates only within blocks of size $k$ qubits. For convenience, we define the block weight $w_k(P)$ for Pauli strings $P$ ($P$ is a tensor product of $n$ Pauli matrices from $\{I, X, Y, Z\}$) as the number of blocks where $P$ acts non-trivially. For example, $w_1(P) (= |\{i \in [n] : P_i \neq I\}|)$ is a widely used weight and $0 \leq w_k(P) \leq \lceil  n/k \rceil$. Throughout, \(\|A\|_p=(\operatorname{Tr}|A|^p)^{1/p}\) denotes the Schatten \(p\)-norm. 

\section{Block shadow}
Previously, studies~\cite{bertoni2024shallow_22, hu2023classical_23, ippoliti2023operator_27} have focused on random circuits with brickwork structures. In contrast, less is known about block shadow~\cite{huggins2022unbiasing_28, tran2023measuring_29}, which also interpolates between RPM and RCM. Both are expected to behave similarly; however, unlike brickwork shadow, block shadow has a structure suitable for computing the shadow channel and its inverse, thereby facilitating post-processing and allowing straightforward extensions to the previously introduced robust shadow~\cite{chen2021robust_18}. We assume that, for block size $k$, a unitary gate for each block is sampled uniformly from Cl($k$). The depth $d$ required to implement the unitary from Cl($k$) is $d = ck$, where $c$ depends on hardware connectivity. Below, we set $\mathbb{U} = \text{Cl}(k)^{\otimes n/k}$ and omit for simplicity.

First, we obtain the shadow channel of block shadow $\mathcal{M}=\mathcal{M}_k^{\otimes n/k}$ where $\mathcal M_k(A)
    =(A+\operatorname{Tr}(A)I_k)/(2^k+1)$ and $I_k$ is $2^k \times 2^k$ identity matrix. Following this, the shadow norm of Pauli strings $P$ can be written as
\begin{equation}\label{main:eq:sh_norm_pauli}
    \lVert P \rVert_{\mathrm{sh}}^2=(2^k+1)^{w_k(P)}.
\end{equation}
Eq.~\eqref{main:eq:sh_norm_pauli} indicates that the shadow norm of a fully-packed Pauli string $P$ becomes $\lVert P \rVert_{\mathrm{sh}}^2= (2^k+1)^{n/k}$ in case $k$ divides $n$. This allows us to observe a transition of  $\lVert P \rVert_{\mathrm{sh}}^2/2^n$ from exponential to constant dependence on $n$ at $k2^k = \Omega(n)$ consistent with the previous findings on brickwork circuits~\cite{bertoni2024shallow_22}. These results indicate that the shadow norm of observables with the bounded \(\|\cdot\|_2\)-norm is constant in typical cases~\cite{bertoni2024shallow_22}. 

However, this fixed-observable argument does not directly apply when the target observable is correlated with the prepared state, as in state verification where the observable may be the ideal target state. 
The average over \(\rho\) then involves correlations between the observable and the prepared state, rather than only the fixed-observable shadow norm.
Using the block structure of the unitary ensemble, for \(k2^k=\Omega(n)\),
\begin{equation}
    \mathbb{E}_{\rho}\!\left[\|\rho\|_{\mathrm{sh},\rho}^{2}\right]
    =
    \mathcal{O}(1),
\end{equation}
where the average is over Haar-random pure states and \(\rho\) is used both as the prepared state and as the rank-one target observable (Appendix~\ref{appx:state_dependent_variance}).

More generally, for any observable \(O\), block shadows with $k2^k=\Omega(n)$ satisfy
\begin{equation}\label{main:eq:worst_observable}
    \|O\|_{\mathrm{sh, \rho}}^{2}
    =
    \mathcal{O}(3^{n/k}\| O\|_2^2).
\end{equation}
This scaling is tight, for instance, when 
\(O=\rho=\bigotimes_{i=1}^{n/k}\ket{\Psi_i}\!\bra{\Psi_i}\). 
Thus, without additional structure, a constant shadow norm—and hence 
\(T=\mathcal{O}(1/\epsilon^2)\)—requires \(k=\Omega(n)\) (Appendix~\ref{appx:worst_shadow_norm}).

A recent approach~\cite{schuster2025random_30} can address this worst-case regime by doubling the block size and applying an additional staggered block layer, at the cost of a deeper measurement circuit. Our result is complementary: it shows that the simpler block circuit in Fig.~\ref{fig1}c already achieves constant average-case scaling for fidelity estimation, and we numerically observe this behavior for states arising in several many-body settings, including ground states and states after quenched dynamics (Appendix~\ref{appx:additional_numerics}).

Besides Pauli strings and low-rank observables, purity estimation via classical shadows has been used as a subroutine in many studies~\cite{elben2020mixed_8, vermersch2024many_31}. 
To explore the potential advantages of block shadows for purity estimation, we prove that, in the single-shot setting, the number of experiments \(T\) required to estimate \(\operatorname{Tr}(\rho^2)\) within additive error \(\epsilon\) satisfies
\begin{equation}
    T=\mathcal{O}\!\left(\max\left\{3^{n/k}/\epsilon^2,\,2^n/\epsilon\right\}\right)
\end{equation}
when \(k2^k=\Omega(n)\). 
The proof is provided in Appendix~\ref{appx:purity_single}. 
This extends the existing analysis for RPM~\cite{elben2020mixed_8, stricker2022experimental_32, rath2023entanglement_33}. 
For constant \(\epsilon\), block shadows with \(k2^k=\Omega(n)\) and \(k\ge 2\) achieve the same sample-complexity scaling as RCM. 
A similar trend also appears in probing quantum chaos via the spectral form factor~\cite{joshi2022probing_34}, as detailed in Appendix~\ref{appx:sff}.

For single-shot purity estimation, applying the estimator used for worst-case fidelity estimation in Ref.~\cite{schuster2025random_30} does not by itself guarantee the same sample-complexity scaling as RCM with logarithmic-depth measurement circuits. 
This is consistent with the numerical observations reported in Ref.~\cite{cioli2025approximate}.

\section{Derandomized block shadow}

\begin{figure}[t]
  \centering
  \includegraphics[
    width=1\linewidth,
    trim={1cm 23cm 6.5cm 0cm},
    clip
    ]{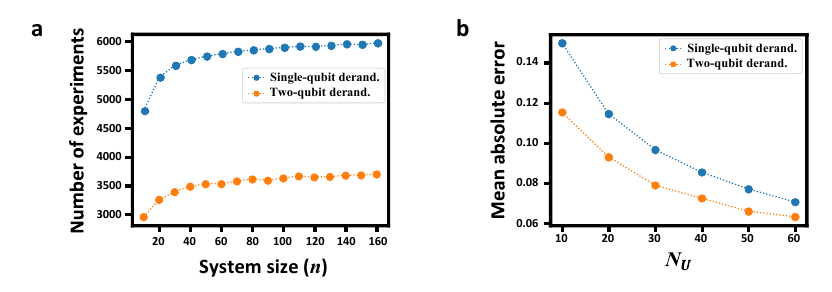}
  \caption{Derandomized block shadow. (a) Comparison with the previous method for measuring each Pauli string $P$ in $H_{\text{CH}}^2$ at least 100 times. Both approaches exhibit a logarithmic dependence on the system size $n$, but using a (block) shallow measurement circuit yields a more favorable multiplicative factor. (b) Experimental results on 20 qubits. As $N_U$ (number of different measurement bases) increases with NS (number of repeated measurements for a given measurement basis) fixed to 10000, we estimate the expectation values $\text{Tr}(\rho P)$ of the Pauli strings $P$ in $H_\text{CH}^2$ and then compute the Mean Absolute Error relative to the ground truth. To separate errors that arise during the optimization process, an approximate ground state is used as the ground truth instead of the exact ground state.}
  \label{fig2}
\end{figure}
Randomized measurement (RM) is an efficient measurement strategy when the physical observables are unknown in advance~\cite{huang2020predicting_13}. However, in cases like variational quantum algorithms~\cite{peruzzo2014variational_35}, where the observables to be measured are predetermined, RM may not be optimal. To overcome this, recent studies introduced a derandomization method, consistently outperforming RM on average~\cite{huang2021efficient_17}. We aim to extend this to a multi-qubit measurement. Although there has been an attempt~\cite{zhang2024minimal_36} to derandomize RCM, the need to implement an $\mathcal{O}(n)$ depth unitary on NISQ devices poses a challenge. To address this, we used block shallow measurements and implemented the derandomization algorithm based on the following two key facts.\\

\begin{fact}\label{fact1}
    Let \(\mathbb{U}_{\mathrm{MUB}}=\{U_\ell\}_{\ell=1}^{2^n+1}\) be a set of Clifford unitaries such that \(\{U_\ell^\dagger\ket{b}: b\in\{0,1\}^n\}\), for \(\ell=1,\ldots,2^n+1\), forms a complete set of mutually unbiased bases on \(n\) qubits. Then \(\mathcal{M}_{\mathbb{U}_{\mathrm{MUB}}}=\mathcal{M}_{\mathrm{Cl}(n)}\)~\cite{zhang2024minimal_36} (Appendix~\ref{appx:fact1}).
\end{fact}

\begin{fact}\label{fact2}
    There exists a Clifford ensemble \(\mathbb{U}_0\subseteq\mathrm{Cl}(n)\) with \(|\mathbb{U}_0|\leq\prod_{i=1}^{n}(2^i+1)\) such that, for every observable \(O\), \(\lVert O\rVert_{\mathrm{sh},\mathbb{U}_0}=\lVert O\rVert_{\mathrm{sh},\mathrm{Cl}(n)}\) (Appendix~\ref{appx:fact2}).
\end{fact}

Let $\mathbf{P}=\{P_l\}_{l=1}^L$ be a set of Pauli strings and $\mathbf{B}=\{U_m\}_{m=1}^{M}$ be a set of measurement bases. Define $M_l(\mathbf{B}) = \sum_m \mathbf{1}\{U_m P_l U_m^{\dagger} \in \pm \mathcal{Z}\}$ and 
\begin{equation}
    \text{CONF}_{\epsilon}(\mathbf{P};\mathbf{B}) \coloneqq \sum_{l=1}^{L}\exp(-\epsilon^2M_l(\mathbf{B})/2).
\end{equation}
If $\text{CONF}_{\epsilon}(\mathbf{P};\mathbf{B})\leq\delta/2$, then the expectation values of all $\{P_l\}_l$ can be estimated within additive error $\epsilon$ with failure probability at most $\delta$~\cite{huang2021efficient_17}. 
By applying the derandomization procedure, we obtain an output set of measurement bases $\mathbf{B}^{\#}$ satisfying
\begin{equation}
    \text{CONF}_{\epsilon}(\mathbf{P};\mathbf{B}^{\#})\leq\mathbb{E}_{\mathbf{B}}[\text{CONF}_{\epsilon}(\mathbf{P};\mathbf{B})].
\end{equation}
At each step of the derandomization, we choose a measurement basis that minimizes the conditional average confidence. 
Using Fact~\ref{fact1} with block size \(k\), the average confidence is
\begin{equation}
    \mathbb{E}_{\mathbf{B}}[\text{CONF}_{\epsilon}(\mathbf{P};\mathbf{B})]
    = \sum_{l=1}^{L}\left(1-\nu(2^k+1)^{-w_{k}(P_l)}\right)^M,
\end{equation}
where \(\nu = 1 - \exp(-\epsilon^2/2)\) (Appendix~\ref{appx:derand_conf}). 
Thus, when Fact~\ref{fact1} is applied blockwise, only \(2^k+1\) candidate bases need to be searched for each block.

In the single-qubit case, the ensemble $\mathbb{U} = \{I, R_Y(-\pi/2), R_X(\pi/2)\}$ already gives the same shadow norm as $\mathrm{Cl}(1)$ for arbitrary observables. 
For \(k>1\), however, the ensemble in Fact~\ref{fact1} reproduces the shadow channel but does not necessarily reproduce the full Clifford shadow norm for arbitrary observables. 
Fact~\ref{fact2} gives a sufficient, but larger, Clifford ensemble for this stronger property. 
In practice, when implementing the derandomization algorithm, it may be useful to use an ensemble whose size lies between the MUB construction in Fact~\ref{fact1} and the sufficient construction in Fact~\ref{fact2}.

To assess the practicality of derandomized block shadows, we performed numerical and hardware experiments. 
We used block shadows with block size \(k=2\) and considered the task of estimating the energy uncertainty \(\langle H^2\rangle-\langle H\rangle^2\) for the cluster-Heisenberg model
\begin{equation}
    H_{\mathrm{CH}}
    =
    \sum_{i=1}^{n-2} Z_iX_{i+1}Z_{i+2}
    +
    \sum_{i=1}^{n-1}
    \left(
        X_iX_{i+1}
        +
        Y_iY_{i+1}
        +
        Z_iZ_{i+1}
    \right),
\end{equation}
with open boundary conditions. In the numerical experiments, since \(\langle H_{\mathrm{CH}}\rangle\) can be measured using a small number of measurement bases, we focus on the number of experiments required to estimate \(\langle H_{\mathrm{CH}}^2\rangle\). 
The operator \(H_{\mathrm{CH}}^2\) contains many nonlocal Pauli strings, making it difficult to manually group simultaneously measurable terms. 
We estimated the total number of experimental runs required to measure each Pauli string appearing in \(H_{\mathrm{CH}}^2\) at least 100 times, for system sizes up to \(n=160\). 
As shown in Fig.~\ref{fig2}a, derandomized block shadows require fewer measurements than single-qubit derandomization~\cite{huang2021efficient_17} across all system sizes considered.

On current noisy quantum computers, however, errors from the two-qubit gates used in shallow measurements may offset the sample-complexity reduction predicted by the noiseless analysis. 
To test whether the advantage persists under hardware noise, we prepared an approximate ground state of \(H_{\mathrm{CH}}\) on a quantum computer with \(n=20\) qubits and estimated the Pauli strings appearing in \(H_{\mathrm{CH}}^2\). 
As shown in Fig.~\ref{fig2}b, derandomized block shadows still reduce the number of measurements required to achieve a given accuracy, even in the presence of hardware noise. 
Experimental details are provided in Appendix~\ref{appx:exp_derand}.

\section{Clifford ensemble}
Assuming a Pauli-invariant unitary ensemble $\mathbb{U}$, the following result holds~\cite{bu2024classical_38}.
\begin{equation}
    \mathcal{M}_{\mathbb{U}}(P)=m_PP,\quad \lVert P \rVert_{\text{sh}}^2=m_P^{-1}
\end{equation}
where $m_P$ is an eigenvalue of $\mathcal{M}_{\mathbb{U}}$, and the Pauli string $P$ is its eigenoperator. Here, we show that the above results still hold when the unitary ensemble is composed only of Clifford gates instead of requiring Pauli-invariance (Appendix~\ref{appx:A1}). There are three motivations for using the Clifford ensemble. First, because Clifford circuits can be simulated efficiently classically.~\cite{aaronson2004improved_39}, post-processing after RM becomes more straightforward, and they can be further combined with tensor network formalism. In particular, we utilize a probabilistic interpretation of $m_P = \mathbb{E}_{U}[\mathbf{1}\{UPU^{\dagger}\in\pm \mathcal{Z} \}] = \text{Pr}_U(UPU^{\dagger} \in \pm \mathcal{Z})$, where $1\{\cdot\}$ is the indicator function~\cite{bertoni2024shallow_22}. Second, through Pauli twirling, also known as randomized compiling~\cite{wallman2016noise_40}, the error channels arising from RM can be transformed into Pauli error channels. Finally, in many quantum error correction codes, Clifford gates are easier to implement than non-Clifford gates~\cite{fowler2012surface_41, piveteau2021error_42}. Building on this, we further demonstrate that it is possible to achieve various advantages of existing variants of classical shadows simultaneously. 

For convenience in this section, we define the number of experiments $T = N_UN_S$, where $N_U$ is the number of unitaries sampled from the unitary ensemble, and $N_S$ represents the number of repeated measurements in the same basis. 

First, we calculated the variance for a Pauli string $P$ using a (brickwork or block) shallow shadow in the multi-shot case ($N_S > 1$) as follows:
\begin{equation}\label{main:eq:ms_pauli}
    \text{Var}(\text{Tr}(\hat{\rho}P))=\frac{1}{N_U}\left(m_P^{-1}\left(\frac{1}{N_S}+\frac{N_S-1}{N_S}\text{Tr}(\rho P)^2\right) - \text{Tr}(\rho P)^2 \right).
\end{equation}
This is consistent with previous results when $\mathbb U=\mathrm{Cl}(n)$~\cite{helsen2023thrifty_19, zhou2023performance_20} and extends the multi-shot analysis to shallow shadows (Appendix~\ref{appx:multi_shots_pauli}). 

When estimating the expectation value of an observable
\(O=\sum_P \alpha_P P\) using a shallow shadow in the multi-shot setting, the variance can be written as
\begin{align}
    \operatorname{Var}\!\left(\operatorname{Tr}(O\hat{\rho})\right)
    &=
    \frac{1}{N_U}
    \left(
        \frac{1}{N_S}V_2(O,\rho)
        +
        \frac{N_S-1}{N_S}V_1(O,\rho)
        -
        \operatorname{Tr}(\rho O)^2
    \right)\\
    &\le
    \frac{1}{N_U}
    \left(
        V_1(O,\rho)+\frac{V_2(O,\rho)}{N_S}
    \right),
\end{align}
where
\begin{equation}
    V_1(O,\rho)
    =
    \sum_{P,Q}
    \alpha_P\alpha_Q\,
    \operatorname{Tr}(\rho P)\operatorname{Tr}(\rho Q)\,
    f(P,Q),
    \quad
    V_2(O,\rho)
    =
    \sum_{P,Q}
    \alpha_P\alpha_Q\,
    \operatorname{Tr}(\rho PQ)\,
    f(P,Q),
\end{equation}
and
$f(P,Q)
=\Pr_U\!\left(
UPU^{\dagger},UQU^{\dagger}\in\pm\mathcal{Z}
\right)
/(m_Pm_Q)$ (Appendix~\ref{appx:multi_shots_observable}).
Since \(V_1(O,\rho)\le V_2(O,\rho)\), increasing \(N_S\) does not increase the variance. 
Moreover, for block shadows and fixed \(O\), the Haar average over \(\rho\) satisfies
\begin{equation}
    \mathbb{E}_{\rho}[V_1(O,\rho)]
    =
    \mathcal{O}\!\left(\|O\|_2^2/D\right),
    \qquad
    \mathbb{E}_{\rho}[V_2(O,\rho)]
    =
    \mathcal{O}(\|O\|_2^2)
\end{equation}
when \(k2^k=\Omega(n)\) (Appendix~\ref{appx:typical_si_V}). 
Thus, for typical states and bounded \(\|O\|_2\), the \(V_2/N_S\) term can be suppressed by reusing each measurement basis, while the limiting term \(V_1\) remains small. 
This makes multi-shot measurements beneficial for fidelity estimation in the average-case regime, even with logarithmic-depth circuits. 
This contrasts with worst-case examples~\cite{helsen2023thrifty_19, zhou2023performance_20}, where \(V_1\) and \(V_2\) are comparable and increasing \(N_S\) gives only little improvement (Appendix~\ref{appx:comp_RCM}).

We next construct a purity estimator compatible with shallow shadows in the multi-shot setting. 
For each sampled measurement basis \(U_i\), we reuse the same basis \(N_S\) times and define
\begin{equation}
    \hat p_2
    =
    \frac{1}{N_U}\sum_{i=1}^{N_U}\hat p_2(U_i),
    \qquad
    \hat p_2(U)
    =
    \binom{N_S}{2}^{-1}
    \sum_{i>j}^{N_S}
    \operatorname{Tr}\!\left(
        U^{\dagger}\ket{b_i}\!\bra{b_i}U\,
        \mathcal{M}^{-1}
        \!\left(
            U^{\dagger}\ket{b_j}\!\bra{b_j}U
        \right)
    \right).
\end{equation}
This estimator is unbiased, \(\mathbb{E}[\hat p_2]=\operatorname{Tr}(\rho^2)\). 
Moreover, 
\begin{equation}
    \operatorname{Var}(\hat p_2) 
    \le 
    \frac{1}{N_U}
    \left(
        V_1(\rho,\rho)
        +
        \frac{4V_2(\rho,\rho)}{N_S}
        +
        \frac{2V_3(\rho)}{(N_S-1)^2}
    \right),
\end{equation}
where $V_3(\rho)=\sum_{P,Q}\operatorname{Tr}(\rho PQ)^2f(P,Q)/D^2$ (Appendix~\ref{appx:purity_multi}).

For block shadows with \(k2^k=\Omega(n)\), the typical-case bounds  \(V_3(\rho)=\mathcal{O}(2^n)\) and \(\mathbb{E}_{\rho}[V_{1,2}(\rho,\rho)]=\mathcal{O}(1)\) (Appendix~\ref{appx:typical_sd_V}) imply that \(N_S=\mathcal{O}(2^{n/2})\) is sufficient for constant-accuracy (single-shot) purity estimation. 
The same analysis extends to the distributed setting with shared randomness~\cite{anshu2022distributed_44}, yielding analogous bounds for estimating the inner product \(\operatorname{Tr}(\rho\sigma)\) between quantum states prepared by two parties~\cite{elben2020cross_46} (Appendix~\ref{appx:inner}).

Although \(N_S\) still grows exponentially and is therefore not scalable to large \(n\), it remains accessible for near-term experiments on systems of around 30 qubits. 
This is useful in settings where purities of local patches are sufficient to infer global properties of interest~\cite{vermersch2024many_31}. 
These results suggest that, in typical cases, local random Clifford circuits can approach the information-theoretically optimal sample complexity without requiring approximate unitary 4-designs~\cite{schuster2025random_30, anshu2022distributed_44, gong2024sample_45}.

For explicit estimates, we focus on block shadows. While brickwork shadows may have comparable sample complexity, block shadows offer a post-processing advantage through their tensor-product structure and exact inverse shadow channel; see Appendices~\ref{appx:fidel_comp} and~\ref{appx:purity_comp} for a detailed comparison.

\begin{figure}[t]
  \centering
  \includegraphics[
    width=0.95\linewidth,
    trim={1.5cm 22.5cm 6.5cm 0cm},
    clip
    ]{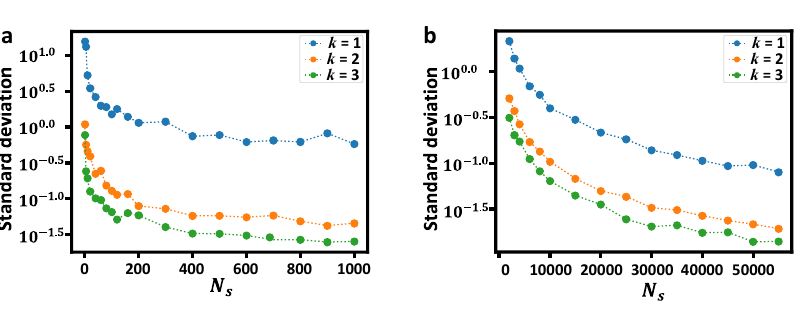}
  \caption{Numerical experiments using block shadows. (a) We prepared a random quantum state by applying a random unitary of depth 8 on $\ket{0}^{\otimes 48}$. We fixed $N_U = 800$ and estimated the standard deviation (std) for fidelity estimation by varying $N_S$ and the block size $k$. (b) We prepared a random quantum state by applying a random unitary of depth 12 on $\ket{0}^{\otimes 24}$. We fixed $N_U$ = 500 and estimated the std for purity estimation by varying $N_S$ and the block size $k$.}
  \label{fig3}
\end{figure}

Through numerical experiments, we show that (block) shallow shadow is helpful in many situations. In Fig.~\ref{fig3}a, we estimate the fidelity of the pure quantum state generated by a random unitary with a two-qubit gate depth of 8 and $n$ = 48. Increasing the block size $k$ allows for more accurate estimations, and using multi-shot measurements is beneficial. In Fig.~\ref{fig3}b, we estimate the purity of the pure quantum state generated by a random unitary with a two-qubit gate depth of 12 and $n$ = 24. Similar to fidelity, increasing $k$ enables more accurate measurements, and multi-shot measurements also aid in purity estimation.

\section{Common randomized measurements using shallow shadows}
When \(\Tr(\rho P)\) is close to \(1\) in Eq.~\eqref{main:eq:ms_pauli}, increasing \(N_S\) provides little benefit. 
If prior information about the prepared state \(\rho\) is available, this limitation can be mitigated using common randomized measurements (CRM)~\cite{vermersch2024enhanced_25}. 
CRM replaces the standard estimator \(\hat{\rho}\) by
\begin{equation}
    \hat{\rho}_{\sigma}
    =
    \hat{\rho}
    -
    \hat{\sigma}
    +
    \sigma,
\end{equation}
where \(\sigma\) is a bias state approximating \(\rho\), and $    \hat{\sigma}=\frac{1}{N_U}\sum_{i=1}^{N_U}\hat{\sigma}^{(U_i)}$
is estimated using the same measurement bases as \(\hat{\rho}\). 
This shared-basis construction can reduce the variance even when multi-shots alone are ineffective. 

Here, we combine CRM with shallow shadows. 
In the original CRM protocol, the bias contribution for a fixed measurement basis \(U\) is evaluated as~\cite{vermersch2024enhanced_25}
\begin{equation}\label{main:eq:old_estimator}
    \hat{\sigma}_{\text{old}}^{(U)}
    =
    \sum_{b\in \{0,1\}^n}
    \bra{b} U\sigma U^\dagger\ket{b} 
    {\cal M}^{ - 1}(U^{\dagger} \ket{b}\bra{b} U).
\end{equation}
We instead allow the bias contribution to be estimated from measurement outcomes of \(\sigma\) obtained in the same basis:
\begin{equation}\label{main:eq:new_estimator}
    \hat \sigma _{{\rm{new}}}^{(U)}
    =
    \frac{1}{N_\sigma}
    \sum_{j = 1}^{N_\sigma}
    {{\cal M}^{ - 1}}({U^{\dagger}}|{b_j}\rangle \langle {b_j}|U),
\end{equation}
where \(b_j\) is sampled from measuring \(U\sigma U^\dagger\). 
With this estimator, the variance for a Pauli string \(P\) satisfies 
\begin{equation}\label{main:eq:multishot_pauli}
    \operatorname{Var}\!\bigl(\operatorname{Tr}(\hat{\rho}_{\sigma} P)\bigr)
    \le
    \frac{m_P^{-1}}{N_U}\left(\frac{1-\operatorname{Tr}(\rho P)^2}{N_\rho}
    +\frac{1-\operatorname{Tr}(\sigma P)^2}{N_\sigma}
    +\operatorname{Tr}((\rho-\sigma)P)^2\right),
\end{equation}
where \(N_{\rho}\) and \(N_{\sigma}\) are the numbers of repeated measurements per basis for \(\rho\) and \(\sigma\), respectively (Appendix~\ref{appx:A16}). 
When \(N_{\sigma}\) is large, Eq.~\eqref{main:eq:multishot_pauli} reduces to the previous CRM bound~\cite{vermersch2024enhanced_25}.

The extended estimator in Eq.~\eqref{main:eq:new_estimator} is useful when the bias state ($\sigma$) is available on another quantum device but does not admit an efficient classical description.
In this case, the bias contribution can be incorporated directly through measurement outcomes of ($\sigma$), without requiring a classical evaluation of the probabilities appearing in Eq.~\eqref{main:eq:old_estimator}. Explicit comparisons of post-processing times are provided in Appendix~\ref{appx:crm_postprocessing_cost}.

On the other hand, when the estimator in Eq.~\eqref{main:eq:old_estimator} is used together with a Clifford ensemble, its expectation value for a Pauli string \(P\) simplifies to
\begin{equation}
    {\rm{Tr}}({\hat \sigma _{{\rm{old}}}}P)
    =
    m_P^{ - 1}{\rm{Tr}}(\sigma P){\bf{1}}\{ UP{U^{\dagger}} \in  \pm {\cal Z}\}.
\end{equation}
Thus, for Pauli observables, the contraction over all bit strings in Eq.~\eqref{main:eq:old_estimator} can be bypassed (Appendix~\ref{appx:old_pauli}). 
In particular, this allows us to incorporate recently introduced Pauli propagation methods for approximately computing \(\text{Tr}(\sigma P)\)~\cite{angrisani2025classically_48, fontana2025classical_49, beguvsic2025simulating_50, nemkov2023fourier_51, rudolph2023classical_52} into randomized measurements. 
Using this Clifford structure, efficient classical post-processing is possible for observables $O=\sum_{i=1}^{r}c_iP_i$ where $r=\mathrm{poly}(n)$, including many local Hamiltonians. 

In the above analysis, we assume that the unitary ensemble consists only of Clifford gates. Consequently, the above results are valid for ensembles that fail to be Pauli-invariant but are composed solely of Clifford gates; one such example~\cite{zhao2021fermionic_21} is $\mathbb{U} = \text{Cl}(n)\cap\text{FGU}(n)$, where $\text{FGU}(n)$ is the set of Fermionic Gaussian Unitaries for $n$ fermionic particles.

\section{Impact of noise on shallow shadows}

We now analyze how errors affect the above results. 
Two noise models for RMs have been studied in previous work~\cite{chen2021robust_18, Rozon2024optimal_53}. 
The first model assumes a noise channel applied after the random unitary $U$, \(\Lambda_U=\Lambda\mathcal{U}\). 
The second model describes layer-wise noise,
\(\Lambda_U=\Lambda_t\mathcal{U}_t\cdots\Lambda_1\mathcal{U}_1\), 
where \(\mathcal{U}_l\) is the unitary channel corresponding to the \(l\)-th layer \(U_l\), \(\Lambda_l\) is the noise channel at that layer, and \(U=U_t\cdots U_1\). 
In both cases, we assume that the noise channels are fixed and independent of the sampled unitary gates.

As mentioned above, when the measurement ensemble consists of Clifford gates, Pauli twirling converts the effective noise into a Pauli channel, which is unital~\cite{wallman2016noise_40}. 
Thus, for a Pauli string \(P\), we can write
\[
    \Lambda_U(P)=\lambda_{U,P}UPU^\dagger .
\]
Using this fact, the noisy shadow channel
\begin{equation}\label{main:err_shadow_1}
    \widetilde {\cal M}(A)
    =
    {\mathbb{E}_U}\left[
    {\sum _b}
    {U^\dagger }|b\rangle \langle b|U\,
    {\rm{Tr}}({\Lambda _U}(A)|b\rangle \langle b|)
    \right]
\end{equation}
still has Pauli strings as eigenoperators (Appendix~\ref{appx:noisy_shallow}). 
Indeed,
\begin{equation}
    \widetilde{\mathcal M}(P)
    =
    \widetilde m_P P,
    \qquad
    {\widetilde m_P}
    =
    {\mathbb{E}_U}
    \left[
    {\lambda _{U,P}}
    {\bf{1}}\{ UP{U^{\dagger}} \in  \pm {\cal Z}\}
    \right].
\end{equation}

Therefore, an error-mitigated estimator can be obtained by using \(\widetilde{\mathcal{M}}^{-1}\) instead of \(\mathcal{M}^{-1}\). 
Extending Eq.~\eqref{main:eq:ms_pauli} to the noisy setting gives, for a Pauli string \(P\),
\begin{equation}\label{main:eq:ms_error_pauli}
    {\rm{Var}}({\rm{Tr}}(\hat \rho P)) 
    =
    \frac{1}{N_U}
    \left[
    \widetilde m_P^{ - 2}
    \left(
        \frac{m_P}{N_S}
        +
        \frac{N_S - 1}{N_S}
        {\rm{Tr}}(\rho P)^2
        {\mathbb{E}_U}
        \!\left[
            \lambda _{U,P}^2
            {\bf{1}}\{ UP{U^{\dagger}} \in  \pm {\cal Z}\}
        \right]
    \right)
    -
    {\rm{Tr}}(\rho P)^2
    \right],
\end{equation}
as derived in Appendix~\ref{appx:noisy_multi}.
In the noiseless limit, where \(\lambda_{U,P}=1\) and \(\widetilde m_P=m_P\), Eq.~\eqref{main:eq:ms_error_pauli} reduces to Eq.~\eqref{main:eq:ms_pauli}. 
The same analysis can also be applied to CRM in the presence of errors (Appendix~\ref{appx:crm_multi_err}). 

\begin{figure}[t]
  \centering
  \includegraphics[
    width=0.55\linewidth,
    trim={1cm 22cm 11.5cm 0cm},
    clip
    ]{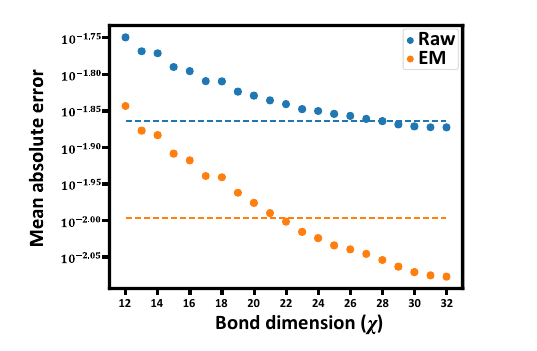}
  \caption{Experimental results of common randomized measurements (CRM). 
  We prepared a random quantum state by applying a random circuit of depth \(9\) to \(\ket{0}^{\otimes 40}\), with maximum bond dimension \(\chi=32\), and estimated the expectation values of contiguous Pauli strings with weights up to \(4\). 
  Classical shadows were generated using \(N_U=900\) measurement bases and \(N_S=1500\) repeated measurements per basis. 
  The blue dots show the mean absolute error from the raw data, while the orange dots show the results after error mitigation (EM). 
  The dotted lines show the corresponding results without CRM.}
  \label{fig4}
\end{figure}

To verify the above analysis, we performed experiments on a noisy quantum device. 
We prepared a random quantum state with circuit depth \(d=9\) on \(n=40\) qubits and generated classical shadows \(S_T(\rho)\) using \(N_U=900\) and \(N_S=1500\). 
The task was to estimate expectation values of contiguous Pauli strings with weights up to \(4\). 
As shown in Fig.~\ref{fig4}, CRM was implemented using a bias state \(\sigma\), represented as an MPS with bond dimension \(\chi_{\sigma}\), obtained by truncating the bond dimension of the ideal state \(\rho_{\mathrm{ideal}}\). 
For error mitigation, we extended robust shadows~\cite{chen2021robust_18} to block shadows (Appendix~\ref{appx:crm_em}). 
We first performed calibration experiments to estimate the amplification factor \(\hat{\alpha}_P(O)\geq 1\) for each observable \(O\), and then used
\begin{equation}
    \hat{o}_{\mathrm{EM}}
    =
    \hat{\alpha}_P(O)\,{\rm Tr}(O\hat{\rho})
\end{equation}
as the error-mitigated estimator. 
The dotted lines in Fig.~\ref{fig4} show the results without CRM. 
The experiment shows that error mitigation improves the accuracy of CRM even when the bias state is less accurate. 
Experimental details are provided in Appendix~\ref{appx:crm_exp}.

\section{Entanglement-enhanced classical ML}

Classical shadows were originally developed for predicting many properties of a quantum state. 
Recent studies~\cite{huang2021power_54} have shown that classical ML models can also learn from quantum experimental data provided in the form of classical shadows, even when classical simulation of the underlying quantum system is computationally difficult.

Since the ML procedure is performed entirely on a classical computer after data acquisition, this approach is well suited for current NISQ devices. 
Moreover, because it does not require specifying observables in advance, it can be useful for studying quantum systems whose relevant physical quantities are not fully known. 
Applications have been demonstrated in theoretical and experimental studies of quantum phase recognition and ground-state learning in many-body systems~\cite{huang2022provably_55, lewis2024improved_56, cho2024machine_57}.

Previous studies~\cite{huang2022provably_55,lewis2024improved_56, cho2024machine_57} have mainly used classical shadows obtained from single-qubit measurements. 
Here, we extend this approach to shallow measurements and study whether the additional information obtained from entangling measurement circuits can improve classical ML performance. 
We use kernel principal component analysis (PCA)~\cite{bishop2006pattern_58} as the classical ML algorithm, employ block shadows with block size \(k=2\), and extend the previously introduced shadow kernel~\cite{huang2022provably_55} as follows:
\begin{equation}
    k^{(\text{shadow})}(S_T(\rho), S_T(\sigma)) =\exp \left( {\frac{\tau }{{{T^2}}}\sum _{t,t' = 1}^T\exp \left( {\frac{\gamma }{{n/k}}\sum _{i = 1}^{n/k}{\rm{Tr}}\left(\hat \rho _i^{(t)}\hat \sigma _i^{(t')}\right)} \right)} \right),
\end{equation}
where \(\hat \rho_i^{(t)}\) is the \(i\)-th block of the \(t\)-th shadow sample for \(\rho\), \(\hat \sigma_i^{(t')}\) is defined similarly, and \(\tau\) and \(\gamma\) are hyperparameters. 
The computational time for evaluating \(k^{(\text{shadow})}\) is \(\mathcal{O}(4^k nT^2)\).

As in previous work~\cite{huang2022provably_55}, the performance guarantee assumes the existence of a local order parameter that distinguishes the phases. 
When this order parameter is geometrically local, shallow shadows can provide an advantage over single-qubit shadows by capturing correlations within each block (Appendix~\ref{appx:shadow_kernel}). 

\begin{figure}[!t]
  \centering
  \includegraphics[
    width=0.8\linewidth,
    trim={1cm 20cm 6.5cm 0cm},
    clip
    ]{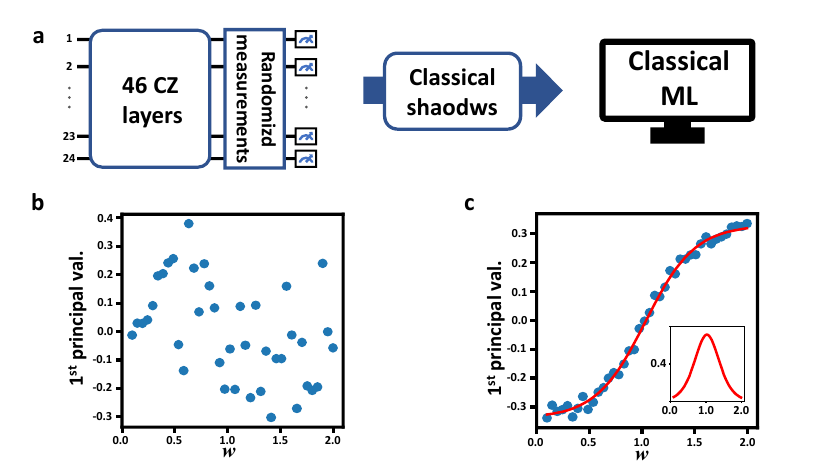}
  \caption{Experimental results of classical machine learning (ML) on quantum experimental data. (a) Schematic diagram. Classical shadows $S_T(\rho)$ are obtained using block sizes $k=1$ (single-qubit measurements) and $k=2$ (two-qubit measurements) during randomized measurements. The classical ML algorithm employed is kernel principal component analysis (PCA). After PCA, the first principal component of each data point is calculated. (b) 1D projection of each $S_T(\rho)$ obtained from $k=1$. (c) 1D projection of each $S_T(\rho)$ obtained from $k=2$. The inset shows the derivative of the fitting function, which peaks at $w=1.026\pm0.0487$. This result is consistent with the theoretically known phase transition point at $w=1$.}
  \label{fig5}
\end{figure}

To test whether shallow measurements provide an advantage on NISQ devices, we performed the experiment illustrated in Fig.~\ref{fig5}a. 
We considered the Su--Schrieffer--Heeger (SSH) Hamiltonian
\begin{equation}
    H_{\text{SSH}}(v, w)
    =
    \sum_i v(a_{2i - 1}^\dagger a_{2i} + \text{h.c.})
    +
    w(a_{2i}^\dagger a_{2i + 1} + \text{h.c.}),
\end{equation}
where the system is in a trivial phase for \(v>w\) and in a topological phase for \(v<w\). 
We prepared ground states of \(H_{\text{SSH}}\) for \(n=24\) qubits and collected training data in the form of classical shadows \(S_T(\rho)\) with \(T=30\). 
Given data from different values of \(w\), with \(v=1\) fixed, we applied kernel PCA and tracked the first principal component as \(w\) was varied from \(0.1\) to \(2\). 
No explicit information about the order parameter or the transition point was used in the training procedure.

As shown in Figs.~\ref{fig5}b and~\ref{fig5}c, the data obtained from two-qubit block measurements exhibit a sharper transition near \(w=1\) than the data obtained from single-qubit measurements. 
The derivative of the fitting function, shown in the inset of Fig.~\ref{fig5}c, has a maximum at \(w=1.026\pm0.0487\), consistent with the known transition point \(w=1\). 
These results indicate that entangling measurements can improve the classical representation of quantum experimental data for this ML task. 
Experimental details and numerical results for larger system sizes are provided in Appendix~\ref{appx:ee_ml}.

\section{Discussion}
We have shown that shallow measurements allow predicting many physical quantities with fewer measurements and obtaining more refined classical data about a quantum state than single-qubit measurements. We developed a new theoretical framework that combines the advantages of various classical shadow variants~\cite{huang2021efficient_17, chen2021robust_18, helsen2023thrifty_19, zhou2023performance_20, zhao2021fermionic_21, bertoni2024shallow_22,hu2023classical_23,hu2025demonstration_24,vermersch2024enhanced_25} and analyzed the impact of errors. We also experimentally confirmed that our approach works effectively on current noisy quantum devices. Ultimately, by leveraging improved gate fidelities, we reduce the required number of experiments and continuously extend the applicability of NISQ devices. Our work opens several pathways for future research. For instance, building on our results, exploring potential enhancements in applications~\cite{huggins2022unbiasing_28, huang2022provably_55, lewis2024improved_56, stilck2024efficient_59} using classical shadows could be a valuable direction for future research. Furthermore, integrating restricted quantum memory~\cite{chen2024optimal_60} into our approach is feasible on near-future quantum hardware, offering another promising avenue for follow-up studies.

\section*{Author contributions}
GC conceived the project, performed cloud-based experiments, and analyzed the data. GC and DK prepared the manuscript. DK supervised the project. The authors used a large language model for language-level editing and submission-checklist assistance. All scientific claims, proofs, and figures were produced and verified by the authors.

\section*{Data and code availability}
The numerical simulation code used in this work is available at \url{https://github.com/gyungmincho/shallow-rm-numerics}.

\section*{Acknowledgments}
This work was supported by a National Research Foundation of Korea (NRF) grant funded by the Korean Government (MSIT) (No. 2019M3E4A1080144, No. 2019M3E4A1080145, No. 2019R1A5A1027055, RS-2023-00283291, SRC Center for Quantum Coherence in Condensed Matter RS-2023-00207732, quantum computing technology development program No. 2020M3H3A1110365, and No. 2023R1A2C2005809, No. RS-2024-00413957, No. RS-2024-00442994), a Korea Basic Science Institute (National Research Facilities and Equipment Center) grant funded by the Ministry of Education (No. 2021R1A6C101B418).

\bibliographystyle{quantum}               
\bibliography{main}

\onecolumn\newpage
\appendix

\section{Preliminaries}\label{appx:preliminaries}

In this section, we collect the notation and elementary identities used in the
appendix. 
Throughout, \(n\) denotes the number of qubits and \(D=2^n\). 
We write \(\mathcal H^D\) for the \(D\)-dimensional Hilbert space and
\(\mathcal L(\mathcal H^D)\) for the space of linear operators on \(\mathcal H^D\).
The Schatten \(p\)-norm of an operator \(X\) is denoted by \(\|X\|_p\).

Let \(\mathcal P_n=\{I,X,Y,Z\}^{\otimes n}\) be the set of \(n\)-qubit Pauli
strings, with phases omitted. 
For \(A\subseteq[n]\), let \(D_A=2^{|A|}\), and for a block size \(k\), let
\(D_k=2^k\). 
We denote by \(\mathrm{Cl}(m)\) the Clifford group on \(m\) qubits. 
For block shadows, \(k\) denotes the block size, and for brickwork shadows,
\(d_{\mathrm{RM}}\) denotes the two-qubit circuit depth of the randomized
measurement circuit.

We also define
\[
    \mathcal Z=\{I,Z\}^{\otimes n},
    \qquad
    (U,P,b)
    :=
    \operatorname{Tr}
    \!\left(
        UPU^\dagger |b\rangle\langle b|
    \right),
\]
where \(P\in\mathcal P_n\), \(U\) is a measurement unitary, and
\(b\in\{0,1\}^n\) is a computational-basis outcome.

\subsection{Projective state designs}

We use standard moment identities for Haar-random pure states. 
Let $\rho=\ket{\psi}\!\bra{\psi}$ be a Haar-random pure state on $\mathcal H^D$. 
For a positive integer $t$, its $t$-th moment is
\begin{equation}
    \mathbb E_\rho[\rho^{\otimes t}]
    =
    \frac{1}{D(D+1)\cdots(D+t-1)}
    \sum_{\sigma\in S_t} U_\sigma .
\end{equation}
Here, $S_t$ is the symmetric group on $t$ elements, and
$U_\sigma\in\mathcal L((\mathcal H^D)^{\otimes t})$ is defined by
\begin{equation}
    U_\sigma
    \ket{i_1}\ket{i_2}\cdots\ket{i_t}
    =
    \ket{i_{\sigma^{-1}(1)}}\ket{i_{\sigma^{-1}(2)}}\cdots
    \ket{i_{\sigma^{-1}(t)}} .
\end{equation}
Equivalently, $\mathbb E_\rho[\rho^{\otimes t}]$ is the normalized projector
onto the symmetric subspace of $(\mathcal H^D)^{\otimes t}$.

More generally, an ensemble of pure states is called a projective $t$-design
if its $t$-th moment agrees with the Haar moment above. 
Therefore, every Haar-average calculation below that only involves
$\rho^{\otimes t}$ remains valid if the average over $\rho$ is replaced by an exact projective $t$-design.

We will use the following trace identity repeatedly. 
For operators $A_1,\ldots,A_t\in\mathcal L(\mathcal H^D)$,
\begin{equation}
    \operatorname{Tr}\!\left[
        U_\sigma(A_1\otimes\cdots\otimes A_t)
    \right]
    =
    \prod_{(i_1\cdots i_\ell)\in\sigma}
    \operatorname{Tr}\!\left(
        A_{i_\ell}\cdots A_{i_1}
    \right),
\end{equation}
where the product is taken over the disjoint cycles of $\sigma$. 
For example, the second moment gives
\begin{equation}
    \mathbb E_\rho[\rho^{\otimes 2}]
    =
    \frac{1}{D(D+1)}
    \sum_{\sigma\in S_2}U_\sigma ,
\end{equation}
and hence
\begin{equation}
    \mathbb E_\rho[
        \operatorname{Tr}(\rho A)
        \operatorname{Tr}(\rho B)
    ]
    =
    \frac{
        \operatorname{Tr}(A)\operatorname{Tr}(B)
        +
        \operatorname{Tr}(AB)
    }{
        D(D+1)
    } .
\end{equation}
Similarly, the third moment used below is
\begin{equation}
    \mathbb E_\rho[\rho^{\otimes 3}]
    =
    \frac{1}{D(D+1)(D+2)}
    \sum_{\sigma\in S_3}U_\sigma ,
\end{equation}
and the fourth moment is
\begin{equation}
    \mathbb E_\rho[\rho^{\otimes 4}]
    =
    \frac{1}{D(D+1)(D+2)(D+3)}
    \sum_{\sigma\in S_4}U_\sigma .
\end{equation}

We can also consider state averages at the density-matrix level by using the
standard induced ensemble of random density matrices. 
Let \(E\) be an environment with \(\dim E=K\), and let
\(\ket{\Psi}\in\mathcal H^D\otimes E\) be a Haar-random pure state. 
The reduced state
\begin{equation}
    \rho
    =
    \operatorname{Tr}_{E}
    \!\left(
        \ket{\Psi}\!\bra{\Psi}
    \right)
\end{equation}
defines an induced ensemble on \(\mathcal H^D\). 
Its moments are obtained by applying the Haar moment formula on the enlarged
Hilbert space and tracing out the environment:
\begin{align}
    \mathbb E_{\rho}^{(K)}[\rho^{\otimes t}]
    &=
    \operatorname{Tr}_{E^{\otimes t}}
    \!\left[
        \mathbb E_{\Psi}
        \left[
            \left(
                \ket{\Psi}\!\bra{\Psi}
            \right)^{\otimes t}
        \right]
    \right]  \\
    &=
    \frac{1}{DK(DK+1)\cdots(DK+t-1)}
    \sum_{\sigma\in S_t}
    K^{c(\sigma)} U_\sigma ,
\end{align}
where \(c(\sigma)\) is the number of cycles in \(\sigma\). 
Here \(U_\sigma\) acts on the \(t\) copies of the system Hilbert space
\(\mathcal H^D\).

\subsection{Probabilistic interpretation of the shadow norm}
\label{appx:A1}

For a random unitary ensemble \(\mathbb U\), the shadow channel is defined by
\begin{equation}
    \mathcal M_{\mathbb U}(\rho)
    =
    \mathbb E_{U\sim\mathbb U}
    \sum_{b\in\{0,1\}^n}
    U^\dagger |b\rangle\langle b|U\,
    \operatorname{Tr}
    \!\left(
        U\rho U^\dagger |b\rangle\langle b|
    \right).
\end{equation}
Given an outcome \(b\) obtained after applying \(U\), the corresponding
classical-shadow estimator is
\begin{equation}
    \hat\rho
    =
    \mathcal M_{\mathbb U}^{-1}
    \!\left(
        U^\dagger |b\rangle\langle b|U
    \right).
\end{equation}
For an observable \(O=\sum_{P\in\mathcal P_n}\alpha_P P\), the single-shot
estimator of \(\operatorname{Tr}(O\rho)\) is
\begin{align}
    \hat o
    &=
    \operatorname{Tr}(O\hat\rho) \\
    &=
    \operatorname{Tr}
    \!\left(
        \mathcal M_{\mathbb U}^{-1}(O)
        U^\dagger |b\rangle\langle b|U
    \right) \\
    &=
    \sum_{P\in\mathcal P_n}
    \alpha_P
    \operatorname{Tr}
    \!\left(
        \mathcal M_{\mathbb U}^{-1}(P)
        U^\dagger |b\rangle\langle b|U
    \right).
\end{align}
Thus, for Pauli observables, it is enough to understand the action of
\(\mathcal M_{\mathbb U}\) on Pauli strings.

Assume now that \(\mathbb U\) consists of Clifford unitaries. 
Then the following identities hold for any fixed \(U\in\mathbb U\).
\begin{itemize}
    \item For any Pauli string \(P\),
    \begin{equation}
        (U,P,b)
        =
        (U,P,b)\,
        \mathbf 1\{UPU^\dagger\in\pm\mathcal Z\}.
    \end{equation}

    \item For Pauli strings \(P_1,\ldots,P_r\),
    \begin{equation}\label{appx:eq:pro2}
        \sum_{b\in\{0,1\}^n}
        \prod_{\ell=1}^r (U,P_\ell,b)
        =
        D\,
        \mathbf 1\!\left\{\prod_{\ell=1}^r P_\ell=\pm I\right\}
        \prod_{\ell=1}^r
        \mathbf 1\{UP_\ell U^\dagger\in\pm\mathcal Z\}.
    \end{equation}
\end{itemize}
The first identity simply states that only Pauli strings diagonal in the
computational basis can contribute to a computational-basis measurement. 
The second identity is the corresponding orthogonality relation after conjugation
by \(U\).

Using these properties, we obtain, for any \(P\in\mathcal P_n\),
\begin{align}
    \mathcal M_{\mathbb U}(P)
    &=
    \mathbb E_U
    \sum_b
    U^\dagger |b\rangle\langle b|U\,
    \operatorname{Tr}
    \!\left(
        UPU^\dagger |b\rangle\langle b|
    \right) \\
    &=
    \mathbb E_U
    \sum_b
    \sum_{Q\in\mathcal P_n}
    \frac{1}{D}
    \operatorname{Tr}
    \!\left(
        UQU^\dagger |b\rangle\langle b|
    \right)
    Q\,
    \operatorname{Tr}
    \!\left(
        UPU^\dagger |b\rangle\langle b|
    \right) \\
    &=
    \mathbb E_U
    \!\left[
        \mathbf 1\{UPU^\dagger\in\pm\mathcal Z\}
    \right]P.
\end{align}
In the last line, we used Eq.~\eqref{appx:eq:pro2}. 
Hence \(P\) is an eigenoperator of the shadow channel with eigenvalue
\begin{equation}
    m_P
    :=
    \mathbb E_U
    \!\left[
        \mathbf 1\{UPU^\dagger\in\pm\mathcal Z\}
    \right]
    =
    \Pr_U(UPU^\dagger\in\pm\mathcal Z).
\end{equation}
In particular,
\begin{equation}
    \mathcal M_{\mathbb U}^{-1}(P)=m_P^{-1}P
\end{equation}
whenever \(m_P\neq0\). 

The state-dependent shadow norm of a Pauli string is then
\begin{align}
    \|P\|_{\mathrm{sh},\rho}^{2}
    &=
    \mathbb E_{U,b}
    \operatorname{Tr}
    \!\left(
        \mathcal M_{\mathbb U}^{-1}(P)
        U^\dagger |b\rangle\langle b|U
    \right)^2 \\
    &=
    \mathbb E_U
    \sum_b
    m_P^{-2}
    (U,P,b)^2
    \operatorname{Tr}
    \!\left(
        U\rho U^\dagger |b\rangle\langle b|
    \right) \\
    &=
    m_P^{-1}.
\end{align}
Thus, the usual Pauli-eigenoperator formula for Pauli-invariant ensembles also
holds for Clifford ensembles. For the block Clifford ensemble \(\mathbb U=\mathrm{Cl}(k)^{\otimes n/k}\), $m_P^{-1}$ is upper-bounded as follows:
\begin{equation}
    m_P^{-1}
    =
    (2^k+1)^{w_k(P)}
    \le
    (2^k+1)^{n/k}
    =
    2^n(1+2^{-k})^{n/k}
    \le
    2^n e^{n/(k2^k)}
    .
\end{equation}
In particular, when \(k2^k=n/\alpha\), we have
\begin{equation}
    m_P^{-1}\le 2^n e^\alpha .
\end{equation}
For Pauli string $P$ with contiguous support of length $\ell$, the shadow norm is $\|P\|_{\text{sh}}^2/2^n \leq (2^k+1)^{\ell/k+1} = \mathcal O(\ell2^\ell)$ where $k2^k = \Omega(\ell)$ and $k \geq 2$, confirming the result predicted by the mean-field approximation in the brickwork shadow case~\cite{ippoliti2023operator_27}.

For a general observable \(O=\sum_P\alpha_P P\), the same calculation gives
\begin{align}
    \|O\|_{\mathrm{sh},\rho}^{2}
    &=
    \sum_{P,Q\in\mathcal P_n}
    \alpha_P\alpha_Q
    \operatorname{Tr}(\rho PQ)
    f(P,Q),
    \label{appx:eq:sh_norm_o}
\end{align}
where
\begin{equation}
    f(P,Q)
    :=
    \frac{
    \Pr_U
    \!\left(
        UPU^\dagger,\,
        UQU^\dagger
        \in\pm\mathcal Z
    \right)}
    {m_Pm_Q}.
\end{equation}
The function \(f(P,Q)\) depends on the unitary ensemble and will be the main quantity controlling the variance calculations below.

For the global Clifford ensemble \(\mathbb U=\mathrm{Cl}(n)\), one obtains
\begin{equation}\label{eq:appx:f}
    f(P,Q)=
    \begin{cases}
    1,
    & \text{if } P=I \text{ or } Q=I, \\[4pt]
    D+1,
    & \text{if } P=Q\neq I, \\[4pt]
    \dfrac{2(D+1)}{D+2},
    & \text{if } P\neq I,\ Q\neq I,\ P\neq Q,\ PQ=QP, \\[6pt]
    0,
    & \text{if } P\neq I,\ Q\neq I,\ P\neq Q,\ PQ\neq QP .
    \end{cases}
\end{equation}

For product ensembles, \(f(P,Q)\) factorizes. 
For example, for \(\mathrm{Cl}(1)^{\otimes n}\),
\begin{equation}
    f(P,Q)
    =
    \prod_{i=1}^{n} f(P_i,Q_i),
\end{equation}
where the same symbol \(f\) is used for the corresponding single-qubit function.
For block shadows, we assume for simplicity that the \(n\) qubits are partitioned
into \(n/k\) blocks of size \(k\). Then
\begin{equation}
    f(P,Q)
    =
    \prod_{i=1}^{n/k} f(P_i,Q_i),
\end{equation}
where \(P_i,Q_i\in\mathcal P_k\) denote the restrictions of \(P,Q\) to the
\(i\)-th block. 
This factorization is one of the main technical advantages of block shadows:
although \(f(P,Q)\) is difficult to compute explicitly for general brickwork
circuits, it is directly computable for block shadows.

We will also use the following elementary sums for block shadows. 
Assume \(\mathbb U=\mathrm{Cl}(k)^{\otimes n/k}\) and \(k2^k=n/\alpha\). Then
\begin{align}
    \bullet\quad
    &\sum_{P\in\mathcal P_n} f(P,I)
    =
    D^2, \\
    \bullet\quad
    &\sum_{P\in\mathcal P_n} f(P,P)
    =
    \sum_{P\in\mathcal P_n}m_P^{-1}
    =
    \sum_{\ell=0}^{n/k}
    \binom{n/k}{\ell}
    (2^k+1)^\ell(4^k-1)^\ell \notag \\
    &\quad
    =
    (8^k+4^k-2^k)^{n/k}
    \le
    D^3
    \left(1+\frac{1}{2^k}\right)^{n/k}
    \le
    D^3e^\alpha, \\
    \bullet\quad
    &\sum_{P,Q\in\mathcal P_n} f(P,Q)
    =
    (16^k)^{n/k} 
    =
    D^4.
\end{align}

Finally, we define the variance quantities used in the following sections. 
For \(O=\sum_P\alpha_P P\), set
\begin{align}
    V_1(O,\rho)
    &:=
    \sum_{P,Q\in\mathcal P_n}
    \alpha_P\alpha_Q\,
    \operatorname{Tr}(\rho P)\operatorname{Tr}(\rho Q)\,
    f(P,Q),\\
    V_2(O,\rho)
    &:=
    \sum_{P,Q\in\mathcal P_n}
    \alpha_P\alpha_Q\,
    \operatorname{Tr}(\rho PQ)\,
    f(P,Q).
\end{align}
For purity and inner-product estimators, we also use
\begin{equation}
    V_3(\rho)
    :=
    \sum_{P,Q\in\mathcal P_n}
    \frac{\operatorname{Tr}(\rho PQ)^2}{D^2}\,
    f(P,Q).
\end{equation}
When \(O=\rho\), we use the shorthand
\(V_1(\rho):=V_1(O=\rho,\rho)\) and
\(V_2(\rho):=V_2(O=\rho,\rho)\). 
Equivalently, using
\begin{equation}
    \rho
    =
    \frac{1}{D}
    \sum_{P\in\mathcal P_n}
    \operatorname{Tr}(\rho P)P,
\end{equation}
we have
\begin{align}
    V_1(\rho)
    &=
    \sum_{P,Q\in\mathcal P_n}
    \frac{
    \operatorname{Tr}(\rho P)^2
    \operatorname{Tr}(\rho Q)^2
    }{D^2}
    f(P,Q),\\
    V_2(\rho)
    &=
    \sum_{P,Q\in\mathcal P_n}
    \frac{
    \operatorname{Tr}(\rho P)
    \operatorname{Tr}(\rho Q)
    \operatorname{Tr}(\rho PQ)
    }{D^2}
    f(P,Q).
\end{align}

For later use, we record the relation
\begin{equation}
    V_1(\rho)\le V_2(\rho)\le V_3(\rho).
\end{equation}
Indeed, the three quantities can be expressed in terms of the same
two-shot overlap. 
For a fixed \(U\), let \(b_1,b_2\) be independent measurement outcomes sampled
from \(U\rho U^\dagger\). Then
\begin{align}
    V_1(\rho)
    &=
    \mathbb E_U
    \left[
    \mathbb E_{b_1,b_2|U}
    \operatorname{Tr}\!\left(
        U^\dagger |b_1\rangle\langle b_1|U\,
        \mathcal M^{-1}
        \!\left(
            U^\dagger |b_2\rangle\langle b_2|U
        \right)
    \right)
    \right]^2,\\
    V_2(\rho)
    &=
    \mathbb E_U
    \mathbb E_{b_1|U}
    \left[
    \mathbb E_{b_2|U}
    \operatorname{Tr}\!\left(
        U^\dagger |b_1\rangle\langle b_1|U\,
        \mathcal M^{-1}
        \!\left(
            U^\dagger |b_2\rangle\langle b_2|U
        \right)
    \right)
    \right]^2,\\
    V_3(\rho)
    &=
    \mathbb E_U
    \mathbb E_{b_1,b_2|U}
    \left[
    \operatorname{Tr}\!\left(
        U^\dagger |b_1\rangle\langle b_1|U\,
        \mathcal M^{-1}
        \!\left(
            U^\dagger |b_2\rangle\langle b_2|U
        \right)
    \right)^2
    \right].
\end{align}
The first two identities differ only in the order in which the square and the
average over \(b_1\) are taken. Hence, by Cauchy--Schwarz inequality,
\begin{align}
    &\left[
    \mathbb E_{b_1|U}
    \mathbb E_{b_2|U}
    \operatorname{Tr}\!\left(
        U^\dagger |b_1\rangle\langle b_1|U\,
        \mathcal M^{-1}
        \!\left(
            U^\dagger |b_2\rangle\langle b_2|U
        \right)
    \right)
    \right]^2 \notag\\
    &\le
    \mathbb E_{b_1|U}
    \left[
    \mathbb E_{b_2|U}
    \operatorname{Tr}\!\left(
        U^\dagger |b_1\rangle\langle b_1|U\,
        \mathcal M^{-1}
        \!\left(
            U^\dagger |b_2\rangle\langle b_2|U
        \right)
    \right)
    \right]^2.
\end{align}
Averaging over \(U\) gives
\begin{equation}
    V_1(\rho)\le V_2(\rho).
\end{equation}
Similarly, for each fixed \(U\) and \(b_1\), Cauchy--Schwarz inequality gives
\begin{align}
    &\left[
    \mathbb E_{b_2|U}
    \operatorname{Tr}\!\left(
        U^\dagger |b_1\rangle\langle b_1|U\,
        \mathcal M^{-1}
        \!\left(
            U^\dagger |b_2\rangle\langle b_2|U
        \right)
    \right)
    \right]^2 \notag\\
    &\le
    \mathbb E_{b_2|U}
    \left[
    \operatorname{Tr}\!\left(
        U^\dagger |b_1\rangle\langle b_1|U\,
        \mathcal M^{-1}
        \!\left(
            U^\dagger |b_2\rangle\langle b_2|U
        \right)
    \right)^2
    \right].
\end{align}
Averaging over \(b_1\) and \(U\) gives
\begin{equation}
    V_2(\rho)\le V_3(\rho).
\end{equation}
Therefore,
\begin{equation}
    V_1(\rho)\le V_2(\rho)\le V_3(\rho).
\end{equation}
With the same procedures as above, we can prove $V_1(O,\rho) \le V_2(O,\rho)$.

\section{Proofs of key results}
\subsection{Proof of the Fact~\ref{fact1}}
\label{appx:fact1}

\begin{fact*}
    Let \(\mathbb{U}_{\mathrm{MUB}}=\{U_\ell\}_{\ell=1}^{2^n+1}\) be a set of Clifford unitaries such that \(\{U_\ell^\dagger\ket{b}: b\in\{0,1\}^n\}\), for \(\ell=1,\ldots,2^n+1\), forms a complete set of mutually unbiased bases on \(n\) qubits. Then \(\mathcal{M}_{\mathbb{U}_{\mathrm{MUB}}}=\mathcal{M}_{\mathrm{Cl}(n)}\)~\cite{zhang2024minimal_36}.
\end{fact*}

\begin{proof}
    It is sufficient to prove that $\mathcal M_{\text{MUB}}(P)=\mathcal M_{\text{Cl}(n)}(P)$ holds for all Pauli strings $P$. For $P = I$, it is trivial. For $P \neq I$, we have
    \begin{equation}
          \mathcal{M}_{\text{Cl}(n)}(P)= (2^n + 1)^{-1}P.
    \end{equation}
    Because $UPU^{\dagger}\in\pm\mathcal{Z}$ must be satisfied by exactly one $U \in \mathbb{U}_{\text{MUB}}$, the $m_P$ from $\mathbb{U}_{\text{MUB}}$ becomes
    \begin{equation}
        m_P = 1/|\mathbb{U}_{\text{MUB}}| = 1/(2^n + 1),
    \end{equation}
    where $|\mathbb{U}_{\text{MUB}}|=2^n+1$. Consequently, $\mathcal M_{\text{MUB}}(P)=\mathcal M_{\mathrm{Cl}(n)}(P)$ holds for all $P$, then $\mathcal M_{\text{MUB}}=\mathcal M_{\text{Cl}(n)}$.
\end{proof}

\subsection{Proof of Fact~\ref{fact2}}
\label{appx:fact2}

\begin{fact*}
    There exists a Clifford ensemble \(\mathbb{U}_0\subseteq\mathrm{Cl}(n)\) with \(|\mathbb{U}_0|\leq\prod_{i=1}^{n}(2^i+1)\) such that, for every observable \(O\), \(\lVert O\rVert_{\mathrm{sh},\mathbb{U}_0}=\lVert O\rVert_{\mathrm{sh},\mathrm{Cl}(n)}\).
\end{fact*}

\begin{proof}
    It is enough to construct an ensemble \(\mathbb{U}_0\) for which \(f(P,Q;\mathbb{U}_0)=f(P,Q;\mathrm{Cl}(n))\) for all Pauli strings \(P,Q\), where \(f(P,Q;\mathbb U)\) denotes the function \(f(P,Q)\) associated with the ensemble \(\mathbb U\).

    Let \(\mathcal{L}\) be the set of maximal commuting Pauli subgroups, with phases omitted. 
    Equivalently, each \(L\in\mathcal{L}\) specifies one stabilizer measurement basis. 
    The number of such subgroups is
    \begin{equation}
        |\mathcal{L}|
        =
        \prod_{i=1}^{n}(2^i+1).
    \end{equation}
    For each \(L\in\mathcal{L}\), choose one Clifford unitary \(U_L\) satisfying
    \begin{equation}
        U_L^\dagger \mathcal{Z} U_L = L,
    \end{equation}
    and define
    \begin{equation}
        \mathbb{U}_0
        =
        \{U_L:L\in\mathcal{L}\}.
    \end{equation}
    Then \(|\mathbb{U}_0|=|\mathcal{L}|\leq\prod_{i=1}^{n}(2^i+1)\).

    We now note that the events entering \(f(P,Q)\) depend on \(U\) only through the subgroup \(U^\dagger\mathcal ZU\). 
    Indeed,
    \begin{align}
        UPU^\dagger\in\pm\mathcal Z
        \quad
        &\Longleftrightarrow
        \quad
        P\in\pm U^\dagger\mathcal ZU,\\
        UPU^\dagger,\,UQU^\dagger\in\pm\mathcal Z
        \quad
        &\Longleftrightarrow
        \quad
        P,Q\in\pm U^\dagger\mathcal ZU.
    \end{align}
    Therefore, the probabilities appearing in \(f(P,Q)\) are determined only by the induced distribution of \(U^\dagger\mathcal ZU\) over \(\mathcal L\).

    For \(U\sim\mathbb{U}_0\), the subgroup \(U^\dagger\mathcal ZU\) is uniform over \(\mathcal L\) by construction. 
    For \(U\sim\mathrm{Cl}(n)\), the same distribution is obtained because the Clifford group acts transitively on stabilizer measurement bases, and each \(L\in\mathcal L\) has the same number of Clifford preimages. 
    Hence the probabilities defining \(f(P,Q)\) are identical for \(\mathbb U_0\) and \(\mathrm{Cl}(n)\). 
    Thus,
    \begin{equation}
        f(P,Q;\mathbb{U}_0)
        =
        f(P,Q;\mathrm{Cl}(n))
    \end{equation}
    for all Pauli strings \(P,Q\).

    Using Eq.~\eqref{appx:eq:sh_norm_o}, we obtain
    \begin{equation}
        \|O\|_{\mathrm{sh},\mathbb{U}_0}
        =
        \|O\|_{\mathrm{sh},\mathrm{Cl}(n)}
    \end{equation}
    for every observable \(O\), as claimed.
\end{proof}

\subsection{An average confidence in block shadow}
\label{appx:derand_conf}

Let ${\bf{P}} = \{ {P_l}\} _{l = 1}^L$ be a set of Pauli strings and  ${\bf{B}} = \{ {U_m}\} _{m = 1}^M$ be a set of measurement basis. Define $M_l(\textbf{B}) = {\sum _m}{\bf{1}}\{ {U_m}{P_l}U_m^\dagger  \in  \pm {\cal Z}\}$ and ${\rm{CON}}{{\rm{F}}_\varepsilon }({\bf{P}};{\bf{B}}) \coloneqq \sum _{l = 1}^L\exp \left( { - \frac{{{\varepsilon ^2}}}{2}{M_l}({\bf{B}})} \right)$, where $\mathcal{Z}=\{I,Z\}^n$. Then following holds
\begin{equation}
    {\rm{CON}}{{\rm{F}}_\varepsilon }({\bf{P}}{\rm{; }}{{\bf{B}}^\# })\leq \mathbb{E}{_{\bf{B}}}[{\rm{CON}}{{\rm{F}}_\varepsilon }({\bf{P}}{\rm{; }}{\bf{B}})]=\sum _{l = 1}^L{(1 - \nu /{({2^k} + 1)^{{w_k}({P_l})}})^M},
\end{equation}
where $\nu = 1 - \exp(-\epsilon^2/2)$.
\begin{proof}
    Closed form expression for  $
    \mathbb{E}{_{\bf{B}}}[{\rm{CON}}{{\rm{F}}_\epsilon }({\bf{P}}{\rm{; }}{\bf{B}})]$ is as follows\\
    \begin{align}
        \mathbb{E}_{\mathbf B}[\mathrm{CONF}_\varepsilon(\mathbf P;\mathbf B)]
        &=\mathbb{E}_{\mathbf B}\!\left[\sum_{l=1}^L
        \exp\!\left(-\frac{\varepsilon^2}{2}M_l(\mathbf B)\right)\right]\\
        &=\mathbb{E}_{\mathbf B}\!\left[\sum_{l=1}^L\prod_{m=1}^M
        \exp\!\left(-\frac{\varepsilon^2}{2}\mathbf1\{U_mP_lU_m^\dagger\in\pm\mathcal Z\}\right)\right]\\
        &=\mathbb{E}_{\mathbf B}\!\left[\sum_{l=1}^L\prod_{m=1}^M
        (1-\nu\,\mathbf1\{U_mP_lU_m^\dagger\in\pm\mathcal Z\})\right]\\
        &=\sum_{l=1}^L\mathbb{E}_{\mathbf B}\!\left[\prod_{m=1}^M
        (1-\nu\,\mathbf1\{U_mP_lU_m^\dagger\in\pm\mathcal Z\})\right]\\
        &=\sum_{l=1}^L\left(1-\frac{\nu}{(2^k+1)^{w_k(P_l)}}\right)^M
    \end{align}
\end{proof}

\subsection{State-averaged variance for state-independent observables}
\label{appx:A5}

\begin{proof}
    Let $O$ be Hermitian, $\rho$ be a pure quantum state, $k2^k=n/\alpha$, and the unitary ensemble
    $\mathbb U=\mathrm{Cl}(k)^{\otimes n/k}$. Let
    $O=\sum_P\alpha_P P$ and define
    $\hat o=\mathrm{Tr}(O\hat\rho)$, then 
    \begin{align}
        {\rm{Var}}(\hat o) 
        &\leq {\mathbb{E}_{U,b}}[{{\hat o}^2}]\\
        &= {\sum _{P,Q}}{\alpha _P}{\alpha _Q}{\rm{Tr}}(\rho PQ)f(P,Q).
    \end{align}
    When taking the average over $\rho$, the resulting mean value becomes the same as the value for $\rho = I/2^n $~\cite{bertoni2024shallow_22}.
    \begin{align}
        \mathbb{E}_\rho[\mathrm{Var}(\hat o)]
        &\le
        \mathbb{E}_\rho\!\left[\sum_{P,Q}\alpha_P\alpha_Q
        \mathrm{Tr}(\rho PQ)\,f(P,Q)\right]\\
        &=\sum_P\alpha_P^2f(P,P)=\sum_P\alpha_P^2m_P^{-1}\\
        &\le\frac{(2^k+1)^{n/k}}{2^n}\,\|O\|_2^2\le e^\alpha\|O\|_2^2
    \end{align}
    Because the circuit depth $d$ required to implement a random Clifford gate from Cl($k$) is $d = ck$, where $c$ depends on the connectivity of the hardware and $k = \mathcal{O}(\log n)$, it follows that $d = \mathcal{O}(\log n)$. This is consistent with previously known results~\cite{bertoni2024shallow_22}.
\end{proof}

\subsection{State-averaged variance for state-dependent observables}
\label{appx:state_dependent_variance}

\begin{proof}
    Let $\rho$ be a pure quantum state, $k2^k = n/\alpha$, and unitary ensemble $\mathbb{U} = \text{Cl}(k)^{\otimes n/k}$. In the case of $O = \rho$, one must exercise caution when computing the average over $\rho$.
    \begin{align}
        \mathrm{Var}(\hat o)
        &\le\sum_{P,Q}\alpha_P\alpha_Q\mathrm{Tr}(\rho PQ)\,f(P,Q)\\
        &=\sum_{P,Q}\mathrm{Tr}(\rho P)\mathrm{Tr}(\rho Q)
        \mathrm{Tr}(\rho PQ)\,f(P,Q)/4^n
    \end{align}
    As shown in the above equation, unlike in~\ref{appx:A5}, we must use a state 3-design to compute the average instead of state 1-design. By employing the property of the projective 3-designs~\cite{mele2024introduction},
    \begin{equation}
        {\mathbb{E}_\rho }[{\rho ^{ \otimes 3}}] = \frac{1}{{D(D + 1)(D + 2)}}{\sum _{\sigma  \in {S_3}}}{U_\sigma }.
    \end{equation}
    We can calculate the average as follows:
    \begin{align}
        \mathbb{E}_\rho[\mathrm{Var}(\hat o)]
        &\le\frac{1}{D^3(D+1)(D+2)}
        \sum_{P,Q}f(P,Q)\,
        \sum_{\sigma\in S_3}
        \mathrm{Tr}\!\big(U_\sigma(P\otimes Q\otimes PQ)\big)\\
        &=\frac{1}{D^3(D+1)(D+2)}
        \sum_{P,Q}f(P,Q)\big[
        \mathrm{Tr}(P)\mathrm{Tr}(Q)\mathrm{Tr}(PQ) \nonumber\\
        &\qquad\qquad\qquad\qquad\qquad
        +\mathrm{Tr}(PQ)^2
        +\mathrm{Tr}(P)^2
        +\mathrm{Tr}(Q)^2
        +2\mathrm{Tr}(P^2Q^2)
        \big]\\
        &\le\frac{1}{D^3(D+1)(D+2)}
        \big(D^3+2D^4+D^5e^\alpha+2D^5\big)\\
        &\leq 2+e^\alpha+\mathcal{O}(1/D) \label{appx:eq:ub_O}.
    \end{align}
    If the quantum state undergoes a state-independent unital channel \(C\), the same bound still holds. 
    Indeed, the above calculation is repeated after replacing each Pauli string by its image under \(C^\dagger\). 
    Since \(C^\dagger\) is also unital, we have \(\|C^\dagger(P)\|_2\le \sqrt{D}\) for every Pauli string \(P\). 
    Thus, each term is bounded in the same way as above, and the averaged variance remains upper-bounded by Eq.~\eqref{appx:eq:ub_O}.   
\end{proof}

\subsection{Proof of the $\left\| O  \right\|_{{\rm{sh},\rho}}^2
= {\cal O}(3^{n/k}\|O\|_2^2$)}
\label{appx:worst_shadow_norm}

\begin{proof}
    Consider the block Clifford ensemble
    \begin{equation}
        \mathbb U=\text{Cl}(k)^{\otimes n/k},
        \qquad
        \mathcal M=\mathcal M_k^{\otimes n/k}.
    \end{equation}
    We prove that, for every state \(\rho\) and every Hermitian observable \(O\),
    \begin{equation}
        \|O\|_{\mathrm{sh,\rho}}^2
        \le
        3^{n/k}\|O\|_2^2 .
    \end{equation}

    We first recall the standard Clifford shadow-norm bound from
    Ref.~\cite{huang2020predicting_13}. For a single \(k\)-qubit block, for
    every Hermitian \(X\in\mathcal L(\mathbb C^{2^k})\) and every
    state \(\sigma\ge0\),
    \begin{equation}\label{appx:eq:single_block_shadow_bound}
        \mathbb E_{U}\sum_b
        {\rm Tr}\!\left(
            \sigma\,U^\dagger |b\rangle\langle b|U
        \right)
        {\rm Tr}\!\left[
            X\,\mathcal M_k^{-1}
            \!\left(
                U^\dagger |b\rangle\langle b|U
            \right)
        \right]^2
        \le
        3\,{\rm Tr}(\sigma)\,\|X\|_2^2 .
    \end{equation}

    We now prove the desired result by induction on the number of blocks.
    The case \(n/k=1\) follows from
    Eq.~\eqref{appx:eq:single_block_shadow_bound} with \(\sigma=\rho\).

    Assume the claim holds for \(n/k-1\) blocks. Separate the first block from
    the remaining blocks. For the remaining blocks, write
    \begin{equation}
        \Phi_R=\bigotimes_{i=2}^{n/k}\phi_i,
        \qquad
        Y_R=\bigotimes_{i=2}^{n/k}\mathcal M_k^{-1}(\phi_i),
    \end{equation}
    where $\phi_i = U_i^{\dagger}\ket{b_i}\bra{b_i}U_i$. Conditioned on \(\Phi_R\), define the subnormalized state on the first block
    by
    \begin{equation}
        \rho_1(\Phi_R)
        =
        {\rm Tr}_R\!\left[
            (I_k\otimes \Phi_R)\rho
        \right].
    \end{equation}
    Then
    \begin{equation}
        {\rm Tr}[\rho(\phi_1\otimes\Phi_R)]
        =
        {\rm Tr}[\rho_1(\Phi_R)\phi_1].
    \end{equation}

    For fixed \(Y_R\), define
    \begin{equation}
        O_1(Y_R)
        =
        \mathrm{Tr}_R(O(I_1\otimes Y_R))
        =
        \frac{1}{2^k}
        \sum_{S\in\mathcal P_k}
        {\rm Tr}[O(S\otimes Y_R)]S .
    \end{equation}
    Then, with \(Y_1=\mathcal M_k^{-1}(\phi_1)\),
    \begin{equation}
        {\rm Tr}[O_1(Y_R)Y_1]
        =
        {\rm Tr}[O(Y_1\otimes Y_R)] .
    \end{equation}
    Applying Eq.~\eqref{appx:eq:single_block_shadow_bound} to the first block,
    we obtain
    \begin{align}
        \|O\|_{\mathrm{sh,\rho}}^2
        &\le
        3\,
        \mathbb E_{U_R}\sum_{b_R}
        {\rm Tr}(\rho_R\Phi_R)
        \|O_1(Y_R)\|_2^2,
    \end{align}
    where \(\rho_R={\rm Tr}_1(\rho)\).

    By Pauli orthogonality on the first block,
    \begin{equation}
        \|O_1(Y_R)\|_2^2
        =
        \frac{1}{2^k}
        \sum_{S\in\mathcal P_k}
        {\rm Tr}[O(S\otimes Y_R)]^2 .
    \end{equation}
    Define
    \begin{equation}
        O_S
        =
        {\rm Tr}_1[(S\otimes I_R)O].
    \end{equation}
    Then
    \begin{equation}
        {\rm Tr}[O(S\otimes Y_R)]
        =
        {\rm Tr}(O_S Y_R).
    \end{equation}
    Therefore,
    \begin{align}
        \|O\|_{\mathrm{sh,\rho}}^2
        &\le
        \frac{3}{2^k}
        \sum_{S\in\mathcal P_k}
        \mathbb E_{U_R}\sum_{b_R}
        {\rm Tr}(\rho_R\Phi_R)
        {\rm Tr}(O_S Y_R)^2 \\
        &=
        \frac{3}{2^k}
        \sum_{S\in\mathcal P_k}
        \|O_S\|_{\mathrm{sh,\rho_R}}^2 .
    \end{align}
    By the induction hypothesis applied to the remaining \(n/k-1\) blocks,
    \begin{equation}
        \|O_S\|_{\mathrm{sh,\rho_R}}^2
        \le
        3^{n/k-1}{\rm Tr}(\rho_R)\|O_S\|_2^2
        =
        3^{n/k-1}\|O_S\|_2^2
        .
    \end{equation}
    Hence
    \begin{equation}
        \|O\|_{\mathrm{sh,\rho}}^2
        \le
        3^{n/k}
        \frac{1}{2^k}
        \sum_{S\in\mathcal P_k}
        \|O_S\|_2^2 .
    \end{equation}
    Finally, from the Pauli expansion on the first block,
    \begin{equation}
        O=\frac{1}{2^k}\sum_{S\in\mathcal P_k}S\otimes O_S,
    \end{equation}
    we have
    \begin{equation}
        \|O\|_2^2
        =
        \frac{1}{2^k}
        \sum_{S\in\mathcal P_k}\|O_S\|_2^2 .
    \end{equation}
    Therefore
    \begin{equation}
        \|O\|_{\mathrm{sh,\rho}}^2
        \le
        3^{n/k}\|O\|_2^2 .
    \end{equation}
\end{proof}

\subsection{Tightness of $\left\| O  \right\|_{{\rm{sh},\rho}}^2 = {\cal O}(3^{n/k})$}
\label{appx:A7}

\begin{proof}    
    Let unitary ensemble $\mathbb{U}$ be $\text{Cl}(k)^{\otimes n/k}$, $k2^k = \Omega(n)$.
    In the case of the $\rho  = O  = \bigotimes_{i = 1}^{n/k}|{\Psi _i}\rangle \langle {\Psi _i}|$, shadow norm $\left\| O  \right\|_{{\rm{sh, \rho}}}^2$ becomes 
    \begin{align}
        \|O\|_{\mathrm{sh, \rho}}^2
        &=\mathbb{E}_{U,b}[\hat o^2]\\
        &=\prod_{i=1}^{n/k}\mathbb{E}_{U_i,b_i}[\hat\Psi_i^2]\\
        &=\left(
        \frac{D_k+1}{D_k+2}
        \left(
        3-\frac{5}{D_k}+\frac{2}{D_k^2}
        \right)
        +\frac{2}{D_k}-\frac{1}{D_k^2}
        \right)^{n/k}\\
        &\to 3^{n/k},\qquad(\text{as }k\to\infty)
    \end{align}
    where \(\hat\Psi_i=\operatorname{Tr}(\ket{\Psi_i}\bra{\Psi_i}\hat\rho_i)\) denotes the single-block fidelity estimator and $U_i$ is a unitary operator on $\mathcal{H}^{D_k}$. 
\end{proof}

\subsection{Variance bound for purity estimation in the single-shot setting}
\label{appx:purity_single}

\begin{proof}
    Let \(\mathbb{U}=\mathrm{Cl}(k)^{\otimes m}\), where \(m=n/k\), and assume
    \(k2^k=n/\alpha\). Consider
    \begin{equation}
        \hat p_2
        =
        \frac{2}{T(T-1)}
        \sum_{i>j}^{T}
        {\rm Tr}(\hat\rho_i\hat\rho_j),
    \end{equation}
    where the \(\hat\rho_i\)'s are independent single-shot shadow estimators.
    The standard variance identity for a second-order \(U\)-statistic gives
    \begin{align}
        {\rm Var}(\hat p_2)
        &=
        4\,\frac{T-2}{T(T-1)}
        {\rm Var}\!\left({\rm Tr}(\rho\hat\rho)\right)
        +
        \frac{2}{T(T-1)}
        {\rm Var}\!\left({\rm Tr}(\hat\rho_1\hat\rho_2)\right).
        \label{appx:eq:p2_ustat}
    \end{align}
    The first term is bounded by the worst-case block-shadow norm:
    \begin{equation}
        {\rm Var}\!\left({\rm Tr}(\rho\hat\rho)\right)
        \le
        \mathbb E\!\left[{\rm Tr}(\rho\hat\rho)^2\right]
        =
        \|\rho\|_{\rm sh,\rho}^2
        \le
        e^{\alpha/3}3^{m}\|\rho\|_2^2
        \le
        e^{\alpha/3}3^{m}.
        \label{appx:eq:p2_first_term}
    \end{equation}

    It remains to bound the second term. Using the Pauli expansion and
    \(\mathcal M^{-1}(P)=m_P^{-1}P\), we have
    \begin{equation}
        {\rm Tr}(\hat\rho_1\hat\rho_2)
        =
        \frac{1}{D}
        \sum_{P\in\mathcal P_n}
        m_P^{-2}
        (U_1,P,b_1)(U_2,P,b_2).
    \end{equation}
    Therefore,
    \begin{align}
        &\mathbb E\!\left[
        {\rm Tr}(\hat\rho_1\hat\rho_2)^2
        \right]  \notag\\
        &=
        \frac{1}{D^2}
        \sum_{P,Q}
        m_P^{-2}m_Q^{-2}
        \mathbb E_{U_1,b_1}\!\left[
            (U_1,P,b_1)(U_1,Q,b_1)
        \right]
        \mathbb E_{U_2,b_2}\!\left[
            (U_2,P,b_2)(U_2,Q,b_2)
        \right].
    \end{align}
    For one copy,
    \begin{align}
        \mathbb E_{U,b}\!\left[
            (U,P,b)(U,Q,b)
        \right]
        &=
        {\rm Tr}(\rho PQ)\,
        p(P,Q),
    \end{align}
    where
    \begin{equation}
        p(P,Q)
        :=
        \Pr_U\!\left(
        UPU^\dagger,UQU^\dagger\in\pm\mathcal Z
        \right).
    \end{equation}
    Thus
    \begin{equation}
        \mathbb E\!\left[
        {\rm Tr}(\hat\rho_1\hat\rho_2)^2
        \right]
        =
        \frac{1}{D^2}
        \sum_{P,Q}
        m_P^{-2}m_Q^{-2}p(P,Q)^2
        \left|{\rm Tr}(\rho PQ)\right|^2 .
        \label{appx:eq:p2_second_exact}
    \end{equation}
    We now keep the factor \(p(P,Q)^2\) and group the sum by the
    phase-free Pauli string \(R\) determined by \(PQ=\pm R\) on the support of
    \(p(P,Q)\). Define
    \begin{equation}
        C_R
        :=
        \sum_{\substack{P,Q\in\mathcal P_n\\ PQ=\pm R}}
        m_P^{-2}m_Q^{-2}p(P,Q)^2 .
    \end{equation}
    Since \(p(P,Q)=0\) unless \(P\) and \(Q\) commute blockwise,
    Eq.~\eqref{appx:eq:p2_second_exact} implies
    \begin{equation}
        \mathbb E\!\left[
        {\rm Tr}(\hat\rho_1\hat\rho_2)^2
        \right]
        =
        \frac{1}{D^2}
        \sum_{R\in\mathcal P_n}
        C_R\,{\rm Tr}(\rho R)^2 .
        \label{appx:eq:p2_grouped}
    \end{equation}

    It remains to bound \(C_R\). By the block product structure, \(C_R\)
    factorizes over the \(m\) blocks. For a single block, define
    \begin{equation}
        c(P,Q)
        := 
        f(P,Q)^2.
    \end{equation}
    By Eq.~\eqref{eq:appx:f},
    \[
        c(P,Q)=
        \begin{cases}
        1,
        & P=I \text{ or } Q=I,\\[4pt]
        (D_k+1)^2,
        & P=Q\neq I,\\[4pt]
        \dfrac{4(D_k+1)^2}{(D_k+2)^2},
        & P,Q\neq I,\ P\neq Q,\ PQ=QP,\\[8pt]
        0,
        & PQ\neq QP .
        \end{cases}
    \]
    Let
    \[
        C_r^{\rm loc}
        :=
        \sum_{\substack{P,Q\in\mathcal P_k\\PQ=\pm r}}
        c(P,Q).
    \]
    Then
    \begin{align}
        C_I^{\rm loc}
        &=
        1+(D_k^2-1)(D_k+1)^2
        \le
        D_k^4\left(1+\frac{2}{D_k}\right),\\
        C_{r(\neq I)}^{\rm loc}
        &=
        2+
        \left(\frac{D_k^2}{2}-2\right)
        \frac{4(D_k+1)^2}{(D_k+2)^2}
        \le
        2D_k^2.
    \end{align}
    Consequently, if \(A=\operatorname{supp}_k(R)\coloneq\{i\in[n/k]:R[i] \neq I_k\}\) is the block support of
    \(R\), then
    \begin{equation}
        C_R
        \le
        \left[
        D_k^4\left(1+\frac{2}{D_k}\right)
        \right]^{m-|A|}
        (2D_k^2)^{|A|}.
        \label{appx:eq:CR_single_shot}
    \end{equation}

    For fixed \(A\), using \(R=R_A\otimes I_{A^c}\) and Pauli orthogonality,
    \begin{equation}
        \sum_{\operatorname{supp}_k(R)=A}
        {\rm Tr}(\rho R)^2
        \le
        \sum_{R_A}
        {\rm Tr}(\rho_A R_A)^2
        =
        D_k^{|A|}{\rm Tr}(\rho_A^2)
        \le
        D_k^{|A|}.
        \label{appx:eq:RA_sum}
    \end{equation}
    Combining Eqs.~\eqref{appx:eq:p2_grouped},
    \eqref{appx:eq:CR_single_shot}, and~\eqref{appx:eq:RA_sum}, we obtain
    \begin{align}
        \mathbb E\!\left[
        {\rm Tr}(\hat\rho_1\hat\rho_2)^2
        \right]
        &\le
        \frac{1}{D^2}
        \sum_{A\subseteq[m]}
        \left[
        D_k^4\left(1+\frac{2}{D_k}\right)
        \right]^{m-|A|}
        (2D_k^2)^{|A|}
        D_k^{|A|} \\
        &=
        \frac{1}{D^2}
        \left[
        D_k^4\left(1+\frac{2}{D_k}\right)
        +
        2D_k^3
        \right]^m \\
        &\le
        D^2
        \left(1+\frac{4}{D_k}\right)^m
        \le
        D^2 e^{4\alpha}.
        \label{appx:eq:p2_second_bound}
    \end{align}
    Therefore,
    \begin{equation}
        {\rm Var}\!\left({\rm Tr}(\hat\rho_1\hat\rho_2)\right)
        \le
        D^2 e^{4\alpha}.
    \end{equation}

    Substituting Eqs.~\eqref{appx:eq:p2_first_term} and
    \eqref{appx:eq:p2_second_bound} into Eq.~\eqref{appx:eq:p2_ustat} gives
    \begin{align}
        {\rm Var}(\hat p_2)
        &\le
        4e^{\alpha/3}
        \frac{T-2}{T(T-1)}
        3^{m}
        +
        \frac{2D^2e^{4\alpha}}{T(T-1)} \\
        &=
        \mathcal O\!\left(
        \frac{3^{n/k}}{T}
        +
        \frac{D^2}{T^2}
        \right).
    \end{align}
    By Chebyshev's inequality, it is sufficient to take
    \begin{equation}
        T
        =
        \mathcal O\!\left(
        \max\left\{
        \frac{3^{n/k}}{\epsilon^2},
        \frac{2^n}{\epsilon}
        \right\}
        \right)
    \end{equation}
    to estimate \({\rm Tr}(\rho^2)\) within additive error \(\epsilon\).
\end{proof}

\subsection{Multi-shot variance for a Pauli observable}
\label{appx:multi_shots_pauli}

For the discussion in this section and below, write \(T=N_U N_S\), where
\(N_U\) is the number of sampled measurement bases and \(N_S\) is the number of
repeated measurements in each basis.

\begin{proof}
    Let
    \begin{equation}
        \hat{\rho}
        =
        \frac{1}{N_U}
        \sum_{i=1}^{N_U}
        \hat{\rho}_{U_i}(N_S),
        \qquad
        \hat{\rho}_{U}(N_S)
        =
        \frac{1}{N_S}
        \sum_{j=1}^{N_S}
        \hat{\rho}_{U,j},
    \end{equation}
    where
    \begin{equation}
        \hat{\rho}_{U,j}
        =
        \mathcal M^{-1}
        \!\left(
            U^\dagger |b_j\rangle\langle b_j|U
        \right).
    \end{equation}
    For a Pauli string \(P\), define
    \[
        \hat{o}_P
        =
        \operatorname{Tr}(P\hat{\rho}).
    \]
    Since the \(N_U\) measurement bases are independent,
    \begin{equation}
        \operatorname{Var}(\hat{o}_P)
        =
        \frac{1}{N_U}
        \operatorname{Var}
        \!\left(
            \operatorname{Tr}(P\hat{\rho}_{U}(N_S))
        \right).
    \end{equation}
    Moreover,
    \begin{align}
        \operatorname{Var}
        \!\left(
            \operatorname{Tr}(P\hat{\rho}_{U}(N_S))
        \right)
        &=
        \mathbb E
        \!\left[
            \operatorname{Tr}(P\hat{\rho}_{U}(N_S))^2
        \right]
        -
        \operatorname{Tr}(\rho P)^2 .
    \end{align}
    Using \(\mathcal M^{-1}(P)=m_P^{-1}P\), we have
    \begin{equation}
        \operatorname{Tr}(P\hat{\rho}_{U,j})
        =
        m_P^{-1}(U,P,b_j),
    \end{equation}
    where $(U,P,b) \coloneqq \operatorname{Tr}\!\left(UPU^\dagger |b\rangle\langle b|\right)$. Therefore,
    \begin{align}
        &\mathbb E
        \!\left[
            \operatorname{Tr}(P\hat{\rho}_{U}(N_S))^2
        \right] \notag \\
        &=
        \frac{1}{N_S^2}
        \sum_{j,\ell=1}^{N_S}
        \mathbb E
        \!\left[
            \operatorname{Tr}(P\hat{\rho}_{U,j})
            \operatorname{Tr}(P\hat{\rho}_{U,\ell})
        \right] \\
        &=
        \frac{1}{N_S}
        m_P^{-2}
        \mathbb E
        \!\left[
            (U,P,b_1)^2
        \right]
        +
        \frac{N_S-1}{N_S}
        m_P^{-2}
        \mathbb E
        \!\left[
            (U,P,b_1)(U,P,b_2)
        \right].
    \end{align}
    The diagonal term is
    \begin{align}
        \mathbb E
        \!\left[
            (U,P,b_1)^2
        \right]
        &=
        \mathbb E_U
        \sum_b
        (U,P,b)^2
        \operatorname{Tr}
        \!\left(
            U\rho U^\dagger |b\rangle\langle b|
        \right) \\
        &=
        \mathbb E_U
        \!\left[
            \mathbf 1\{UPU^\dagger\in\pm\mathcal Z\}
        \right] \\
        &=
        m_P .
    \end{align}
    The off-diagonal term is
    \begin{align}
        \mathbb E
        \!\left[
            (U,P,b_1)(U,P,b_2)
        \right]
        &=
        \mathbb E_U
        \left[
            \sum_b
            (U,P,b)
            \operatorname{Tr}
            \!\left(
                U\rho U^\dagger |b\rangle\langle b|
            \right)
        \right]^2 \\
        &=
        \mathbb E_U
        \!\left[
            \mathbf 1\{UPU^\dagger\in\pm\mathcal Z\}
            \operatorname{Tr}(\rho P)^2
        \right] \\
        &=
        m_P\operatorname{Tr}(\rho P)^2 .
    \end{align}
    Combining these identities gives
    \begin{align}
        \operatorname{Var}
        \!\left(
            \operatorname{Tr}(P\hat{\rho}_{U}(N_S))
        \right)
        &=
        m_P^{-1}
        \left(
            \frac{1}{N_S}
            +
            \frac{N_S-1}{N_S}
            \operatorname{Tr}(\rho P)^2
        \right)
        -
        \operatorname{Tr}(\rho P)^2 .
    \end{align}
    Hence,
    \begin{align}
        \operatorname{Var}
        \!\left(
            \operatorname{Tr}(P\hat{\rho})
        \right)
        =
        \frac{1}{N_U}
        \left[
            m_P^{-1}
            \left(
                \frac{1}{N_S}
                +
                \frac{N_S-1}{N_S}
                \operatorname{Tr}(\rho P)^2
            \right)
            -
            \operatorname{Tr}(\rho P)^2
        \right].
    \end{align}
    This proves the claim.
\end{proof}

\subsection{Multi-shot variance for a general observable}
\label{appx:multi_shots_observable}

For the discussion in this section, write \(T=N_U N_S\), where
\(N_U\) is the number of sampled measurement bases and \(N_S\) is the number of
repeated measurements in each basis. Let \(O=\sum_P \alpha_P P\).

\begin{proof}
    As in Appendix~\ref{appx:multi_shots_pauli}, define
    \begin{equation}
        \hat{\rho}
        =
        \frac{1}{N_U}
        \sum_{i=1}^{N_U}
        \hat{\rho}_{U_i}(N_S),
        \qquad
        \hat{\rho}_{U}(N_S)
        =
        \frac{1}{N_S}
        \sum_{j=1}^{N_S}
        \hat{\rho}_{U,j},
    \end{equation}
    where
    \begin{equation}
        \hat{\rho}_{U,j}
        =
        \mathcal M^{-1}
        \!\left(
            U^\dagger |b_j\rangle\langle b_j|U
        \right).
    \end{equation}
    Since the \(N_U\) measurement bases are sampled independently,
    \begin{equation}
        \operatorname{Var}\!\left(\operatorname{Tr}(O\hat{\rho})\right)
        =
        \frac{1}{N_U}
        \operatorname{Var}\!\left(
            \operatorname{Tr}(O\hat{\rho}_{U}(N_S))
        \right).
    \end{equation}
    Moreover, using \(\mathcal M^{-1}(P)=m_P^{-1}P\), we have
    \begin{equation}
        \operatorname{Tr}(O\hat{\rho}_{U,j})
        =
        \sum_P
        \alpha_P m_P^{-1}(U,P,b_j),
    \end{equation}
    where
    \((U,P,b)=\operatorname{Tr}(UPU^\dagger |b\rangle\langle b|)\).
    Therefore,
    \begin{align}
        &\mathbb E\!\left[
            \operatorname{Tr}(O\hat{\rho}_{U}(N_S))^2
        \right] \notag\\
        &=
        \frac{1}{N_S}
        \mathbb E\!\left[
            \operatorname{Tr}(O\hat{\rho}_{U,1})^2
        \right]
        +
        \frac{N_S-1}{N_S}
        \mathbb E\!\left[
            \operatorname{Tr}(O\hat{\rho}_{U,1})
            \operatorname{Tr}(O\hat{\rho}_{U,2})
        \right].
    \end{align}

    We first compute the diagonal term. Expanding
    \(\rho = D^{-1}\sum_R \operatorname{Tr}(\rho R)R\), we obtain
    \begin{align}
        &\mathbb E\!\left[
            \operatorname{Tr}(O\hat{\rho}_{U,1})^2
        \right] \notag\\
        &=
        \mathbb E_U
        \sum_b
        \sum_{P,Q}
        \alpha_P\alpha_Q
        m_P^{-1}m_Q^{-1}
        (U,P,b)(U,Q,b)
        \operatorname{Tr}\!\left(
            U\rho U^\dagger |b\rangle\langle b|
        \right) \\
        &=
        \mathbb E_U
        \sum_b
        \sum_{P,Q,R}
        \alpha_P\alpha_Q
        m_P^{-1}m_Q^{-1}
        (U,P,b)(U,Q,b)(U,R,b)
        \frac{\operatorname{Tr}(\rho R)}{D} \\
        &=
        \sum_{P,Q}
        \alpha_P\alpha_Q
        \operatorname{Tr}(\rho PQ)
        \frac{
        \Pr_U\!\left(
            UPU^\dagger,\,
            UQU^\dagger
            \in\pm\mathcal Z
        \right)
        }{
        m_Pm_Q
        } \\
        &=
        \sum_{P,Q}
        \alpha_P\alpha_Q
        \operatorname{Tr}(\rho PQ)
        f(P,Q) \\
        &=
        V_2 .
    \end{align}
    In the third equality, we used the Pauli orthogonality identity in
    Eq.~\eqref{appx:eq:pro2}. In particular, if \(UPU^\dagger\) and
    \(UQU^\dagger\) are both diagonal, then the only nonzero contribution comes
    from \(R=PQ\).

    We next compute the off-diagonal term. Since \(b_1\) and \(b_2\) are
    independent conditioned on the same \(U\),
    \begin{align}
        &\mathbb E\!\left[
            \operatorname{Tr}(O\hat{\rho}_{U,1})
            \operatorname{Tr}(O\hat{\rho}_{U,2})
        \right] \notag\\
        &=
        \mathbb E_U
        \sum_{b_1,b_2}
        \sum_{P,Q}
        \alpha_P\alpha_Q
        m_P^{-1}m_Q^{-1}
        (U,P,b_1)(U,Q,b_2) \notag\\
        &\qquad\qquad\qquad\qquad\qquad
        \times
        \operatorname{Tr}\!\left(
            U\rho U^\dagger |b_1\rangle\langle b_1|
        \right)
        \operatorname{Tr}\!\left(
            U\rho U^\dagger |b_2\rangle\langle b_2|
        \right) \\
        &=
        \sum_{P,Q}
        \alpha_P\alpha_Q
        \operatorname{Tr}(\rho P)
        \operatorname{Tr}(\rho Q)
        \frac{
        \Pr_U\!\left(
            UPU^\dagger,\,
            UQU^\dagger
            \in\pm\mathcal Z
        \right)
        }{
        m_Pm_Q
        } \\
        &=
        \sum_{P,Q}
        \alpha_P\alpha_Q
        \operatorname{Tr}(\rho P)
        \operatorname{Tr}(\rho Q)
        f(P,Q) \\
        &=
        V_1 .
    \end{align}
    Combining the diagonal and off-diagonal contributions gives
    \begin{align}
        \mathbb E\!\left[
            \operatorname{Tr}(O\hat{\rho}_{U}(N_S))^2
        \right]
        =
        \frac{1}{N_S}V_2
        +
        \frac{N_S-1}{N_S}V_1 .
    \end{align}
    Since the estimator is unbiased,
    \begin{equation}
        \mathbb E\!\left[
            \operatorname{Tr}(O\hat{\rho}_{U}(N_S))
        \right]
        =
        \operatorname{Tr}(\rho O).
    \end{equation}
    Hence,
    \begin{align}
        \operatorname{Var}\!\left(\operatorname{Tr}(O\hat{\rho})\right)
        &=
        \frac{1}{N_U}
        \left(
            \frac{1}{N_S}V_2
            +
            \frac{N_S-1}{N_S}V_1
            -
            \operatorname{Tr}(\rho O)^2
        \right)\\
        &\le
        \frac{1}{N_U}
        \left(
            V_1+\frac{V_2}{N_S}
        \right).
    \end{align}
\end{proof}

\subsection{Variance bound for purity estimation in the multi-shot setting}
\label{appx:purity_multi}

\begin{proof}
    Let
    $\hat p_2 = \frac{1}{N_U} \sum_{i=1}^{N_U} \hat p_2(U_i)$,
    where
    \[
    \hat p_2(U)
    = \binom{N_S}{2}^{-1}
    \sum_{i>j}
    \operatorname{Tr}\!\left(
    U^\dagger |b_i\rangle \langle b_i| U\,
    \mathcal{M}^{-1}
    \bigl(
    U^\dagger |b_j\rangle \langle b_j| U
    \bigr)
    \right).
    \]
    Then, $\hat p_2$ is an unbiased estimator, as shown below.
    \begin{align}
        \mathbb{E}\!\left[\hat p_2\right]
        &= \mathbb{E}\!\left[\hat p_2(U)\right] \\
        &= \mathbb{E}_{U,b_1,b_2}\!\left[
        {\rm Tr}\!\left(
        U^{\dagger}|b_1\rangle\langle b_1|U\,
        {\cal M}^{-1}
        \!\left(
        U^{\dagger}|b_2\rangle\langle b_2|U
        \right)
        \right)
        \right] \\
        &= \mathbb{E}_{U,b_1,b_2}\!\left[
        \sum_P m_P^{-1}(U,P,b_1)(U,P,b_2)/D
        \right] \\
        &= \mathbb{E}_{U,b_1,b_2}\!\left[
        \sum_{b_1,b_2}\sum_{P,Q,R}
        m_P^{-1}
        (U,P,b_1)(U,Q,b_1)
        (U,P,b_2)(U,R,b_2)
        \frac{{\rm Tr}(\rho Q)\,{\rm Tr}(\rho R)}{D^3}
        \right] \\
        &= \sum_P \frac{{\rm Tr}(\rho P)^2}{D} \\
        &= {\rm Tr}(\rho^2)
    \end{align}
    Building on this, $\text{Var}[\hat{p}_2]$ can be expressed as follows:
    \begin{align}
        {\rm Var}(\hat p_2)
        &= \frac{1}{N_U}\,{\rm Var}(\hat p_2(U)) \\
        &= \frac{1}{N_U}
        \left(
        \mathbb{E}[\hat p_2(U)^2]
        -
        \mathbb{E}[\hat p_2(U)]^2
        \right) \\
        &= \frac{1}{N_U}
        \left(
        \mathbb{E}[\hat p_2(U)^2]
        -
        {\rm Tr}(\rho^2)^2
        \right).
    \end{align}
    Then, $\mathbb{E}[{\hat p_2}{(U)^2}]$ can be calculated as 
    \begin{align}
        &\mathbb{E}[\hat p_2(U)^2]\nonumber\\
        &= \frac{(N_S-2)(N_S-3)}{N_S(N_S-1)}
        \mathbb{E}\Bigl[
        {\rm Tr}\!\left(
        U^{\dagger}|b_1\rangle\langle b_1|U\,
        {\cal M}^{-1}
        \!\left(
        U^{\dagger}|b_2\rangle\langle b_2|U
        \right)
        \right) 
        {\rm Tr}\!\left(
        U^{\dagger}|b_3\rangle\langle b_3|U\,
        {\cal M}^{-1}
        \!\left(
        U^{\dagger}|b_4\rangle\langle b_4|U
        \right)
        \right)
        \Bigr]  \nonumber\\
        &\quad
        + \frac{N_S-2}{N_S(N_S-1)}
        \mathbb{E}\Bigl[
        {\rm Tr}\!\left(
        U^{\dagger}|b_1\rangle\langle b_1|U\,
        {\cal M}^{-1}
        \!\left(
        U^{\dagger}|b_2\rangle\langle b_2|U
        \right)
        \right)
        {\rm Tr}\!\left(
        U^{\dagger}|b_1\rangle\langle b_1|U\,
        {\cal M}^{-1}
        \!\left(
        U^{\dagger}|b_3\rangle\langle b_3|U
        \right)
        \right)
        \Bigr]  \nonumber\\
        &\quad
        + \frac{2}{N_S(N_S-1)}
        \mathbb{E}\Bigl[
        {\rm Tr}\!\left(
        U^{\dagger}|b_1\rangle
        \langle b_1|U\,
        {\cal M}^{-1}
        \!\left(
        U^{\dagger}|b_2\rangle\langle b_2|U
        \right)
        \right)^2
        \Bigr]
        \label{eq:p2}
    \end{align}
    The first term in Eq.~\eqref{eq:p2} can be calculated as
    \begin{align}
        &\mathbb{E}_{U,b_1,b_2,b_3,b_4}\Bigl[
        \operatorname{Tr}\!\left(
        U^{\dagger}|b_1\rangle\langle b_1|U\,
        \mathcal{M}^{-1}
        \!\left(
        U^{\dagger}|b_2\rangle\langle b_2|U
        \right)
        \right)
        \operatorname{Tr}\!\left(
        U^{\dagger}|b_3\rangle\langle b_3|U\,
        \mathcal{M}^{-1}
        \!\left(
        U^{\dagger}|b_4\rangle\langle b_4|U
        \right)
        \right)
        \Bigr] \notag\\
        &=
        \mathbb{E}_{U,b_1,b_2,b_3,b_4}
        \left[
        \sum_{P,Q}
        m_P^{-1}m_Q^{-1}
        (U,P,b_1)(U,P,b_2)
        (U,Q,b_3)(U,Q,b_4)
        \Big/ D^2
        \right] \\
        &=
        \mathbb{E}_{U}
        \Bigg[
        \sum_{b_1,b_2,b_3,b_4}
        \sum_{P,Q,R,S,T,L}
        m_P^{-1}m_Q^{-1}
        (U,P,b_1)(U,R,b_1)
        (U,P,b_2)(U,S,b_2)
        \notag\\
        &\hspace{2.0cm}\times
        (U,Q,b_3)(U,T,b_3)
        (U,Q,b_4)(U,L,b_4)\,
        \operatorname{Tr}(\rho R)
        \operatorname{Tr}(\rho S)
        \operatorname{Tr}(\rho T)
        \operatorname{Tr}(\rho L)
        \Big/ D^6
        \Bigg] \\
        &=
        \sum_{P,Q}
        \frac{
        \operatorname{Tr}(\rho P)^2
        \operatorname{Tr}(\rho Q)^2
        }{D^2}
        \frac{
        \Pr_U
        \!\left(
            UPU^\dagger,\,
            UQU^\dagger
            \in\pm\mathcal Z
        \right)
        }{
        m_Pm_Q
        } \\
        &=
        \sum_{P,Q}
        \frac{
        \operatorname{Tr}(\rho P)^2
        \operatorname{Tr}(\rho Q)^2
        }{D^2}
        f(P,Q) \\
        &=
        V_1(\rho).
    \end{align}
    The second term in Eq.~\eqref{eq:p2} can be calculated as
    \begin{align}
        &\mathbb{E}_{U,b_1,b_2,b_3}\!\Bigl[
        {\rm Tr}\!\left(
        U^{\dagger}|b_1\rangle\langle b_1|U\,
        {\cal M}^{-1}
        \!\left(
        U^{\dagger}|b_2\rangle\langle b_2|U
        \right)
        \right)
        {\rm Tr}\!\left(
        U^{\dagger}|b_1\rangle\langle b_1|U\,
        {\cal M}^{-1}
        \!\left(
        U^{\dagger}|b_3\rangle\langle b_3|U
        \right)
        \right)
        \Bigr]\\
        &= \mathbb{E}_{U,b_1,b_2,b_3}\!\Bigl[
        \sum_{P,Q}
        m_P^{-1} m_Q^{-1}
        (U,P,b_1)(U,P,b_2)
        (U,Q,b_1)(U,Q,b_3)
        \Big/ D^2
        \Bigr]  \\
        &= \mathbb{E}_U\!\Bigl[
        \sum_{b_1,b_2,b_3}
        \sum_{P,Q,R,S,T}
        m_P^{-1} m_Q^{-1}
        (U,P,b_1)(U,Q,b_1)(U,R,b_1)  \nonumber\\
        &\qquad\qquad\qquad\quad
        \times (U,P,b_2)(U,S,b_2)
        (U,Q,b_3)(U,T,b_3)
        \frac{
        {\rm Tr}(\rho R)\,
        {\rm Tr}(\rho S)\,
        {\rm Tr}(\rho T)
        }{D^5}
        \Bigr]  \\
        &= \sum_{P,Q}
        \frac{
        {\rm Tr}(\rho P)\,
        {\rm Tr}(\rho Q)\,
        {\rm Tr}(\rho PQ)
        }{D^2}
        \, f(P,Q)  \\
        &= V_2(\rho).
    \end{align}
    The third term in Eq.~\eqref{eq:p2} can be calculated as
    \begin{align}
        &\mathbb{E}_{U,b_1,b_2}\!\Bigl[
        {\rm Tr}\!\left(
        U^{\dagger}|b_1\rangle
        \langle b_1|U\,
        {\cal M}^{-1}
        \!\left(
        U^{\dagger}|b_2\rangle\langle b_2|U
        \right)
        \right)^2
        \Bigr]\\
        &= \mathbb{E}_{U,b_1,b_2}\!\Bigl[
        \sum_{P,Q}
        m_P^{-1} m_Q^{-1}
        (U,P,b_1)(U,Q,b_1)
        (U,P,b_2)(U,Q,b_2)
        \Big/ D^2
        \Bigr] \\
        &= \mathbb{E}_U\!\Bigl[
        \sum_{b_1,b_2}
        \sum_{P,Q,R,S}
        m_P^{-1} m_Q^{-1}
        (U,P,b_1)(U,Q,b_1)
        (U,P,b_2)(U,Q,b_2)(U,R,b_1)(U,S,b_2)
        \Big/ D^4
        \Bigr]  \\
        &= \mathbb{E}_U\!\Bigl[
        \sum_{b_1,b_2}
        \sum_{P,Q,R,S}
        m_P^{-1} m_Q^{-1}
        (U,P,b_1)(U,Q,b_1)
        (U,P,b_2)(U,Q,b_2)(U,R,b_1)(U,S,b_2)\,
        {\rm Tr}(\rho R)\,{\rm Tr}(\rho S)/D^4
        \Bigr]  \\
        &= \sum_{P,Q}
        \frac{{\rm Tr}(\rho PQ)^2}{D^2}\,
        f(P,Q)  \\
        &= V_3(\rho).
    \end{align}
    Combining the three estimates above with Eq.~\eqref{eq:p2}, we obtain
    \begin{align}
        \mathbb{E}[\hat p_2(U)^2]
        &=
        \frac{(N_S-2)(N_S-3)}{N_S(N_S-1)} V_1(\rho)
        +
        \frac{4(N_S-2)}{N_S(N_S-1)} V_2(\rho)
        +
        \frac{2}{N_S(N_S-1)} V_3(\rho) \\
        &\le
        V_1(\rho)
        +
        \frac{4}{N_S}V_2(\rho)
        +
        \frac{2}{(N_S-1)^2}V_3(\rho).
    \end{align}
    Therefore,
    \begin{align}
        \operatorname{Var}(\hat p_2)
        &=
        \frac{1}{N_U}
        \left(
            \mathbb{E}[\hat p_2(U)^2]
            -
            \operatorname{Tr}(\rho^2)^2
        \right) \\
        &\le
        \frac{1}{N_U}
        \left(
            V_1(\rho)
            +
            \frac{4}{N_S}V_2(\rho)
            +
            \frac{2}{(N_S-1)^2}V_3(\rho)
        \right).
    \end{align}
    This proves the claim.
\end{proof}

\subsection{Inner product estimation in the distributed setting}
\label{appx:inner}

Let $\hat f = \frac{1}{N_U}{\sum _{i = 1}}\hat f({U_i})$ and $\hat f(U) = \frac{1}{{N_S^2}}\sum _{i,j = 1}^{{N_S}}{\rm{Tr}}({U^{\dagger}}|{b_i}\rangle \langle {b_i}|U{{\cal M}^{ - 1}}({U^{\dagger}}|{c_j}\rangle \langle {c_j}|U))$ , where $b_i$ and $c_j$ are measurement outcomes of $U\rho U^{\dagger}$ and $U\sigma U^{\dagger}$, respectively. Then, following the same procedures in~\ref{appx:purity_multi} and Cauchy-Schwarz inequality, we can compute the following
\begin{align}
\mathbb{E}[\hat f]
    &= {\rm Tr}(\rho \sigma),  \\[4pt]
    {\rm Var}(\hat f)
    &= \frac{1}{N_U}
    \Biggl(
    \left(\frac{N_S-1}{N_S}\right)^2
    \mathbb{E}_U\!\Bigl[
    {\rm Tr}\!\left(
    U^{\dagger}|b_1\rangle\langle b_1|U\,
    {\cal M}^{-1}
    \!\left(
    U^{\dagger}|c_1\rangle\langle c_1|U
    \right)
    \right)  \nonumber\\
    &\qquad\qquad\qquad\qquad\quad
    \times
    {\rm Tr}\!\left(
    U^{\dagger}|b_2\rangle\langle b_2|U\,
    {\cal M}^{-1}
    \!\left(
    U^{\dagger}|c_2\rangle\langle c_2|U
    \right)
    \right)
    \Bigr]  \nonumber\\
    &\quad
    + \frac{N_S-1}{N_S^2}
    \mathbb{E}_U\!\Bigl[
    {\rm Tr}\!\left(
    U^{\dagger}|b_1\rangle\langle b_1|U\,
    {\cal M}^{-1}
    \!\left(
    U^{\dagger}|c_1\rangle\langle c_1|U
    \right)
    \right)  \nonumber\\
    &\qquad\qquad\qquad\qquad\quad
    \times
    {\rm Tr}\!\left(
    U^{\dagger}|b_1\rangle\langle b_1|U\,
    {\cal M}^{-1}
    \!\left(
    U^{\dagger}|c_2\rangle\langle c_2|U
    \right)
    \right)
    \Bigr]  \nonumber\\
    &\quad
    + \frac{N_S-1}{N_S^2}
    \mathbb{E}_U\!\Bigl[
    {\rm Tr}\!\left(
    U^{\dagger}|b_1\rangle\langle b_1|U\,
    {\cal M}^{-1}
    \!\left(
    U^{\dagger}|c_1\rangle\langle c_1|U
    \right)
    \right)  \nonumber\\
    &\qquad\qquad\qquad\qquad\quad
    \times
    {\rm Tr}\!\left(
    U^{\dagger}|b_2\rangle\langle b_2|U\,
    {\cal M}^{-1}
    \!\left(
    U^{\dagger}|c_1\rangle\langle c_1|U
    \right)
    \right)
    \Bigr]  \nonumber\\
    &\quad
    + \frac{1}{N_S^2}
    \mathbb{E}_U\!\Bigl[
    {\rm Tr}\!\left(
    U^{\dagger}|b_1\rangle\langle b_1|U\,
    {\cal M}^{-1}
    \!\left(
    U^{\dagger}|c_1\rangle\langle c_1|U
    \right)
    \right)^2
    \Bigr]  
    - {\rm Tr}(\rho\sigma)^2
    \Biggr)  \\[6pt]
    &\le \frac{1}{N_U}
    \Biggl(
    \frac{1}{N_S^2}\sqrt{V_3(\rho)V_3(\sigma)}
    + \frac{N_S-1}{N_S^2}
    \bigl(
    \sqrt{V_1(\rho)V_3(\sigma)}
    + \sqrt{V_3(\rho)V_1(\sigma)}
    \bigr)  \nonumber\\
    &\qquad\qquad
    + \left(\frac{N_S-1}{N_S}\right)^2
    \sqrt{V_1(\rho)V_1(\sigma)}
    \Biggr).
\end{align}
This gives the multi-shot variance bound for distributed inner-product estimation.
In the typical regime where \(V_1(\rho),V_1(\sigma)=\mathcal O(1)\) and
\(V_3(\rho),V_3(\sigma)=\mathcal O(2^n)\), choosing
\(N_S=\mathcal O(2^{n/2})\) suppresses the \(V_3\)-dependent term to a constant.

\subsection{Typical values of $V_1(O,\rho)$ and $V_2(O,\rho)$}
\label{appx:typical_si_V}

Let \(O=\sum_P \alpha_P P\) be a fixed observable independent of \(\rho\). 
Let \(\mathbb U=\mathrm{Cl}(k)^{\otimes n/k}\), and assume that \(k2^k=n/\alpha\). 
For Haar-random pure states, the second moment identity gives
\begin{equation}
    \mathbb E_\rho[
        \operatorname{Tr}(\rho P)
        \operatorname{Tr}(\rho Q)
    ]
    =
    \frac{
        \operatorname{Tr}(P)\operatorname{Tr}(Q)
        +
        \operatorname{Tr}(PQ)
    }{
        D(D+1)
    } .
\end{equation}
Using this identity in \(V_1(O,\rho)\), we obtain
\begin{align}
    \mathbb E_\rho[V_1(O,\rho)]
    &=
    \sum_{P,Q}
    \alpha_P\alpha_Q\,
    \mathbb E_\rho[
        \operatorname{Tr}(\rho P)
        \operatorname{Tr}(\rho Q)
    ]\,
    f(P,Q)\\
    &=
    \frac{\operatorname{Tr}(O)^2}{D(D+1)}
    +
    \frac{1}{D+1}
    \sum_P \alpha_P^2 f(P,P)\\
    &=
    \mathcal{O}\!\left(\frac{\|O\|_2^2}{D}\right).
\end{align}
Here we used \(f(P,P)=m_P^{-1}\le D e^\alpha\) and
\(\sum_P\alpha_P^2=\|O\|_2^2/D\).

Similarly, since \(\mathbb E_\rho[\rho]=I/D\), we have
\begin{align}
    \mathbb E_\rho[V_2(O,\rho)]
    &=
    \sum_{P,Q}
    \alpha_P\alpha_Q\,
    \mathbb E_\rho[
        \operatorname{Tr}(\rho PQ)
    ]\,
    f(P,Q)\\
    &=
    \sum_P \alpha_P^2 f(P,P)\\
    &=
    \mathcal{O}(\|O\|_2^2).
\end{align}
Therefore, for observables with bounded \(\|O\|_2\), the typical values satisfy
\begin{equation}
    \mathbb E_\rho[V_1(O,\rho)]
    =
    \mathcal{O}(1/D),
    \qquad
    \mathbb E_\rho[V_2(O,\rho)]
    =
    \mathcal{O}(1).
\end{equation}

\subsection{Typical values of $V_1(\rho)$, $V_2(\rho)$, and $V_3(\rho)$}
\label{appx:typical_sd_V}

Let \(\mathbb U=\mathrm{Cl}(k)^{\otimes n/k}\), and assume that
\(k2^k=n/\alpha\). 
Using the standard fourth-moment identity for Haar-random pure states,
\begin{equation}
    \mathbb E_\rho[\rho^{\otimes 4}]
    =
    \frac{1}{D(D+1)(D+2)(D+3)}
    \sum_{\sigma\in S_4} U_\sigma,
\end{equation}
we can upper-bound \(\mathbb E_\rho[V_1(\rho)]\) as follows.
\begin{align}
    \mathbb{E}_\rho\!\left[ V_1(\rho) \right]
    &=
    \mathbb{E}_\rho\!\left[
    \sum_{P,Q}
    \frac{
    {\rm Tr}(\rho P)^2{\rm Tr}(\rho Q)^2
    }{D^2}
    \, f(P,Q)
    \right]  \\
    &=
    \sum_{P,Q}
    \frac{
    {\rm Tr}\!\left(
    \mathbb{E}_\rho\!\left[\rho^{\otimes 4}\right]
    (P \otimes P \otimes Q \otimes Q)
    \right)
    }{D^2}
    \, f(P,Q)\\
    &=
    \frac{1}{D^3(D+1)(D+2)(D+3)}
    \sum_{P,Q}
    f(P,Q)
    \sum_{\sigma\in S_4}
    {\rm Tr}\!\left(
    U_\sigma(P\otimes P\otimes Q\otimes Q)
    \right)\\
    &\le
    \frac{
    D^2\sum_{P,Q}f(P,Q)
    +{\cal O}\!\left(
        D^2\sum_P f(P,P)
        +D^2\sum_P f(P,I)
        +D\sum_{P,Q}f(P,Q)
    \right)
    }{
    D^3(D+1)(D+2)(D+3)
    }\\
    &\le
    1+{\cal O}(1/D).
\end{align}
The same third-moment calculation used in Appendix~\ref{appx:state_dependent_variance} gives
\begin{equation}
    \mathbb E_\rho[V_2(\rho)]=\mathcal{O}(1).
\end{equation}
It remains to bound \(V_3(\rho)\). 
Using the standard second-moment identity for Haar-random pure states,
\begin{equation}
    \mathbb E_\rho[\rho^{\otimes 2}]
    =
    \frac{1}{D(D+1)}
    \sum_{\sigma\in S_2}U_\sigma,
\end{equation}
we obtain
\begin{align}
    \mathbb{E}_\rho
    \sum_{P,Q}
    \frac{{\rm Tr}(\rho P Q)^2}{D^2}
    \, f(P,Q)
    &=
    \sum_{P,Q}
    \frac{
    {\rm Tr}\!\left(
    \mathbb{E}_\rho\!\left[\rho^{\otimes 2}\right]
    (PQ \otimes PQ)
    \right)
    }{D^2}
    \, f(P,Q) \\
    &=
    \frac{1}{D^3(D+1)}
    \left(
    \sum_{P,Q}
    {\rm Tr}(PQ)^2\, f(P,Q)
    +
    \sum_{P,Q}
    f(P,Q)
    \right)  \\
    &\le D e^{\alpha} + {\cal O}(1).
\end{align}
In the calculation above, the average for the typical case was taken only for pure states. To include mixed states in the averaging, we can purify the system by introducing an environment $E$ and work with the composite pure state $\rho_{\text{SE}}$. In that setting, every Pauli string $P$ used in the proof is replaced by $P\otimes I_E$, where $I_E$ is the identity operator in the environment $E$. Repeating the same steps under this extension shows that all resulting values are strictly lower (or at most equal) than those obtained in the pure state case.

\subsection{Worst-case values of \(V_1(\rho)\), \(V_2(\rho)\), and \(V_3(\rho)\)}
\label{appx:worst_V}

Let \(\mathbb{U}=\mathrm{Cl}(k)^{\otimes m}\), where \(m=n/k\), and assume \(k2^k=n/\alpha\). In the worst case, both \(V_1(\rho)\) and \(V_2(\rho)\) scale as \(\mathcal O(3^m)\). However, the conditions under which this scaling is attained are different. The upper bound for \(V_2(\rho)\) is already attained, up to constant factors, by block-product pure states
\[
    \rho=\ket{\Psi}\!\bra{\Psi},
    \qquad
    \ket{\Psi}=\bigotimes_{i=1}^{m}\ket{\Psi_i},
    \qquad
    \ket{\Psi_i}\in\mathcal H^{D_k}.
\]
By contrast, for \(V_1(\rho)\) to attain the same worst-case scaling, the local
states \(\ket{\Psi_i}\) must additionally be stabilizer states.

We now bound \(V_3(\rho)\). Recall that
\[
    V_3(\rho)
    =
    \frac{1}{D^2}
    \sum_{P,Q}
    {\rm Tr}(\rho PQ)^2 f(P,Q).
\]
Since \(f(P,Q)=0\) unless \(P\) and \(Q\) commute blockwise, the contributing
terms satisfy \(PQ=\pm R\) for a phase-free Pauli string \(R\). As in the
single-shot purity calculation, we group the sum by \(R=PQ\), but here the
coefficient is \(f(P,Q)\) rather than \(f(P,Q)^2\). Define
\[
    B_R
    :=
    \sum_{\substack{P,Q\in\mathcal P_n\\PQ=\pm R}}
    f(P,Q).
\]
Then
\begin{equation}
    V_3(\rho)
    =
    \frac{1}{D^2}
    \sum_{R\in\mathcal P_n}
    B_R\,{\rm Tr}(\rho R)^2 .
    \label{appx:eq:V3_grouped}
\end{equation}

The coefficient \(B_R\) factorizes over the \(m\) blocks. For a single block,
let
\[
    B_r^{\rm loc}
    :=
    \sum_{\substack{P,Q\in\mathcal P_k\\PQ=\pm r}}
    f(P,Q).
\]
Using Eq.~\eqref{eq:appx:f},
we obtain
\begin{align}
    B_I^{\rm loc}
    &=
    1+(D_k^2-1)(D_k+1)
    =
    D_k^3+D_k^2-D_k,\\
    B_{r(\neq I)}^{\rm loc}
    &=
    2+
    \left(\frac{D_k^2}{2}-2\right)
    \frac{2(D_k+1)}{D_k+2}
    =
    D_k^2-D_k.
\end{align}
Therefore, if \(A=\operatorname{supp}_k(R)\), then
\begin{equation}
    B_R
    =
    (D_k^3+D_k^2-D_k)^{m-|A|}
    (D_k^2-D_k)^{|A|}.
    \label{appx:eq:BR_bound}
\end{equation}

Let \(\mathcal P_A\) be the phase-free Pauli basis on the blocks in \(A\), and
let \(D_A=D_k^{|A|}\). Since the exact block-support condition only restricts
the Pauli strings in \(\mathcal P_A\), Pauli orthogonality gives
\begin{equation}
    \sum_{\operatorname{supp}_k(R)=A}
    {\rm Tr}(\rho R)^2
    \le
    \sum_{R_A\in\mathcal P_A}
    {\rm Tr}(\rho_A R_A)^2
    =
    D_A{\rm Tr}(\rho_A^2)
    \le
    D_k^{|A|}.
    \label{appx:eq:V3_RA_sum}
\end{equation}
Combining Eqs.~\eqref{appx:eq:V3_grouped},
\eqref{appx:eq:BR_bound}, and~\eqref{appx:eq:V3_RA_sum}, we obtain
\begin{align}
    V_3(\rho)
    &\le
    \frac{1}{D^2}
    \sum_{A\subseteq[m]}
    (D_k^3+D_k^2-D_k)^{m-|A|}
    (D_k^2-D_k)^{|A|}
    D_k^{|A|}\\
    &=
    \frac{1}{D^2}
    \left[
        D_k^3+D_k^2-D_k
        +
        D_k(D_k^2-D_k)
    \right]^m\\
    &=
    \frac{1}{D^2}
    (2D_k^3-D_k)^m\\
    &=
    \left(2D_k-D_k^{-1}\right)^m\\
    &\le
    (2D_k)^m
    =
    D\,2^m
    =
    2^n2^{n/k}.
\end{align}
Thus, in the worst case,
\begin{equation}
    V_3(\rho)
    =
    \mathcal O\!\left(2^n2^{n/k}\right).
\end{equation}

This upper bound is tight up to constant factors. For a block-product pure
state
\[
    \rho=\bigotimes_{i=1}^{m}\ket{\psi_i}\!\bra{\psi_i},
\]
the expression for \(V_3(\rho)\) factorizes over blocks. Using
\(\sum_{r\in\mathcal P_k}{\rm Tr}(\psi_i r)^2=D_k\) and
\({\rm Tr}(\psi_i I)^2=1\), the single-block contribution is
\[
    \frac{1}{D_k^2}
    \left[
        B_I^{\rm loc}
        +
        (D_k-1)B_{r\neq I}^{\rm loc}
    \right]
    =
    2D_k-1.
\]
Hence
\begin{equation}
    V_3(\rho)=(2D_k-1)^m,
\end{equation}
showing that the worst-case scaling is
\(V_3(\rho)=\Theta(D\,2^m)\).

\subsection{Upper bound on 
\texorpdfstring{\(\operatorname{Var}(\operatorname{Tr}(\hat{\rho}_{\sigma}P))\)}
{Var(Tr(rhohat_sigma P))}}
\label{appx:A16}

\begin{proof}
    Let \(U_1,\ldots,U_{N_U}\) be independently sampled measurement bases. 
    For each basis \(U_i\), let \(b_{ij}\) and \(c_{ij}\) denote the \(j\)-th measurement
    outcomes obtained from \(\rho\) and \(\sigma\), respectively. 
    We define
    \begin{equation}
        \hat{\rho}_{\sigma}
        =
        \hat{\rho}
        -
        \hat{\sigma}
        +
        \sigma,
    \end{equation}
    where
    \begin{equation}
        \hat{\rho}
        =
        \frac{1}{N_U}
        \sum_{i=1}^{N_U}
        \hat{\rho}_{U_i}(N_\rho),
        \qquad
        \hat{\sigma}
        =
        \frac{1}{N_U}
        \sum_{i=1}^{N_U}
        \hat{\sigma}_{U_i}(N_\sigma),
    \end{equation}
    with
    \begin{equation}
        \hat{\rho}_{U}(N_\rho)
        =
        \frac{1}{N_\rho}
        \sum_{j=1}^{N_\rho}
        \mathcal M^{-1}
        \!\left(
            U^\dagger |b_{ij}\rangle\langle b_{j}| U
        \right),
    \end{equation}
    and
    \begin{equation}
        \hat{\sigma}_{U}(N_\sigma)
        =
        \frac{1}{N_\sigma}
        \sum_{j=1}^{N_\sigma}
        \mathcal M^{-1}
        \!\left(
            U^\dagger |c_{j}\rangle\langle c_{j}| U
        \right).
    \end{equation}
    For a fixed basis \(U\), write
    \begin{equation}
        \hat{\rho}_{\sigma}(U)
        =
        \hat{\rho}_{U}(N_\rho)
        -
        \hat{\sigma}_{U}(N_\sigma)
        +
        \sigma .
    \end{equation}
    Since the \(N_U\) measurement bases are independent,
    \begin{equation}
        \operatorname{Var}
        \!\left(
            \operatorname{Tr}(\hat{\rho}_{\sigma}P)
        \right)
        =
        \frac{1}{N_U}
        \operatorname{Var}
        \!\left(
            \operatorname{Tr}(\hat{\rho}_{\sigma}(U)P)
        \right).
    \end{equation}
    
    Define
    \[
        A_U = \operatorname{Tr}(\hat{\rho}_{U}(N_\rho)P),
        \qquad
        B_U = \operatorname{Tr}(\hat{\sigma}_{U}(N_\sigma)P).
    \]
    Then
    \begin{equation}
        \operatorname{Tr}(\hat{\rho}_{\sigma}(U)P)
        =
        A_U - B_U + \operatorname{Tr}(\sigma P).
    \end{equation}
    Therefore,
    \begin{align}
        \operatorname{Var}
        \!\left(
            \operatorname{Tr}(\hat{\rho}_{\sigma}(U)P)
        \right)
        &=
        \operatorname{Var}(A_U-B_U) \\
        &=
        \mathbb E[(A_U-B_U)^2]
        -
        \operatorname{Tr}\!\left((\rho-\sigma)P\right)^2 .
    \end{align}
    We now compute the second moment. 
    By the multi-shot variance calculation in Appendix~\ref{appx:multi_shots_pauli},
    \begin{align}
        \mathbb E[A_U^2]
        &=
        m_P^{-1}
        \left(
            \frac{1}{N_\rho}
            +
            \frac{N_\rho-1}{N_\rho}
            \operatorname{Tr}(\rho P)^2
        \right),\\
        \mathbb E[B_U^2]
        &=
        m_P^{-1}
        \left(
            \frac{1}{N_\sigma}
            +
            \frac{N_\sigma-1}{N_\sigma}
            \operatorname{Tr}(\sigma P)^2
        \right).
    \end{align}
    It remains to compute the cross term. We have
    \begin{align}
        \mathbb E[A_U B_U]
        &=
        \frac{1}{N_\rho N_\sigma}
        \sum_{j=1}^{N_\rho}
        \sum_{\ell=1}^{N_\sigma}
        \mathbb E
        \!\left[
            m_P^{-2}
            (U,P,b_j)(U,P,c_\ell)
        \right] \\
        &=
        m_P^{-2}
        \mathbb E_U
        \sum_{b,c}
        (U,P,b)(U,P,c)\,
        \operatorname{Tr}
        \!\left(
            U\rho U^\dagger |b\rangle\langle b|
        \right)
        \operatorname{Tr}
        \!\left(
            U\sigma U^\dagger |c\rangle\langle c|
        \right) \\
        &=
        m_P^{-2}
        \mathbb E_U
        \sum_{b,c}
        \sum_{Q,R}
        (U,P,b)(U,Q,b)
        (U,P,c)(U,R,c)\,
        \frac{\operatorname{Tr}(\rho Q)\operatorname{Tr}(\sigma R)}{D^2} \\
        &=
        m_P^{-2}
        \mathbb E_U
        \!\left[
            \mathbf 1\{UPU^\dagger\in\pm\mathcal Z\}
        \right]
        \operatorname{Tr}(\rho P)\operatorname{Tr}(\sigma P) \\
        &=
        m_P^{-1}
        \operatorname{Tr}(\rho P)\operatorname{Tr}(\sigma P).
    \end{align}
    Combining the above identities gives 
    \begin{align}
        &\operatorname{Var}
        \!\left(
            \operatorname{Tr}(\hat{\rho}_{\sigma}P)
        \right) \notag \\
        &=
        \frac{1}{N_U}
        \Bigg[
        m_P^{-1}
        \left(
            \frac{1}{N_\rho}
            +
            \frac{N_\rho-1}{N_\rho}\operatorname{Tr}(\rho P)^2
            +
            \frac{1}{N_\sigma}
            +
            \frac{N_\sigma-1}{N_\sigma}\operatorname{Tr}(\sigma P)^2
            -
            2\operatorname{Tr}(\rho P)\operatorname{Tr}(\sigma P)
        \right) \nonumber\\
        &\qquad\qquad\qquad\qquad\qquad\qquad\qquad\qquad\qquad\qquad-
        \operatorname{Tr}\!\left((\rho-\sigma)P\right)^2
        \Bigg]\\
        &\le
        \frac{m_P^{-1}}{N_U}
        \left(
            \frac{1-\operatorname{Tr}(\rho P)^2}{N_\rho}
            +
            \frac{1-\operatorname{Tr}(\sigma P)^2}{N_\sigma}
            +
            \Tr((\rho-\sigma)P)^2
        \right)\label{appx:eq:crm_pauli}.
    \end{align}
    This proves the desired bound.
\end{proof}

We now consider a variant in which the bias contribution is not evaluated exactly,
but is estimated using an additional independent classical shadow of \(\sigma\).
Let
\begin{equation}
    \hat{\rho}_{\sigma}
    =
    \hat{\rho}
    -
    \hat{\sigma}_1
    +
    \hat{\sigma}_2,
\end{equation}
where \(\hat{\rho}\) and \(\hat{\sigma}_1\) are constructed using the same
\(N_U\) measurement bases, while \(\hat{\sigma}_2\) is constructed independently
from \(N_U'\) additional measurement bases applied to \(\sigma\). 
For a Pauli string \(P\), we have
\begin{align}
    \operatorname{Var}
    \!\left(
        \operatorname{Tr}(P\hat{\rho}_{\sigma})
    \right)
    &=
    \operatorname{Var}
    \!\left(
        \operatorname{Tr}
        \!\left(
            P(\hat{\rho}-\hat{\sigma}_1)
        \right)
        +
        \operatorname{Tr}(P\hat{\sigma}_2)
    \right) \\
    &=
    \operatorname{Var}
    \!\left(
        \operatorname{Tr}
        \!\left(
            P(\hat{\rho}-\hat{\sigma}_1)
        \right)
    \right)
    +
    \operatorname{Var}
    \!\left(
        \operatorname{Tr}(P\hat{\sigma}_2)
    \right).
\end{align}
Here, the second equality follows from the independence of \(\hat{\sigma}_2\)
from \(\hat{\rho}\) and \(\hat{\sigma}_1\). 
Since \(\hat{\sigma}_2\) is averaged over \(N_U'\) independent measurement bases,
\begin{equation}
    \operatorname{Var}
    \!\left(
        \operatorname{Tr}(P\hat{\sigma}_2)
    \right)
    =
    \frac{1}{N_U'}
    \operatorname{Var}
    \!\left(
        \operatorname{Tr}(P\hat{\sigma}_{2,U})
    \right).
\end{equation}
Thus, when \(N_U'\gg N_U\), the second term is negligible compared with the
shared-basis contribution. In this limit, the variance reduces to
Eq.~\eqref{appx:eq:crm_pauli}.

\subsection{Proof of the ${\rm{Tr}}(P{\hat \sigma _{{\rm{old}}}}) = m_P^{ - 1}{\rm{Tr}}(\sigma P){\bf{1}}\{ UP{U^{\dagger}} \in  \pm {\cal Z}\}$}
\label{appx:old_pauli}

\begin{proof}
    Let ${\hat \sigma _{{\rm{old}}}} = {\sum _b}{\rm{Tr}}(U\sigma {U^{\dagger}}|b\rangle \langle b|){{\cal M}^{ - 1}}({U^{\dagger}}|b\rangle \langle b|U)$, then 
    \begin{align}
        {\rm Tr}(P\hat\sigma_{\rm old})
        &=
        \sum_b
        {\rm Tr}\!\left(
        U\sigma U^{\dagger}|b\rangle\langle b|
        \right)
        {\rm Tr}\!\left(
        P{\cal M}^{-1}
        \!\left(
        U^{\dagger}|b\rangle\langle b|U
        \right)
        \right) \\
        &=
        \sum_b
        m_P^{-1}
        {\rm Tr}\!\left(
        U\sigma U^{\dagger}|b\rangle\langle b|
        \right)
        {\rm Tr}\!\left(
        UPU^{\dagger}|b\rangle\langle b|
        \right) \\
        &=
        \sum_b
        \sum_Q
        m_P^{-1}
        {\rm Tr}\!\left(
        UQU^{\dagger}|b\rangle\langle b|
        \right)
        {\rm Tr}\!\left(
        UPU^{\dagger}|b\rangle\langle b|
        \right)
        \frac{{\rm Tr}(\sigma Q)}{D} \\
        &=
        \sum_b
        \sum_Q
        m_P^{-1}
        (U,Q,b)(U,P,b)\,
        {\rm Tr}(\sigma Q)\,
        \mathbf{1}\{UPU^{\dagger}\in\pm{\cal Z}\}\,
        \mathbf{1}\{UQU^{\dagger}\in\pm{\cal Z}\}
        /D \\
        &=
        \sum_Q
        m_P^{-1}
        \delta_{P,Q}\,
        {\rm Tr}(\sigma Q)\,
        \mathbf{1}\{UPU^{\dagger}\in\pm{\cal Z}\}\,
        \mathbf{1}\{UQU^{\dagger}\in\pm{\cal Z}\} \\
        &=
        m_P^{-1}
        {\rm Tr}(\sigma P)\,
        \mathbf{1}\{UPU^{\dagger}\in\pm{\cal Z}\}.
    \end{align}
\end{proof}

\subsection{Shadow channel $\widetilde{\mathcal{M}}$ in the presence of errors.}
\label{appx:noisy_shallow}

In~\ref{appx:A1}, we have shown that, in the absence of errors, the eigenoperators of the shadow channel $\mathcal{M}$ are the Pauli strings $P$, and their corresponding eigenvalues are  $m_P=\mathbb{E}{_U}[{\bf{1}}\{ UP{U^\dagger } \in  \pm {\cal Z}\} ] = {\Pr _U}(UP{U^\dagger } \in  \pm {\cal Z})$. In this section, we will extend this to the regime with the presence of errors. First, since the unitaries we used in RM consist solely of Clifford gates, we can transform any arbitrary error channel in RM into a Pauli error channel through Pauli Twirling~\cite{wallman2016noise_40}. Previously, two main models of error channel have been studied: the first~\cite{chen2021robust_18}, $\Lambda_U = \Lambda\cdot\mathcal{U}$, and the second~\cite{Rozon2024optimal_53}, $\Lambda_U = \Lambda_t\cdot\mathcal{U}_t … \Lambda_1\cdot\mathcal{U}_1$ where, where $\mathcal{U}_{(l)}$ represents superoperator of $U_{(l)}$ and $\Lambda_{(l)}$ is the quantum error channel (at the $l$th layer) and $U = U_t ...U_1$. Because two error channels are Pauli error channel, denote $\Lambda_U(P) = \lambda_{U,P}UPU^{\dagger}$ (For the first error model $\lambda_{U,P}=\lambda_{UPU^{\dagger}}$ and for the second error model $\lambda_{U,P}={\lambda _{{U_1}P{U_1}^\dagger }}...{\lambda _{{U_t}...{U_1}P{U_1}^\dagger ...{U_t}^\dagger }}$). 
\begin{align}
\widetilde{\cal M}(P)
&=
    \mathbb{E}_U\!\left[
    \sum_b
    U^{\dagger}|b\rangle\langle b|U \cdot
    {\rm Tr}\!\left(\Lambda_U(P)|b\rangle\langle b|\right)
    \right] \\
    &=
    \mathbb{E}_U\!\left[
    \sum_b
    U^{\dagger}|b\rangle\langle b|U \cdot
    \lambda_{U,P}\,
    {\rm Tr}\!\left(
    UPU^{\dagger}|b\rangle\langle b|
    \right)
    \right] \\
    &=
    \mathbb{E}_U\!\left[
    \sum_b
    \sum_Q
    \lambda_{U,P}\,
    {\rm Tr}\!\left(
    UPU^{\dagger}|b\rangle\langle b|
    \right)
    {\rm Tr}\!\left(
    UQU^{\dagger}|b\rangle\langle b|
    \right)
    Q/D
    \right] \\
    &=
    \mathbb{E}_U\!\left[
    \sum_b
    \sum_Q
    \lambda_{U,P}\,
    {\rm Tr}\!\left(
    UPU^{\dagger}|b\rangle\langle b|
    \right)
    {\rm Tr}\!\left(
    UQU^{\dagger}|b\rangle\langle b|
    \right)
    \right. \nonumber\\
    &\qquad\qquad\left.
    \cdot
    \mathbf{1}\{UPU^{\dagger}\in\pm{\cal Z}\}
    \cdot
    \mathbf{1}\{UQU^{\dagger}\in\pm{\cal Z}\}
    Q/D
    \right] \\
    &=
    \mathbb{E}_U\!\left[
    \sum_Q
    \lambda_{U,P}\,
    \delta_{P,Q}\,
    \mathbf{1}\{UPU^{\dagger}\in\pm{\cal Z}\}
    \cdot
    \mathbf{1}\{UQU^{\dagger}\in\pm{\cal Z}\}
    \right] Q \\
    &=
    \mathbb{E}_U\!\left[
    \lambda_{U,P}\,
    \mathbf{1}\{UPU^{\dagger}\in\pm{\cal Z}\}
    \right] P \\
    &=
    \widetilde m_P P
\end{align}
Because $\lambda_{U,P} \leq 1$, $\widetilde{m}_P$ is always less than or equal to $m_P$.

\subsection{Upper bound on 
\texorpdfstring{\(\operatorname{Var}(\operatorname{Tr}(\hat{\rho}P))\)}
{Var(Tr(rhohat P))} in the presence of errors}
\label{appx:noisy_multi}

We consider the noisy shadow channel
\begin{equation}
    \widetilde{\mathcal M}(A)
    =
    \mathbb{E}_{U}
    \sum_{b\in\{0,1\}^n}
    U^\dagger |b\rangle\langle b| U\,
    \operatorname{Tr}\!\left(
        \Lambda_U(A)|b\rangle\langle b|
    \right).
\end{equation}
After Pauli twirling, we assume that the effective noisy measurement channel
acts on Pauli operators as
\begin{equation}
    \Lambda_U(P)
    =
    \lambda_{U,P} UPU^\dagger .
\end{equation}
In particular, the channel is unital, so \(\lambda_{U,I}=1\). 
As shown in the previous section, each Pauli string \(P\) remains an eigenoperator of
\(\widetilde{\mathcal M}\), with eigenvalue
\begin{equation}
    \widetilde m_P
    =
    \mathbb{E}_{U}
    \!\left[
        \lambda_{U,P}
        \mathbf 1\{UPU^\dagger\in\pm\mathcal Z\}
    \right],
\end{equation}
where \(\mathcal Z=\{I,Z\}^n\).

For the multi-shot estimator, let \(U_1,\ldots,U_{N_U}\) be independently sampled
measurement bases. 
For each basis \(U_i\), let \(b_{ij}\) be the \(j\)-th measurement outcome from
the noisy measurement of \(\rho\). 
We define
\begin{equation}
    \hat\rho
    =
    \frac{1}{N_U}
    \sum_{i=1}^{N_U}
    \hat\rho_{U_i}(N_S),
    \qquad
    \hat\rho_{U_i}(N_S)
    =
    \frac{1}{N_S}
    \sum_{j=1}^{N_S}
    \hat\rho_{U_i,j},
\end{equation}
where
\begin{equation}
    \hat\rho_{U_i,j}
    =
    \widetilde{\mathcal M}^{-1}
    \!\left(
        U_i^\dagger |b_{ij}\rangle\langle b_{ij}| U_i
    \right).
\end{equation}
Then
\begin{equation}
    \operatorname{Tr}(\hat\rho_{U,j}P)
    =
    \widetilde m_P^{-1}(U,P,b_j).
\end{equation}
Since the \(N_U\) measurement bases are independent, we have
\begin{equation}
    \operatorname{Var}\!\left(\operatorname{Tr}(\hat\rho P)\right)
    =
    \frac{1}{N_U}
    \operatorname{Var}\!\left(
        \operatorname{Tr}(\hat\rho_U(N_S)P)
    \right).
\end{equation}
Moreover,
\begin{align}
    &\operatorname{Var}\!\left(
        \operatorname{Tr}(\hat\rho_U(N_S)P)
    \right) \notag \\
    &=
    \mathbb{E}
    \!\left[
        \operatorname{Tr}(\hat\rho_U(N_S)P)^2
    \right]
    -
    \operatorname{Tr}(\rho P)^2 .
\end{align}
The first term is
\begin{align}
    &\mathbb{E}
    \!\left[
        \operatorname{Tr}(\hat\rho_U(N_S)P)^2
    \right] \notag \\
    &=
    \frac{1}{N_S}
    \mathbb{E}
    \!\left[
        \operatorname{Tr}(\hat\rho_{U,1}P)^2
    \right]
    +
    \frac{N_S-1}{N_S}
    \mathbb{E}
    \!\left[
        \operatorname{Tr}(\hat\rho_{U,1}P)
        \operatorname{Tr}(\hat\rho_{U,2}P)
    \right].
\end{align}
For the diagonal contribution,
\begin{align}
    \mathbb{E}
    \!\left[
        \operatorname{Tr}(\hat\rho_{U,1}P)^2
    \right]
    &=
    \widetilde m_P^{-2}
    \mathbb{E}_{U}
    \sum_{b}
    (U,P,b)^2
    \operatorname{Tr}
    \!\left(
        \Lambda_U(\rho)|b\rangle\langle b|
    \right) \\
    &=
    \widetilde m_P^{-2}
    \mathbb{E}_{U}
    \sum_{b}
    \sum_Q
    \lambda_{U,Q}
    (U,P,b)^2
    (U,Q,b)
    \frac{\operatorname{Tr}(\rho Q)}{2^n} \\
    &=
    \widetilde m_P^{-2}
    \mathbb{E}_{U}
    \!\left[
        \mathbf 1\{UPU^\dagger\in\pm\mathcal Z\}
    \right] \\
    &=
    \widetilde m_P^{-2} m_P .
\end{align}
Here we used \(\lambda_{U,I}=1\) in the third line. 
For the off-diagonal contribution,
\begin{align}
    &\mathbb{E}
    \!\left[
        \operatorname{Tr}(\hat\rho_{U,1}P)
        \operatorname{Tr}(\hat\rho_{U,2}P)
    \right] \notag \\
    &=
    \widetilde m_P^{-2}
    \mathbb{E}_{U}
    \sum_{b_1,b_2}
    (U,P,b_1)(U,P,b_2)
    \operatorname{Tr}
    \!\left(
        \Lambda_U(\rho)|b_1\rangle\langle b_1|
    \right)
    \operatorname{Tr}
    \!\left(
        \Lambda_U(\rho)|b_2\rangle\langle b_2|
    \right) \\
    &=
    \widetilde m_P^{-2}
    \operatorname{Tr}(\rho P)^2
    \mathbb{E}_{U}
    \!\left[
        \lambda_{U,P}^2
        \mathbf 1\{UPU^\dagger\in\pm\mathcal Z\}
    \right].
\end{align}
Combining these identities gives
\begin{align}
    \operatorname{Var}\!\left(
        \operatorname{Tr}(\hat\rho P)
    \right)
    =
    \frac{1}{N_U}
    \Bigg[
    \widetilde m_P^{-2}
    \Bigg(
        \frac{m_P}{N_S}
        +
        \frac{N_S-1}{N_S}
        \operatorname{Tr}(\rho P)^2
        \mathbb{E}_{U}
        \!\left[
            \lambda_{U,P}^2
            \mathbf 1\{UPU^\dagger\in\pm\mathcal Z\}
        \right]
    \Bigg)
    -
    \operatorname{Tr}(\rho P)^2
    \Bigg].
\end{align}
In the noiseless case, \(\lambda_{U,P}=1\) and \(\widetilde m_P=m_P\), so this reduces to
\begin{equation}
    \operatorname{Var}\!\left(
        \operatorname{Tr}(\hat\rho P)
    \right)
    =
    \frac{1}{N_U}
    \left[
        m_P^{-1}
        \left(
            \frac{1}{N_S}
            +
            \frac{N_S-1}{N_S}
            \operatorname{Tr}(\rho P)^2
        \right)
        -
        \operatorname{Tr}(\rho P)^2
    \right],
\end{equation}
which agrees with the noiseless multi-shot formula.

\subsection{Alternative noise shadow channel}
In the main text, we used the noise shadow channel defined by
\begin{equation}
    {\widetilde {\cal M}_1}(\rho ) = {\mathbb{E}_{U,b}}[{U^\dagger }|b\rangle \langle b|U] = {\mathbb{E}_U}\left[{\sum _b}{U^{\dagger}}|b\rangle \langle b|U{\rm{Tr}}({\Lambda _U}(\rho )|b\rangle \langle b|)\right].
\end{equation}
However, in the presence of errors, it is possible to utilize new forms of shadow channels in addition to the $\widetilde{\mathcal{M}}_1$, such as the one described by 
\begin{equation}
    {\widetilde {\cal M}_2}(\rho ) = {\mathbb{E}_{U,b}}[\Lambda _U^\dagger (|b\rangle \langle b|)] = {\mathbb{E}_U}\left[{\sum _b}\Lambda _U^\dagger (|b\rangle \langle b|){\rm{Tr}}({\Lambda _U}(\rho )|b\rangle \langle b|)\right],
\end{equation}
where  $\Lambda _U^{\dagger} $ is defined by 
\begin{equation}
    {\rm{Tr}}(A{\Lambda _U}(B)) = {\rm{Tr}}(\Lambda _U^\dagger (A)B).
\end{equation}
When using the shadow channel $\widetilde{\mathcal{M}}_2$, the unbiased estimator $\hat{\rho}$ for the quantum state $\rho$ can be written as 
\begin{equation}
    \hat \rho  = \widetilde {\cal M}_2^{ - 1}(\Lambda _U^\dagger (|b\rangle \langle b|)).
\end{equation}

\begin{proof}
    The unbiasedness of $\hat{\rho}$ can be proven as follow:
    \begin{align}
        \mathbb{E}_{U,b}[\hat\rho]
        &=
        \mathbb{E}_U\!\left[
        \sum_b
        \widetilde{\cal M}_2^{-1}
        \!\left(
        \Lambda_U^{\dagger}(|b\rangle\langle b|)
        \right)
        {\rm Tr}\!\left(
        \Lambda_U(\rho)\,|b\rangle\langle b|
        \right)
        \right]  \\
        &=
        \mathbb{E}_U\!\left[
        \sum_b
        \sum_P
        {\rm Tr}\!\left(
        \widetilde{\cal M}_2^{-1}
        \!\left(
        \Lambda_U^{\dagger}(|b\rangle\langle b|)
        \right) P
        \right)
        {\rm Tr}\!\left(
        \Lambda_U(\rho)\,|b\rangle\langle b|
        \right)
        P/D
        \right]  \\
        &=
        \mathbb{E}_U\!\left[
        \sum_b
        \sum_P
        {\rm Tr}\!\left(
        \Lambda_U^{\dagger}(|b\rangle\langle b|)
        \widetilde{\cal M}_2^{-1}(P)
        \right)
        {\rm Tr}\!\left(
        \Lambda_U(\rho)\,|b\rangle\langle b|
        \right)
        P/D
        \right]  \\
        &=
        \sum_P
        {\rm Tr}\!\left(
        \widetilde{\cal M}_2(\rho)\,
        \widetilde{\cal M}_2^{-1}(P)
        \right)
        P/D  \\
        &=
        \sum_P
        {\rm Tr}(\rho P)\,
        P/D  \\
        &=
        \rho.
    \end{align}By using this, $\hat o = {\rm{Tr}}(O\hat \rho ) = {\rm{Tr}}(O\widetilde {\cal M}_2^{ - 1}(\Lambda _U^\dagger (|b\rangle \langle b|)))$ is an unbiased estimator of observable $O$. Similarly, in $\widetilde{\mathcal{M}}_2$, it can be proven that the Pauli string $P$ is an eigenoperator as follows:
    \begin{align}
        \widetilde{\cal M}_2(P)
        &=
        \mathbb{E}_U\!\left[
        \sum_b
        \Lambda_U^{\dagger}(|b\rangle\langle b|)\,
        {\rm Tr}\!\left(\Lambda_U(P)\,|b\rangle\langle b|\right)
        \right] \\
        &=
        \mathbb{E}_U\!\left[
        \sum_b
        \lambda_{U,P}\,
        \Lambda_U^{\dagger}(|b\rangle\langle b|)\,
        {\rm Tr}\!\left(UPU^{\dagger}|b\rangle\langle b|\right)
        \right]  \\
        &=
        \mathbb{E}_U\!\left[
        \sum_b
        \sum_Q
        \lambda_{U,P}\,
        {\rm Tr}\!\left(
        \Lambda_U^{\dagger}(|b\rangle\langle b|)\,Q
        \right)
        {\rm Tr}\!\left(
        UPU^{\dagger}|b\rangle\langle b|
        \right)
        \frac{Q}{D}
        \right]  \\
        &=
        \mathbb{E}_U\!\left[
        \sum_b
        \sum_Q
        \lambda_{U,P}\,
        {\rm Tr}\!\left(
        |b\rangle\langle b|\,\Lambda_U(Q)
        \right)
        {\rm Tr}\!\left(
        UPU^{\dagger}|b\rangle\langle b|
        \right)
        \frac{Q}{D}
        \right]  \\
        &=
        \mathbb{E}_U\!\left[
        \sum_b
        \sum_Q
        \lambda_{U,P}\lambda_{U,Q}\,
        {\rm Tr}\!\left(
        UQU^{\dagger}|b\rangle\langle b|
        \right)
        {\rm Tr}\!\left(
        UPU^{\dagger}|b\rangle\langle b|
        \right)
        \frac{Q}{D}
        \right]  \\
        &=
        \mathbb{E}_U\!\left[
        \lambda_{U,P}^2\,
        \mathbf{1}\{UPU^{\dagger}\in\pm{\cal Z}\}
        \right] P \\
        &=
        \widetilde m_{P,2}\,P.
    \end{align}
    In the noise shadow channel $\widetilde{\cal M}_1$, the Pauli string $P$ is an eigenoperator, albeit with eigenvalue  $\mathbb{E}{_U}[{\lambda _{U,P}}{\bf{1}}\{ UP{U^\dagger } \in  \pm {\cal Z}\} ]$.

    Shadow norm computed by the noise shadow channel $\widetilde{\cal M}_2$ is given by
    \begin{align}
    \left\| P \right\|_{{\rm sh},2}^2
    &=
    \mathbb{E}_{U,b}\Bigl[
    {\rm Tr}\!\left(
    P\,\widetilde{\cal M}_2^{-1}
    \!\left(
    \Lambda_U^{\dagger}(|b\rangle\langle b|)
    \right)
    \right)^2
    \Bigr] \\
    &=
    \mathbb{E}_{U,b}\Bigl[
    \widetilde m_{P,2}^{-2}\,
    {\rm Tr}\!\left(
    \Lambda_U(P)\,|b\rangle\langle b|
    \right)^2
    \Bigr] \\
    &=
    \widetilde m_{P,2}^{-2}\,
    \mathbb{E}_U\!\left[
    \sum_b
    {\rm Tr}\!\left(
    \Lambda_U(P)\,|b\rangle\langle b|
    \right)^2
    {\rm Tr}\!\left(
    \Lambda_U(\rho)\,|b\rangle\langle b|
    \right)
    \right] \\
    &=
    \widetilde m_{P,2}^{-2}\,
    \mathbb{E}_U\!\left[
    \sum_b
    \sum_Q
    {\rm Tr}\!\left(
    \Lambda_U(P)\,|b\rangle\langle b|
    \right)^2
    {\rm Tr}\!\left(
    \Lambda_U(Q)\,|b\rangle\langle b|
    \right)
    \frac{{\rm Tr}(\rho Q)}{D}
    \right] \\
    &=
    \widetilde m_{P,2}^{-2}\,
    \mathbb{E}_U\!\left[
    \sum_b
    \sum_Q
    \lambda_{U,P}^2\lambda_{U,Q}\,
    (U,P,b)(U,P,b)(U,Q,b)\,
    \frac{{\rm Tr}(\rho Q)}{D}
    \right] \\
    &=
    \widetilde m_{P,2}^{-2}\,
    \mathbb{E}_U\!\left[
    \lambda_{U,P}^2\lambda_{U,I}\,
    \mathbf{1}\{UPU^{\dagger}\in\pm{\cal Z}\}
    \right] \\
    &=
    \widetilde m_{P,2}^{-1}.
    \end{align}
    When compared with the shadow norm $||P||_{{\rm{sh,1}}}^2 = \widetilde m_{P,1}^{-2}{m_{P,1}}$ obtained through ${\widetilde {\cal M}_1}$, following inequality always holds 
    \[\widetilde m_{P,2}^{ - 1} \le \widetilde m_{P,1}^{ - 2}{m_{P,1}}.\]
    It means that using ${\widetilde {\cal M}_2}$ has lower variance than ${\widetilde {\cal M}_1}$. The difference between two shadow channels decreases as  ${\lambda _{U,P}}$s become uniform across all Pauli string $P$. In actual experiments, since ${\widetilde m_{P,2}}$ is smaller than ${\widetilde m_{P,1}}$, more calibration experiments are needed to accurately estimate ${\widetilde m_{P,2}}$  than ${\widetilde m_{P,1}}$. Therefore, when using  ${\widetilde {\cal M}_2}$ is better than using  ${\widetilde {\cal M}_1}$ is unclear, we will leave this for future research.
\end{proof}

\subsection{Matrix Bernstein inequality for block shadow}
\label{appx:shadow_kernel}

Although block shadow does not always provide more accurate measurements than random Pauli measurement (RPM) for every physical observable—for instance, consider a Pauli string in which, within each block, only one Pauli matrix is nontrivial while all others are identities; in this case, the variance under RPM is $\mathcal{O}(3^{n/k})$, whereas for block shadow, it reaches the maximum value of $(2^k + 1)^{n/k}$ —using shallow shadow can offer favorable scaling compared to RPM if the region of interest is contiguous. For example, to analyze how few measurements are required for quantum state tomography on a contiguous region A of $r (\leq n)$ qubits using block shadow, one can invoke the following matrix Bernstein inequality~\cite{guctua2020fast}.

Let $X_1, X_2, ... X_T$ defined by 
\begin{equation}
    {X_t} =  \bigotimes _{i = 1}^{r/k}\left(({2^k} + 1){\rm{ }}U_i^{\left( t \right)\dagger }|b_i^{(t)}\rangle \langle b_i^{(t)}|U_i^{(t)} - I\right)
\end{equation}
be independent random $D_r$-dimensional matrices with  $\mathbb{E}[{X_t}] = {\rho _A}$ and $\|{X_t} - E[{X_t}]\|_\infty \leq R$ almost surely, then the following holds
\begin{equation}
    \Pr\!\left[
    \left\|
    \frac{1}{T}\sum_t \bigl(X_t - \mathbb{E}[X_t]\bigr)
    \right\|_\infty
    \ge \epsilon
    \right]
    \le
    2D_r \exp\!\left(
    -\frac{T\epsilon^2/2}{\sigma^2 + R\epsilon/3}
    \right)
\end{equation}
where ${\sigma ^2} = {\rm{ }}||1/T{\sum _t}\mathbb{E}[X_t^2]|{|_\infty }$. First, it is straightforward to compute $R = 2^r+1$. $\sigma^2$ can be determined as follows:
Let $X =  \otimes _{i = 1}^{r/k}[({2^k} + 1)U_i^{\dagger}|{b_i}\rangle \langle {b_i}|{U_i}) - I]$, then ${X^2} =  \otimes _{i = 1}^{r/k}[({4^k} - 1)U_i^{\dagger}|{b_i}\rangle \langle {b_i}|{U_i}) + I]$ and 
\begin{equation}
    {\sigma ^2} = {\left\| {\mathbb{E}[{X^2}]} \right\|_\infty } \le \sum _{l = 0}^{r/k}\binom{r/k}{l}{2^{kl}}{({2^k} - 1)^{r/k - l}} = {({2^{k + 1}} - 1)^{r/k}}
\end{equation}
When $k = 1$, $\sigma^2 \leq 3^r$, which is consistent with previous results~\cite{guctua2020fast, huang2022provably_55}. For $k = 2$, $\sigma^2 \leq 7^{r/2} = (2.646...)^r$, and as $k$ increases, it gradually approaches to $\mathcal{O}(2^r)$. Building on this, we can obtain a tail bound for trace norm $\|\cdot\|_1$:
\begin{align}
    \Pr\!\left[
    \left\|
    \frac{1}{T}\sum_t \bigl(X_t - \mathbb{E}[X_t]\bigr)
    \right\|_1
    \ge \epsilon
    \right]
    &\le
    2D_r \exp\!\left(
    -\frac{T\epsilon^2/2}{4^r(\sigma^2 + R\epsilon/3)}
    \right) \\
    &\le
    2D_r \exp\!\left(
    -\frac{3T\epsilon^2}{8 \times 4^r \bigl((2^{k+1}-1)^{r/k}\bigr)}
    \right)
\end{align}
Based on these results, setting $T = \mathcal{O}(4^r(2^{k+1} - 1)^{r/k}/\epsilon^2)$ ensures that the empirical average converges to $\rho_A$ with high probability. Moreover, by following the remaining steps outlined in the Appendix of~\cite{huang2022provably_55}, one can prove that the required sample complexity $T$ for training can be reduced accordingly.

\subsection{Estimation of spectral form factor}
\label{appx:sff}
It has been suggested that the spectral form factor (SFF), denoted as 
\begin{equation}
    K(t)= \frac{1}{4^n} \mathbb{E}_T[\text{Tr}(T(t))\text{Tr}(T(t)^{\dagger})],
\end{equation}
can be utilized to distinguish whether quantum dynamics driven by a Hamiltonian is chaotic, consistent with predictions from random matrix theory, or many-body localized. Most of the previous analyses in our work involved calculations dependent on specific quantum states, whereas the SFF differs in that it is determined not by particular quantum states $\rho$ but by the underlying quantum dynamics $U =\exp(-iHt)$. Nevertheless, we show that by employing the block shadow method introduced in the main text, it is possible to achieve the same sample complexity scaling as obtained using global Clifford circuits, even with circuits of $\mathcal{O}(\log n)$ depth. We used the same experimental sequences as described in~\cite{joshi2022probing_34}. The key difference is that, instead of $U \sim \text{Cl}(1)^{\otimes n}$, we used  $U \sim \text{Cl}(k)^{\otimes n/k}$. For further details regarding the measurement protocol, we refer the reader to~\cite{joshi2022probing_34}. We can use an unbiased estimator for $K(t)$ as follows:
\begin{equation}\label{eq:sff}
    \hat K(t) = \frac{1}{M}\sum\limits_{r = 1}^M {{{( - {2^k})}^{ - |{s^{(r)}}{|_k}}}},
\end{equation}
where $|s|_k$ is the number of blocks for which the measurement bitstrings of that block differs from the all-zero string 0…0. It can be readily shown that Eq.~\eqref{eq:sff} is an unbiased estimator, and its variance is calculated as follows:

\begin{align}
    {\rm Var}[\hat K(t)]
    &=
    \frac{{\rm Var}\!\left[( - 2^k)^{-\,|s|_k}\right]}{M} \\
    &=
    \frac{1}{M}
    \left(
    \frac{1}{2^n}
    \left(
    1 + \frac{1}{2^k} - \frac{1}{4^k}
    \right)^{n/k}
    - K(t)^2
    \right)
\end{align}

Thus, from the above calculation, we see that when $k2^k = \Omega(n)$, the scaling matches that of the global Clifford case ($k = n$).

\section{Experiment details}
In this section, we describe the details of the experiments presented in the main text. To verify whether theoretically guaranteed advantages still persist despite the additional errors in two-qubit gates introduced by shallow measurements, we conducted experiments on a currently available quantum computer. All experiments were performed on a superconducting qubit quantum computer accessible through the IBM Cloud, specifically using the \texttt{ibm\_marrakesh} device.

\begin{table}[t]
    \centering
    \begin{tabular}{|c|c|}
    \hline
    Quantity & Median value \\
    \hline
    $T_1$ & $188.55~\mu\mathrm{s}$ \\
    $T_2$ & $113.97~\mu\mathrm{s}$ \\
    SX error & $2.532\times 10^{-4}$ \\
    X error & $2.532\times 10^{-4}$ \\
    CZ error & $2.227\times 10^{-3}$ \\
    Readout error & $1.465\times 10^{-2}$ \\
    \hline
    \end{tabular}
    \caption{\texttt{ibm\_marrakesh}. Median values of the \(T_1\) and \(T_2\) times, gate errors (SX, X, CZ), and readout error.}
    \label{tab:ibm_marrakesh}
\end{table}

The \texttt{ibm\_marrakesh} device features a heavy-hexagonal 2D lattice geometry, in which each qubit is connected to two or three neighboring qubits. Its native gate set consists of \{CZ, SX, X, RZ\}. To implement the unitaries required in our experiments, we decomposed them into sequences of native gates. The error rates of each hardware component are listed in Table~\ref{tab:ibm_marrakesh}. Because the hardware performance fluctuates over time, we selected the best-performing qubit layout at the time of each run, utilizing the Python package \texttt{mapomatic}~\cite{nation2023suppressing} with minor changes. In all hardware experiments reported in the main text, we applied dynamical decoupling~\cite{viola1999dynamical} (DD) and Pauli twirling (also known as randomized compiling)~\cite{wallman2016noise_40} to every circuit. Additional error mitigation methods specific to each experiment are described in the corresponding sections. All experiments were carried out using the Python package \texttt{Qiskit}~\cite{javadi2024quantum}.

\subsection{Derandomized block shadow (Fig.~\ref{fig2} in the main text)}
\label{appx:exp_derand}
\subsubsection{A set of measured Pauli strings}
In various variational algorithms~\cite{peruzzo2014variational_35}, the physical quantity that needs to be measured is often predetermined, and in such cases, using Randomized Measurement (RM) may not be optimal. By employing a technique called derandomization~\cite{hu2025demonstration_24}, it has been shown that, on average, one can achieve better performance than standard RM. Our goal in the main text was to extend this approach from single-qubit measurements to shallow measurements. Among the various methods for implementing shallow measurements, we selected the block shadow technique. By doing so, we reduced the number of unitaries that need to be searched during the optimization process compared to approaches based on the well-known Clifford group.

To verify whether this extended measurement scheme is practically useful in real-world scenarios, we carried out numerical experiments comparing it with the single-qubit case~\cite{hu2025demonstration_24}. In these experiments, we targeted the problem of calculating the energy uncertainty $\langle H^2 \rangle - \langle H \rangle^2$ for the cluster-Heisenberg model 
\begin{equation}
    H_{\text{CH}} = \sum _{i = 1}^{n - 2}{Z_i}{X_{i + 1}}{Z_{i + 2}}+ \sum _{i = 1}^{n - 1}\lambda ({X_i}{X_{i + 1}} + {Y_i}{Y_{i + 1}} + {Z_i}{Z_{i + 1}}).
\end{equation}
Since $\langle H_{\text{CH}} \rangle$ can be measured using only four measurement bases, we solely focused on the number of experiments required to estimate $\langle H_{\text{CH}}^2 \rangle$. The operator $H_{\text{CH}}^2$ contains many nonlocal Pauli strings, making it difficult to manually group the Pauli strings that can be measured simultaneously. The terms in $H_{\text{CH}}^2$ can be categorized as follows:
\begin{align}\label{eq:b1}
    (P {\text{ in }} H_{CH}^2) 
    &= 
    (P{\text{ in }}H_{ZXZ}^2) + (P{\text{ in }}H_{XX}^2) + (P{\text{ in }}H_{YY}^2) + (P{\text{ in }}H_{ZZ}^2)\nonumber\\
    &+ 
    (P{\text{ in }}{H_{ZXZ}}{H_{XX}}) + (P{\text{ in }}{H_{ZXZ}}{H_{YY}}) + (P{\text{ in }}{H_{ZXZ}}{H_{ZZ}})\nonumber\\
    &+ 
    (P{\text{ in }}{H_{XX}}{H_{YY}}) + (P{\text{ in }}{H_{XX}}{H_{ZZ}}) + (P{\text{ in }}{H_{YY}}{H_{ZZ}}),
\end{align}
where ${H_{ZXZ}} = \sum _{i = 1}^{n - 2}{Z_i}{X_{i + 1}}{Z_{i + 2}}, {H_{XX}} = \sum _{i = 1}^{n - 1}{X_i}{X_{i + 1}}, {H_{YY}} = \sum _{i = 1}^{n - 1}{Y_i}{Y_{i + 1}}$, and ${H_{ZZ}} = \sum _{i = 1}^{n - 1}{Z_i}{Z_{i + 1}}$.

The first four terms in Eq.~\eqref{eq:b1} can be measured easily using the same measurement bases employed for measuring $\langle H_{\text{CH}} \rangle$. The remaining part consists of terms containing the Pauli strings $P$, for which it is difficult to manually devise a measurement strategy. Consequently, in our numerical experiments, we focused on measuring only these Pauli strings, leaving aside the other portions that can be measured easily.

\subsubsection{Derandomization algorithm }
In the derandomization algorithm, we replace a randomly chosen measurement basis with a deterministic one at each step. In this study, we specifically employ a unitary ensemble satisfying Fact~\ref{fact1} or Fact~\ref{fact2} in the main text. By doing so, we obtain a measurement basis that provides, on average, more accurate measurements than the standard RM approach. To achieve this, we first need to determine the average confidence for a partially determined measurement basis $\mathbb{E}{_{\bf{B}}}[{\rm{CON}}{{\rm{F}}_\varepsilon }({\bf{P}}{\rm{; }}{\bf{B}})|{{\bf{B}}^\# }]$, which can be expressed as follows:
\begin{align}\label{eq:b2}
    &\mathbb{E}_{\mathbf B}\!\left[
      \left.
        \mathrm{CONF}_{\varepsilon}(\mathbf P;\mathbf B)
      \,\right|\,
      \mathbf B^\#
    \right]
    =
    \mathbb{E}_{\mathbf B}\!\left[
      \sum_{l=1}^L
      \exp\!\left(
        -\frac{\varepsilon^2}{2} M_l(\mathbf B)
      \right)
      \,\middle|\,
      \mathbf B^\#
    \right]  \\
    &=
    \mathbb{E}_{\mathbf B}\!\left[
      \sum_{l=1}^L
      \prod_{m'=1}^M
      \exp\!\left(
        -\frac{\varepsilon^2}{2}
        \mathbf 1\!\left\{
          U_{m'} P_l U_{m'}^{\dagger}
          \in \pm \mathcal Z
        \right\}
      \right)
      \,\middle|\,
      \mathbf B^\#
    \right]  \\
    &=
    \mathbb{E}_{\mathbf B}\!\left[
      \sum_{l=1}^L
      \prod_{m'=1}^M
      \left(
        1
        - \nu
        \mathbf 1\!\left\{
          U_{m'} P_l U_{m'}^{\dagger}
          \in \pm \mathcal Z
        \right\}
      \right)
      \,\middle|\,
      \mathbf B^\#
    \right]  \\
    &=
    \mathbb{E}_{\mathbf B}\!\Biggl[
      \sum_{l=1}^L
      \prod_{m'=1}^{m-1}
      \left(
        1
        - \nu
        \mathbf 1\!\left\{
          U_{m'}^\# P_l U_{m'}^{\#\dagger}
          \in \pm \mathcal Z
        \right\}
      \right) \notag \\
    &\qquad\times
      \Bigl(
        1
        - \nu
        \mathbf 1\!\left\{
          U_m^\#[1] P_l[1] U_m^{\#\dagger}[1]
          \in \pm \mathcal Z_k
        \right\}
        \;\dots\;
        \left\{
          U_m[l] P_l[l] U_m^{\dagger}[l]
          \in \pm \mathcal Z
        \right\} \notag \\
    &\qquad\qquad
        \dots\;
        \left\{
          U_m[n/k] P_l[n/k] U_m^{\dagger}[n/k]
          \in \pm \mathcal Z
        \right\}
      \Bigr) \notag \\
    &\qquad\times
      \prod_{m'=m+1}^{M}
      \left(
        1
        - \nu
        \mathbf 1\!\left\{
          U_{m'} P_l U_{m'}^{\dagger}
          \in \pm \mathcal Z
        \right\}
      \right)
    \Biggr]  \\
    &=
    \sum_{l=1}^L
    \prod_{m'=1}^{m-1}
    \left(
      1
      - \nu
      \mathbf 1\!\left\{
        U_{m'}^\# P_l U_{m'}^{\#\dagger}
        \in \pm \mathcal Z
      \right\}
    \right) \notag \\
    &\quad\times
    \Biggl(
      1
      - \nu
      \mathbf 1\!\left\{
        U_m^\#[1] P_l[1] U_m^{\#\dagger}[1]
        \in \pm \mathcal Z_k
      \right\}
      \;\dots\;
      \left\{
        U_m[r] P_l[r] U_m^{\dagger}[r]
        \in \pm \mathcal Z_k
      \right\}
      \big/
      \left(
        2^k + 1
      \right)^{w_k(P_l[r+1:])}
    \Biggr) \notag \\
    &\quad\times
    \left(
      1
      - \nu
      \big/
      \left(
        2^k + 1
      \right)^{w_k(P_l)}
    \right)^{M-m}.
\end{align}
By applying the results presented above together with the subsequent lemma, a mathematical proof of the derandomization algorithm becomes attainable.

\begin{lemma}
    Let $f = f(X, Y, Z) \geq 0$, where$ X = (X_1, X_2, ...), Z = (Z_1, Z_2, ...)$, and each $\{X_i\}_i, Y, \{Z_i\}_i$ are random variables. Then, the following inequality
    \[{\rm{mi}}{{\rm{n}}_y}({\mathbb{E}_{\bf{z}}}[f|{\bf{X}} = {\bf{x}},Y = y]) \le {\mathbb{E}_{Y,{\bf{Z}}}}[f|{\bf{X}} = {\bf{x}}]\] 
    holds. In the inequality, $X$ correspond to the already determined measurement bases (= $\textbf{B}^{\#} [:m - 1] + \textbf{B}^{\#} [m][:r-1]$), $Y$ corresponds to the measurement basis available at the $(m, r)$ – step in the algorithm and $Z$ correspond to the yet-to-be-determined measurement bases $(= \textbf{B}[m][r + 1:] + \textbf{B}[m+1:])$. 
\end{lemma}
\begin{proof}
    \begin{align}
        \mathbb{E}_{Y,\mathbf Z}[f \mid \mathbf X = \mathbf x]
        &=
        \sum_{y,\mathbf z}
        \Pr(y,\mathbf z \mid \mathbf X = \mathbf x)\,
        f(\mathbf x,y,\mathbf z)  \\
        &=
        \sum_y \sum_{\mathbf z}
        \frac{\Pr(\mathbf x,y,\mathbf z)}{\Pr(\mathbf x)}\,
        f(\mathbf x,y,\mathbf z)  \\
        &=
        \sum_y
        \frac{\Pr(\mathbf x,y)}{\Pr(\mathbf x)}
        \sum_{\mathbf z}
        \frac{\Pr(\mathbf x,y,\mathbf z)}{\Pr(\mathbf x,y)}\,
        f(\mathbf x,y,\mathbf z)  \\
        &=
        \sum_y
        \frac{\Pr(\mathbf x,y)}{\Pr(\mathbf x)}
        \,\mathbb{E}_{\mathbf z}[f \mid \mathbf X=\mathbf x, Y=y]  \\
        &\ge
        \min_y\!\left(
          \mathbb{E}_{\mathbf z}[f \mid \mathbf X=\mathbf x, Y=y]
        \right)
        \sum_y
        \frac{\Pr(\mathbf x,y)}{\Pr(\mathbf x)} \\
        &=
        \min_y\!\left(
          \mathbb{E}_{\mathbf z}[f \mid \mathbf X=\mathbf x, Y=y]
        \right)
    \end{align}
\end{proof}
Therefore, by utilizing $\mathbb{E}{_{\bf{B}}}[{\rm{CON}}{{\rm{F}}_\varepsilon }({\bf{P}}{\rm{; }}{\bf{B}})|{{\bf{B}}^\# }]$ in Eq.~\eqref{eq:b2} at each step, the derandomization algorithm yields a measurement basis $\textbf{B}^{\#}$ that, on average, outperforms RM. In numerical and hardware experiments, where M is often not predetermined, we conducted the derandomization procedure by omitting ${(1 - \nu /{{\rm{(}}{{\rm{2}}^k} + 1{\rm{)}}^{{w_k}({P_l})}})^{M - m}}$ from ${\mathbb{E}_{\bf{B}}}[{\rm{CON}}{{\rm{F}}_\varepsilon }({\bf{P}}{\rm{; }}{\bf{B}})|{{\bf{B}}^\# }]$~\cite{hu2025demonstration_24}.

\subsubsection{Procedure for obtaining approximate ground state}
In our experiments, we employed a hardware-efficient ansatz~\cite{kandala2017hardware} consisting of a system size of $n$ = 20 qubits and 9 CZ layers to approximate the ground state of the cluster-Heisenberg model ($H_{\text{CH}}$). After each CZ gate, we applied a sequence of parameterized single-qubit gates $R_Z(\theta_2)R_Y(\theta_1)$. For the parameter optimization process, we utilized the Simultaneous Perturbation Stochastic Approximation~\cite{spall1998overview} algorithm with 20,000 steps. The optimization was conducted on classical devices, and the optimized parameters were subsequently used in quantum hardware experiment.

\subsubsection{Details on the hardware experiment}
In the hardware experiment (Fig.~\ref{fig2}(b) in the main text), we employed a derandomization algorithm to select 60 measurement bases and performed 10,000 measurements for each. In practice, it is more efficient to perform multiple measurements using a fixed measurement basis rather than repeatedly changing bases. To account for this practical consideration, we selected a relatively large value of $\varepsilon$ in our derandomization algorithm, allowing us to cover as many of the sets $\{P_l\}_l$ as possible using the fewest possible measurement bases.

\subsection{Fidelity estimation (Fig.~\ref{fig3}(a) in the main text)}
Estimating the fidelity with respect to a target quantum state has been extensively studied~\cite{huang2020predicting_13, flammia2011direct}. Such estimations are essential in evaluating how accurately desired quantum states are prepared on noisy quantum computers. Previous studies~\cite{huang2020predicting_13} have established that the RCM effectively estimates expectation values of low-rank observables, including fidelity. Here, we investigate the advantages and limitations of employing block shadow with block size $k$, where $k = \mathcal{O}(\log n)$.

In the worst case, block shadow requires a sample complexity that increases exponentially $\mathcal{O}(3^{n/k})$ for system size $n$. This worst-case saturation occurs when the quantum state has a simple block tensor product structure $|\Psi \rangle  =  \bigotimes _{i = 1}^{n/k}|{\Psi _i}\rangle$ where all blocks are not entangled with each other, which might differ from the quantum states typically encountered in practice. In contrast, we proved that, for typical cases, the block shadow allows for sample complexity scaling similar to that of RCM~\cite{huang2020predicting_13}. To support this, we conducted numerical experiments presented in the main text, showing that accurate fidelity estimation is possible even for approximately 50 qubits.

\subsubsection{State preparation}
We prepared a random quantum state for a system of size $n$ = 48 qubits by applying 8 CZ layers and interleaving random single-qubit gates sampled from $\mathbb{U}(2)$ before and after each CZ gate.

\subsubsection{Comparison with RCM}
\label{appx:comp_RCM}

Previous analyses of multi-shot classical shadows with RCM reported that reusing the same measurement basis gives little improvement in certain regimes~\cite{helsen2023thrifty_19, zhou2023performance_20}. 
This behavior can be understood from the variance decomposition for fidelity estimation. 
For a rank-one target observable \(\Psi=\ket{\Psi}\!\bra{\Psi}\), the multi-shot variance is
\begin{equation}
    {\rm{Var}}({\rm{Tr}}(\hat \rho \Psi ))
    =
    \frac{1}{{N_U}}
    \left(
        \frac{1}{{N_S}} V_2(\Psi,\rho)
        +
        \frac{{N_S}-1}{{N_S}} V_1(\Psi,\rho)
        -
        {\rm{Tr}}(\rho\Psi)^2
    \right).
\end{equation}
Thus, increasing \(N_S\) is useful only when the limiting term \(V_1(\Psi,\rho)\) is small compared with \(V_2(\Psi,\rho)\).

For \(\mathbb{U}=\mathrm{Cl}(n)\), one has
\begin{equation}
    V_1(\Psi,\rho)\le 3+\mathcal{O}(1/D),
    \qquad
    V_2(\Psi,\rho)\le 3+\mathcal{O}(1/D).
\end{equation}
In the worst case, this bound is saturated. 
For example, when \(\rho=\Psi\) is a stabilizer state, both \(V_1(\Psi,\rho)\) and \(V_2(\Psi,\rho)\) are \(3+\mathcal{O}(1/D)\). 
In this case, increasing \(N_S\) does not reduce the variance beyond the \(V_1\) floor, which explains why multi-shot measurements give little benefit in such worst-case examples~\cite{helsen2023thrifty_19, zhou2023performance_20}.

By contrast, the typical-case behavior is different. 
As shown in Appendix~\ref{appx:typical_si_V}, for a fixed rank-one target \(\Psi\) and Haar-random \(\rho\),
\begin{equation}
    \mathbb{E}_{\rho}[V_1(\Psi,\rho)]
    =
    \mathcal{O}(1/D),
    \qquad
    \mathbb{E}_{\rho}[V_2(\Psi,\rho)]
    =
    \mathcal{O}(1).
\end{equation}
Therefore, in the typical-case regime, the term \(V_2/N_S\) can be suppressed by reusing each measurement basis, while the remaining term \(V_1\) stays small. 
This shows that multi-shot measurements can be beneficial for fidelity estimation, even though they may provide little improvement in worst-case stabilizer examples. Importantly, this typical-case advantage persists even for logarithmic-depth block-shadow circuits.

\subsubsection{Post-processing procedure using a tensor network}
\label{appx:fidel_comp}
In many cases, shallow shadow methods offer the advantage of estimating fidelity with a measurement circuits of depth $\mathcal{O}(\log n)$ and sample complexity scaling comparable to RCM; however, they introduce additional complexity in the post-processing step. In this section, we describe the post-processing procedure and compare the computational costs associated with brickwork and block circuits.

we can re-express the estimator ${\rm{Tr}}(\Psi \hat \rho )$, where $\Psi = \ket{\Psi}\!\bra{\Psi}$ as follows:
\begin{align}
    \operatorname{Tr}\!\left(
      \lvert \Psi \rangle \langle \Psi \rvert \hat{\rho}
    \right)
    &=
    \operatorname{Tr}\!\left(
      \lvert \Psi \rangle \langle \Psi \rvert
      \mathcal M^{-1}
      \bigl(
        U^{\dagger}
        \lvert b \rangle \langle b \rvert
        U
      \bigr)
    \right)  \\
    &=
    \langle\!\langle
      \Psi
      \,\big|\,
      \mathcal M^{-1}
      \mathcal U^{\dagger}
      \,\big|\,
      b
    \rangle\!\rangle,
\end{align}
where $|A\rangle\!\rangle$ is a vectorization of $A$. First, we describe the post-processing procedure using a brickwork circuit. 
Let \(d_{\mathrm{RM}}\) denote the two-qubit gate depth used in RM, and assume that \(\ket{\Psi}\) can be expressed as a one-dimensional matrix product state (MPS) with bond dimension \(\chi_{\Psi}\). 
In the shallow regime considered here, both the brickwork depth \(d_{\mathrm{RM}}\) and the block size \(k\) scale logarithmically with \(n\). 
In brickwork circuits, since a closed-form expression for the inverse shadow channel \(\mathcal{M}^{-1}\) is unavailable, an approximate inverse must be obtained numerically~\cite{bertoni2024shallow_22}. 
If the approximate inverse has bond dimension \(\chi_{\mathrm{inv}}\), then the tensor-contraction cost for a single term with open boundary conditions scales as 
\begin{equation}\label{appx:eq:time_fidel_brick}
    \mathrm{Time_{\mathrm{brickwork}}}=\mathcal{O}\!\left(\chi_{\Psi}^{4}\chi_{\mathrm{inv}}^{2}2^{4d_{\mathrm{RM}}} n\right).
\end{equation}
Alternatively, for block shadows, the inverse shadow channel can be obtained exactly. 
The corresponding tensor-contraction cost is 
\begin{equation}\label{appx:eq:time_fidel_block}
    \mathrm{Time_{\mathrm{block}}}=\mathcal{O}\!\left((\chi_{\Psi}^{4}+\chi_{\Psi}^{2}2^{2k})\frac{n}{k}
    \right).
\end{equation}
The cost in Eq.~\eqref{appx:eq:time_fidel_brick} grows rapidly with the brickwork depth \(d_{\mathrm{RM}}\). 
Moreover, because the inverse shadow channel is not available in closed form for brickwork circuits, it must be approximated unless one allows an exponentially large bond dimension. 
This introduces an additional overhead through \(\chi_{\mathrm{inv}}\). 
Partly due to this post-processing cost, previous numerical and hardware demonstrations of fidelity estimation with shallow shadows have mainly considered target states with small MPS bond dimension, such as \(\chi_{\Psi}=1\) or \(2\)~\cite{hu2025demonstration_24, bertoni2024shallow_22}. 
In our simulations, by using the block structure and the exact inverse shadow channel, we could treat target states with \(\chi_{\Psi}=16\).

\subsubsection{Estimation of $V_1(\rho)$ and $V_2(\rho)$ by fitting}
Calculating quantities such as $V_1(\rho)$ and $V_2(\rho)$ requires a significant increase in memory and computational time, even with only a slight growth in the bond dimension of $\rho$. Therefore, we tried to estimate approximate values of $V_1(\rho)$ and $V_2(\rho)$ for a random quantum state $\rho$ with $n$ = 48 and two-qubit circuit depth $d$ = 8 through fitting.

\begin{figure}[t]
  \centering
  \includegraphics[
    width=0.6\linewidth,
    trim={1.5cm 22cm 12cm 0cm},
    clip
    ]{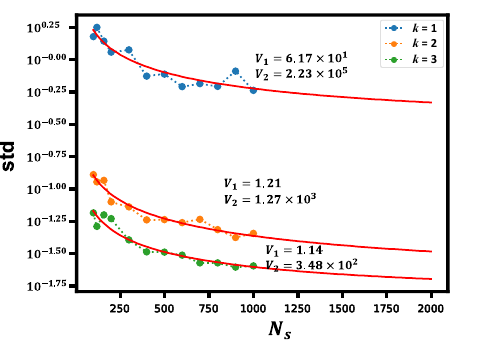}
  \caption{
    Estimated \(V_1(\rho)\) and \(V_2(\rho)\) from fitting the standard deviation (std).
    The target state is a random quantum state with \(n=48\) and two-qubit circuit depth \(d=8\).
    The data are shown as a function of the number of repeated measurements \(N_S\) for block sizes \(k=1,2,3\).
    }
  \label{fig:s1}
\end{figure}

As shown in Fig.~\ref{fig:s1}, even with $k$ = 2, the value of $V_1(\rho)$ approaches nearly 1, indicating that performing multi-shots offers substantial advantages. A statistical analysis associated with estimating $V_1(\rho)$ and $V_2(\rho)$ through fitting is left for future research.

\subsubsection{Additional numerical experiments on fidelity estimations}
\label{appx:additional_numerics}
\begin{figure}[t]
  \centering
  \includegraphics[
    width=1\linewidth,
    trim={1cm 21.3cm 3cm 0cm},
    clip
    ]{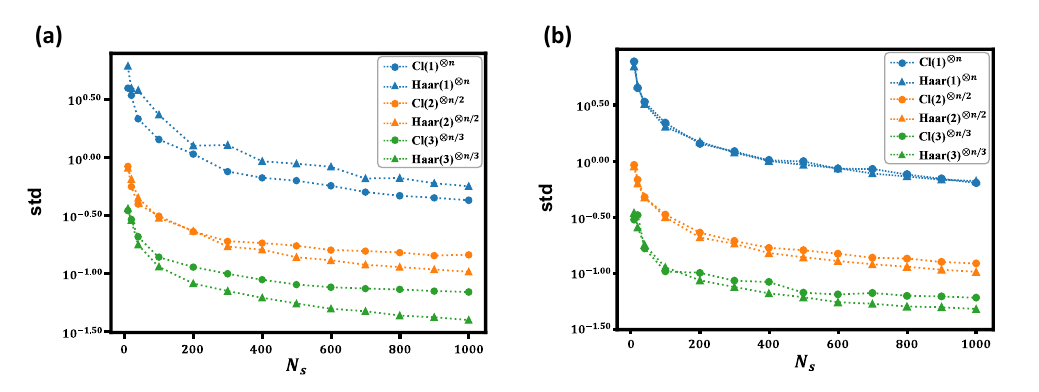}
  \caption{
    Additional numerical experiments for fidelity estimation.
    (a) Fidelity estimation for the ground state of the cluster-Heisenberg model \(H_{\mathrm{CH}}\) with \(n=36\).
    (b) Fidelity estimation for the time-evolved state generated by a quench dynamics.
    The standard deviation (std) is shown as a function of the number of repeated measurements \(N_S\).
    }
  \label{fig:s2}
\end{figure}
In Appendix~\ref{appx:state_dependent_variance}, we prove that, in typical cases, fidelity estimation for a target observable that is identical to the prepared state (e.g., state verification) is sample-efficient as long as the block size $k$ satisfies $k2^k = \Omega(n)$. To verify this in the main text, we used a randomly generated quantum state (Fig.~\ref{fig3}a). Here, we conducted additional numerical experiments to estimate fidelity for various other quantum states.

First, we estimated the fidelity with the ground state of the cluster-Heisenberg Hamiltonian $H_{\text{CH}} = \sum _{i = 1}^{n - 2}{Z_i}{X_{i + 1}}{Z_{i + 2}} + \sum _{i = 1}^{n - 1}\lambda ({X_i}{X_{i + 1}} + {Y_i}{Y_{i + 1}} + {Z_i}{Z_{i + 1}})$ with system size $n$ = 36. We obtained this ground state using the density matrix renormalization group (DMRG) algorithm with a one-dimensional MPS of bond dimension $\chi$ = 16. After preparing the ground state, we applied the local random unitaries needed for randomized measurements (without any bond-dimension truncations) and collected the measurement outcomes as bitstrings through a perfect sampling algorithm~\cite{ferris2012perfect}. Second, we estimated the fidelity with the quantum state obtained by the quench dynamic of Hamiltonian $H = \sum _{i = 1}^{n - 1}1.3{X_i}{X_{i + 1}} + 1.7{Y_i}{Y_{i + 1}} + 1.6{Z_i}{Z_{i + 1}}$ from initial state $\ket{t=0} = \ket{0}^{\otimes 18}\otimes\ket{1}^{\otimes 18}$ to $\ket{t=1.8} = e^{-iHt}\ket{t = 0}$. To simulate the time dynamics, we used the time-dependent density matrix renormalization group (tDMRG)~\cite{daley2004time} with bond dimension $\chi$ = 16 and $\delta t$ = 0.03 with a second-order trotterization method. The results are shown below.

As shown in Fig.~\ref{fig:s2}, we observe a scaling similar to the main text. However, in these cases, $V_1$ is larger than that of a typical random quantum state. Since the variance saturates to $(V_1 - \text{Tr}(O\rho)^2)/N_U$ in the large $N_S$ limit, reducing $V_1$ is crucial for accurate fidelity estimation. We therefore propose three methods to achieve this reduction. First, one may increase $N_U$ and decrease $N_S$ while keeping the total number of experiments $T = N_UN_S$ fixed, thereby reducing the contribution of $V_1$ in the overall variance. However, this approach necessitates a greater number of measurement bases, which can increase the time required for circuit compilation. Second, if one has some prior knowledge about the prepared quantum state $\rho$, this bias can be exploited as in CRM. By employing CRM, we obtain a variance as follows: 
\begin{equation}
    {\rm{Var}}({\rm{Tr}}(O\hat \rho )) = \frac{1}{{{N_U}}}\left[ {\frac{1}{{{N_S}}}\left( {{V_2}(O,\rho ) - {V_1}(O,\rho )} \right) + {V_1}(O,\rho  - \sigma ) - {\rm{Tr}}{{(O(\rho  - \sigma ))}^2}} \right]
\end{equation}
in the large $N_{\sigma}$ limit and $N_{\rho}=N_S$. As the bias state $\sigma$ approaches $\rho$, $V_1(O, \rho-\sigma)$ decreases, indicating that the saturation level in the large $N_S$ limit can be lowered. Lastly, instead of employing $\mathbb{U} = \text{Cl}(k)^{\otimes n/k}$ in the block-shadow protocol, one can use the $\mathbb{U} = \text{Haar}(k)^{\otimes n/k}$ as the unitary ensemble. While $V_2$ remains the same regardless of the chosen ensemble, $V_1$ may differ. In general, using the $\mathbb{U} = \text{Haar}(k)^{\otimes n/k}$ helps to avoid worst-case scenarios. We compare the estimation results from $\text{Cl}(k)^{\otimes n/k}$ and $\text{Haar}(k)^{\otimes n/k}$ in Fig.~\ref{fig:s2}. All three methods are compatible and can be used independently or in tandem, depending on the specific situation.

\subsection{Purity estimation (Fig.~\ref{fig3}(b) in the main text)}
In this section, we examine how utilizing multi-shots and shallow measurements can facilitate accurate purity measurement. By measuring the purity $\text{Tr}(\rho_A^2)$, we can determine whether a quantum system $A$ is entangled with its environment or verify the presence of a topologically ordered phase by computing quantities such as the Rényi topological entanglement entropy~\cite{satzinger2021realizing_6}. Moreover, a recent study~\cite{vermersch2024many_31} suggests that, in certain cases, to calculate the purity of the entire quantum state, it suffices to measure the purity of only a few patches. Since all of the aforementioned research requires precise purity measurements, devising an efficient method for measuring purity can immediately benefit a variety of fields. With this in mind, we demonstrate that, in many cases, favorable sample complexity previously known to require a measurement circuit depth of $\mathcal{O}(n)$~\cite{anshu2022distributed_44} can be obtained using shallow circuits with a depth of $\mathcal{O}(\log n)$.

\subsubsection{State preparation}
We prepared a random quantum state on a system of $n$ = 24 qubits by applying 12 layers of CZ gates, with random single-qubit gates sampled from $\mathbb{U}(2)$ applied before and after each CZ layer.

\subsubsection{Post-processing procedure for purity estimation}
\label{appx:purity_comp}

A similar distinction appears in the post-processing cost for purity estimation. 
In the shallow regime considered here, we take \(d_{\mathrm{RM}}=\mathcal{O}(\log n)\) for brickwork circuits and \(k=\mathcal{O}(\log n)\) for block circuits. 
For brickwork shadows, if the approximate inverse shadow channel is represented as Matrix product operator (MPO) with bond dimension \(\chi_{\mathrm{inv}}\), the tensor-contraction cost for evaluating a single term $\Tr\!\left(
    U^{\dagger} \ket{b}\!\bra{b} U\,
    \mathcal{M}^{-1}
    \!\left(
    U^{\dagger} \ket{c}\!\bra{c} U
    \right)
    \right)$
scales as
\begin{equation}
    \mathrm{Time_{\mathrm{brickwork}}}=\mathcal{O}\!\left(\chi_{\mathrm{inv}}^{2}2^{8d_{\mathrm{RM}}} n\right).
\end{equation}
In contrast, for block shadows, the inverse shadow channel is available exactly, and the corresponding cost scales as
\begin{equation}
    \mathrm{Time_{\mathrm{block}}}=\mathcal{O}(n).
\end{equation}

This reduction comes from the tensor-product structure of the block-shadow channel. 
Indeed, for \(U=\bigotimes_{i=1}^{n/k}U_i\), the above overlap factorizes as
\begin{equation}
    \Tr\!\left(
    U^{\dagger} \ket{b}\!\bra{b} U\,
    \mathcal{M}^{-1}
    \!\left(
    U^{\dagger} \ket{c}\!\bra{c} U
    \right)
    \right)
    =
    \prod_{i=1}^{n/k}
    \left[
        (2^k+1)\delta_{b_i,c_i}-1
    \right],
\end{equation}
where \(b_i,c_i\in\{0,1\}^k\) denote the bit strings restricted to the \(i\)-th block. 
Thus, the required post-processing reduces to evaluating a product over blocks, rather than contracting a tensor network.

\subsubsection{Estimation of $V_1(\rho)$ and $V_3(\rho)$ from fitting}
\begin{figure}[t]
  \centering
  \includegraphics[
    width=0.7\linewidth,
    trim={1cm 22.2cm 12cm 0cm},
    clip
    ]{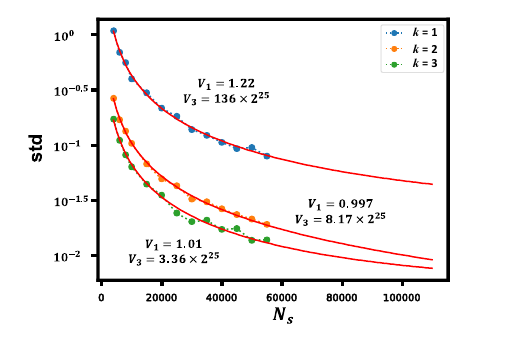}
  \caption{
    Estimated \(V_1(\rho)\) and \(V_3(\rho)\) from fitting the standard deviation (std).
    The target state is a random quantum state with \(n=24\) and two-qubit circuit depth \(d=12\).
    The data are shown as a function of the number of repeated measurements \(N_S\) for block sizes \(k=1,2,3\).
    }
  \label{fig:s3}
\end{figure}
Calculating quantities such as $V_1(\rho)$ and $V_3(\rho)$ requires substantial memory and computational time, even with only the small bond dimension of $\rho$. Therefore, we estimated approximate values of $V_1(\rho)$ and $V_3(\rho)$ for a random quantum state $\rho$ with $n$ = 24 and two-qubit circuit depth $d$ = 12 through fitting (the contribution of $V_2(\rho)$ to the variance is small, causing the fitted value of $V_2(\rho)$  to become unstable).

As shown in Fig.~\ref{fig:s3}, even with $k$ = 2, the value of $V_1(\rho)$ approaches approximately 1, indicating that performing multi-shots offers significant advantages. 

\subsubsection{Comparison between block shadow and $\epsilon$-approximate unitary design}
\begin{figure}[t]
  \centering
  \includegraphics[
    width=0.6\linewidth,
    trim={1cm 16.5cm 6cm 0cm},
    clip
    ]{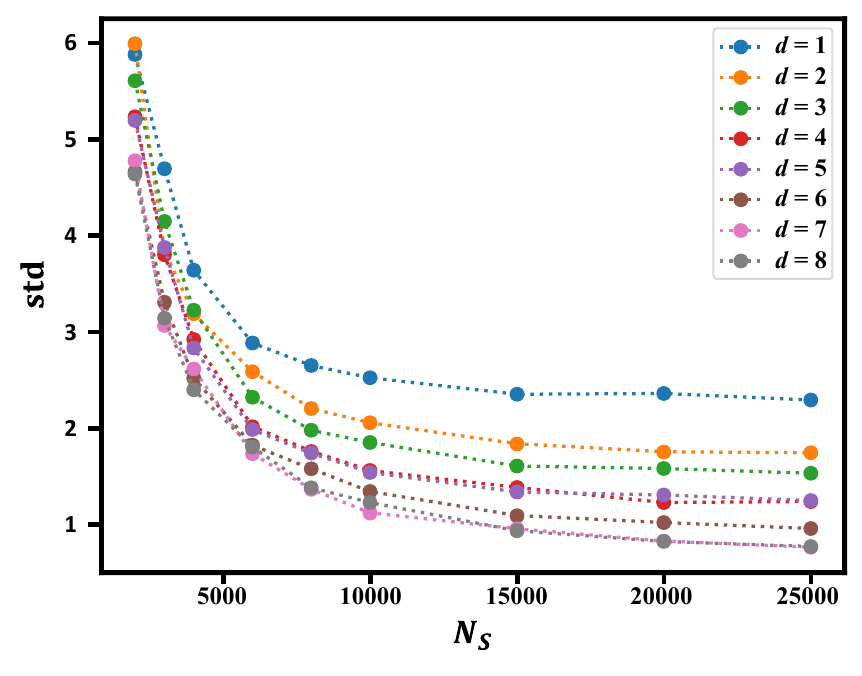}
    \caption{
    Purity estimation using \(\epsilon\)-approximate unitary designs.
    The standard deviation (std) is shown as a function of the number of repeated measurements \(N_S\) for an \(n=24\) quantum state.
    The depth \(d\) denotes the number of CX gate layers interleaved with random single-qubit gates.
    }
  \label{fig:s4}
\end{figure}
Unlike the single-shot setting ($N_S = 1$) introduced in the main text, a multi-shot approach allows us to estimate the purity using previously studied $\epsilon$-approximate unitary designs~\cite{schuster2025random_30}. When employing $\epsilon$-approximate unitary designs, an approximate inverse channel derived from the Haar measure, $\mathcal{M}^{-1}(A) = (2^n + 1)A - \text{Tr}(A)I$, is utilized instead of the exact one to simplify post-processing. This approach results in a biased estimator; however, the bias—which equals $\epsilon(1 + \rm{Tr}(\rho^2))$—decreases as the $\epsilon$-approximate unitary ensemble becomes more accurate. We conducted numerical experiments (Fig.~\ref{fig:s4}) to estimate the purity under the same conditions as those used in Fig.~\ref{fig3}b of the main text. We gradually increased the depth $d$ of the CX gate layers, each interspersed with a random single-qubit gate layer, to construct an approximate unitary ensemble~\cite{schuster2025random_30}.

Considering that block shadow methods typically utilize about 0.489$n$ two-qubit gates (assuming a block size of $k$ = 3 with nearest-neighbor connectivity), $\epsilon$-approximate unitary designs require more two-qubit gates. This increased gate count can lead to additional error accumulation during RM, potentially causing the estimated purity to deviate further from its true value.

\subsection{CRM with shallow shadow (Fig.~\ref{fig4} in the main text)}
\label{appx:crm_exp}

\subsubsection{Post-processing cost for CRM bias estimators}
\label{appx:crm_postprocessing_cost}

We compare the classical post-processing cost for evaluating the CRM bias contribution,
\begin{equation}
    \operatorname{Tr}\left(O\hat{\sigma}_{\mathrm{old}}^{(U)}\right)
    \quad\text{and}\quad
    \operatorname{Tr}\left(O\hat{\sigma}_{\mathrm{new}}^{(U)}\right),
\end{equation}
with open boundary conditions.
Here $O$ is an MPO with bond dimension $\chi_O$, and the bias state is $\sigma=\ket{\psi_\sigma}\bra{\psi_\sigma}$, where $\ket{\psi_\sigma}$ is an MPS with bond dimension $\chi_\sigma$.
For block shadows, $k$ is the block size and there are $n/k$ blocks.
For brickwork shadows, $d_{\mathrm{RM}}$ denotes the number of two-qubit-block layers in the brickwork circuit.
In the brickwork case, we assume that the inverse shadow channel $\mathcal M^{-1}$ is already represented as an MPO with bond dimension $\chi_{\mathrm{inv}}$; the cost of constructing this MPO is not included.

\begin{table}[h]
    \centering
    \renewcommand{\arraystretch}{1.8}
    \setlength{\extrarowheight}{2pt}
    \begin{tabular}{c|c|c}
    \hline
    Measurement circuit & Bias estimator & Post-processing time \\
    \hline

    Block & \(\hat\sigma_{\mathrm{old}}\) [Eq.~\eqref{main:eq:old_estimator}] &
    \(
    \displaystyle
    \mathcal{O}\!\left(
    \frac{n}{k}
    \left(
    4^k\chi_\sigma^2
    +
    8^k\chi_O^2
    +
    2^k
    \left(
    \chi_\sigma^3\chi_O
    +
    \chi_\sigma^2\chi_O^2
    \right)
    \right)
    \right)
    \)
    \\
    \hline

    Block & \(\hat\sigma_{\mathrm{new}}\) [Eq.~\eqref{main:eq:new_estimator}] &
    \(
    \displaystyle
    \mathcal{O}\!\left(
    N_\sigma
    4^k
    \chi_O^2\frac{n}{k}
    \right)
    \)
    \\
    \hline

    Brickwork & \(\hat\sigma_{\mathrm{old}}\) [Eq.~\eqref{main:eq:old_estimator}] &
    \(
    \displaystyle
    \mathcal{O}\!\left(
    n
    \chi_\sigma^4
    \chi_O^2
    \chi_{\mathrm{inv}}^2
    2^{8d_{\mathrm{RM}}}
    \right)
    \)
    \\
    \hline

    Brickwork & \(\hat\sigma_{\mathrm{new}}\) [Eq.~\eqref{main:eq:new_estimator}] &
    \(
    \displaystyle
    \mathcal{O}\!\left(
    N_\sigma
    n
    \chi_O^2
    \chi_{\mathrm{inv}}^2
    2^{4d_{\mathrm{RM}}}
    \right)
    \)
    \\
    \hline
    \end{tabular}

    \caption{
    Post-processing costs for CRM bias estimators.
    }
    \label{tab:crm_postprocessing_cost}
\end{table}

\subsubsection{State preparation}
We prepared a random quantum state for a system size $n$ = 40 qubits by applying 9 CZ layers, interleaved with random single-qubit gates sampled from $\mathbb{U}(2)$ before and after each CZ gate. In our experiments, we utilized the $\textit{ibm\_marrakesh}$ provided through IBM Cloud, using the following qubits: [132, 131, 138, 151, 150, 149, 148, 147, 146, 145, 144, 143, 136, 123, 124, 125, 117, 105, 104, 103, 96, 83, 84, 85, 77, 65, 66, 67, 68, 69, 70, 71, 58, 51, 52, 53, 54, 55, 59, 75].

\subsubsection{Error mitigation (EM)}
\label{appx:crm_em}

In the experiment of the main text, we applied EM to the block shadow. To do this, we assume the error channel can be modeled by ${\Lambda _U} = {\cal U} \cdot {\cal E}$, where $\cal U$ is a superoperator of unitary channel $U(\cdot)U^{\dagger}$. When using the unitary ensemble $\mathbb{U} = \text{Cl}(k)^{\otimes n/k}$, the following equation holds~\cite{chen2021robust_18, zhao2024group}.
\begin{equation}
     \mathbb{E}[\,\mathcal U^{\dagger} A \mathcal U\,]
        = 
        \sum_{\lambda \in R}\frac{\operatorname{Tr}(A \Pi_\lambda)}{\operatorname{Tr}(\Pi_\lambda)}\,\Pi_\lambda
        = \sum_{\lambda \in R} f_\lambda \Pi_\lambda,
\end{equation}
where superoperator are projector onto the irreducible subspace labeled by $\lambda$, and R is the set of labels for the irreducible representation. In our case of $\mathbb{U} = \text{Cl}(k)^{\otimes n/k}$, the number of irreps is $2^{n/k}$. Noiseless and noise shadow channels can be written as 
\begin{align}
    \mathcal M
    &=
    \mathbb{E}_{\mathcal U}\!\left[
      \mathcal U^{\dagger} \mathcal M_Z \mathcal U
    \right]
    =
    \sum_{\lambda \in R}
    \frac{
      \operatorname{Tr}(\mathcal M_Z \Pi_\lambda)
    }{
      \operatorname{Tr}(\Pi_\lambda)
    }
    \,\Pi_\lambda
    =
    \sum_{\lambda \in R}
    f_\lambda \Pi_\lambda
     \\
    \widetilde{\mathcal M}
    &=
    \mathbb{E}_{\mathcal U}\!\left[
      \mathcal U^{\dagger} \mathcal M_Z \mathcal E \mathcal U
    \right]
    =
    \sum_{\lambda \in R}
    \frac{
      \operatorname{Tr}(\mathcal M_Z \mathcal E \Pi_\lambda)
    }{
      \operatorname{Tr}(\Pi_\lambda)
    }
    \,\Pi_\lambda
    =
    \sum_{\lambda \in R}
    \widetilde f_\lambda \Pi_\lambda
\end{align}
where ${{\cal M}_Z} = {\sum _{b \in {{\{ 0,1\} }^n}}}|b\rangle\!\rangle \langle\!\langle b|$. Let $\hat \rho  = {\widetilde {\cal M}^{ - 1}}{{\cal U}^\dagger }|b\rangle\!\rangle $, then, the estimator of the expectation value of a Pauli string $P$ is as follows:
\begin{align}\label{eq:b6}
    \operatorname{Tr}(P \hat{\rho})
    &=
    \langle\!\langle P \mid \hat{\rho} \rangle\!\rangle  \\
    &=
    \langle\!\langle
      P
      \mid
      \widetilde{\mathcal M}^{-1}
      \mathcal U^{\dagger}
      \mid
      b
    \rangle\!\rangle  \\
    &=
    \widetilde f_\lambda^{-1}
    \langle\!\langle
      P
      \mid
      \mathcal U^{\dagger}
      \mid
      b
    \rangle\!\rangle  \\
    &=
    \widetilde f_\lambda^{-1}
    f_\lambda
    f_\lambda^{-1}
    \langle\!\langle
      P
      \mid
      \mathcal U^{\dagger}
      \mid
      b
    \rangle\!\rangle  \\
    &=
    \alpha_\lambda
    f_\lambda^{-1}
    \langle\!\langle
      P
      \mid
      \mathcal U^{\dagger}
      \mid
      b
    \rangle\!\rangle  \\
    &=
    \alpha_\lambda
    f_\lambda^{-1}
    \,(U,P,b),
\end{align}
where $P$ is in the subspace labeled by $\lambda$ (${\Pi _\lambda }P = P{\Pi _\lambda } = P$) and ${\alpha _\lambda }: = \widetilde f_\lambda ^{ - 1}{f_\lambda }$ . By utilizing the fact that they share the same projector $\Pi_{\lambda}$  even in the presence of errors, we can estimate $\alpha_{\lambda}$. Assume $Q$ is in the same subspace $\Pi_{\lambda}$  as $P$ and $\text{Tr}(\sigma Q)$ is known in advance, then $\alpha_{\lambda}$ can be estimated as 
\begin{equation}\label{appx:eq:robust_pauli}
    {\hat \alpha _\lambda } 
    = 
    \frac{{{\rm{Tr}}{{(\sigma Q)}_{{\rm{exact}}}}}}{{{\rm{Tr}}{{(\sigma Q)}_{{\rm{exp}}}}}} 
    = 
    \frac{\mathrm{Tr}(\sigma Q)_{\text{exact}}}{\frac{f_{\lambda}^{-1}}{N_{\text{cal}}}\sum_{i=1}^{N_{\text{cal}}}(U_i,Q,b_i)}
\end{equation}
By inserting the obtained ${\hat{\alpha}_{\lambda}}$ into Eq.~\eqref{appx:eq:robust_pauli}, we can calculate the error-mitigated expectation value.

\begin{figure}[t]
  \centering
  \includegraphics[
    width=0.6\linewidth,
    trim={1cm 22.5cm 12.5cm 0cm},
    clip
    ]{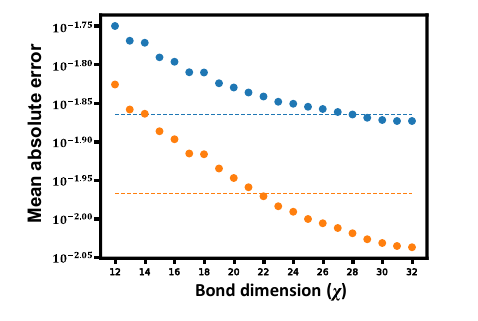}
  \caption{Experimental results from raw (blue dots) and error-mitigated data (orange dots) without using bias-variance tradeoff. Each dotted line represents the mean absolute error without using common randomized measurement (CRM).}
  \label{fig:s5}
\end{figure}

When performing error mitigation, employing the bias-variance tradeoff can sometimes be helpful~\cite{cai2025biased}. For instance, if the amplification factor ${\alpha _\lambda }(P)$  estimated from the calibration experiment is large, the variance can increase, leading to higher measurement uncertainty. Therefore, in Fig.~\ref{fig4} presented in the main text, whenever ${\alpha _\lambda }(P)$  exceeded 1.5, we used $0.8\times{\alpha _\lambda }(P)$  as the amplification value. Fig.~\ref{fig:s5} shows the result obtained without using the bias-variance tradeoff (the overall trend remains the same, but the error is relatively larger).

\subsubsection{The relationship between bond dimension truncation and the relative error}
\label{appx:crm_multi_err}

\begin{figure}[t]
  \centering
  \includegraphics[
    width=0.6\linewidth,
    trim={1cm 22cm 12cm 0cm},
    clip
    ]{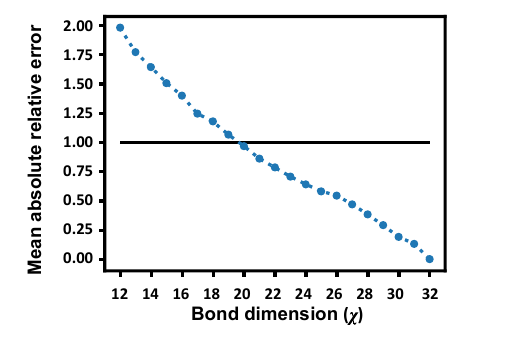}
  \caption{
    Mean absolute relative error of the bias state \(\sigma\) as a function of the bond dimension \(\chi\).
    The horizontal line marks the threshold below which common randomized measurements (CRM) can reduce the variance for Pauli-string estimation.
    }
  \label{fig:s6}
\end{figure}
When performing Common Randomized Measurement~\cite{vermersch2024enhanced_25} (CRM), the closer the bias state $\sigma$ is to the prepared quantum state $\rho$, the more accurately physical quantities can be estimated. In the main text, our goal was to estimate the expectation value for a Pauli string $P$, and in this case, CRM is only beneficial when $|\text{Tr}(\sigma P) - \text{Tr}(\rho P)|/|\text{Tr}(\rho P)| \leq 1$. In the experiments presented in the main text, we used the bias state σ obtained by truncating the bond dimension of $\rho$. The mean absolute relative error ($\mathbb{E}_P|\text{Tr}(\sigma P) - \text{Tr}(\rho P)|/|\text{Tr}(\rho P)|$) resulting from truncating the bond dimension is illustrated in Fig.~\ref{fig:s6}.

Therefore, in the noiseless case, CRM becomes beneficial starting from $\chi=20$, where the mean absolute relative error is smaller than 1. In hardware experiments without error mitigation (EM), CRM only becomes beneficial starting from $\chi=29$. This implies that CRM is not helpful in many situations because it requires relatively accurate knowledge of the prepared quantum state $\rho$. However, as we demonstrated in the main text, by mitigating some of the errors, we not only obtain results with reduced errors but also show that CRM is effective even for a relatively less accurate ($\chi=22$) bias state. To perform a quantitative analysis of this, we calculate the variance of the estimator for the Pauli string P when CRM is applied in the presence of noise.

For the noisy CRM analysis, we use the following notation. 
Let \(U_1,\ldots,U_{N_U}\) be the measurement bases shared by the estimators for
\(\rho\) and the bias state \(\sigma\). 
For each \(U_i\), let \(b_{ij}\) denote the \(j\)-th measurement outcome obtained
from the noisy measurement of \(\rho\). 
We define
\begin{equation}
    \hat{\rho}
    =
    \frac{1}{N_U}
    \sum_{i=1}^{N_U}
    \hat{\rho}_{U_i}(N_S),
    \qquad
    \hat{\rho}_{U_i}(N_S)
    =
    \frac{1}{N_S}
    \sum_{j=1}^{N_S}
    \widetilde{\mathcal M}^{-1}
    \!\left(
        U_i^\dagger |b_{ij}\rangle\langle b_{ij}| U_i
    \right).
\end{equation}
The corresponding CRM estimator is
\begin{equation}
    \hat{\rho}_{\sigma}
    =
    \hat{\rho}
    -
    \hat{\sigma}
    +
    \sigma,
    \qquad
    \hat{\sigma}
    =
    \frac{1}{N_U}
    \sum_{i=1}^{N_U}
    \hat{\sigma}_{U_i}.
\end{equation}
Here, for a fixed basis \(U\), the bias-state estimator is defined by
\begin{equation}
    \hat{\sigma}_{U}
    =
    \sum_{b\in\{0,1\}^n}
    \operatorname{Tr}
    \!\left(
        U\sigma U^\dagger |b\rangle\langle b|
    \right)
    \mathcal M^{-1}
    \!\left(
        U^\dagger |b\rangle\langle b| U
    \right).
\end{equation}
We assume that, after Pauli twirling, the noisy measurement channel acts on Pauli
operators as
\begin{equation}
    \Lambda_U(P)
    =
    \lambda_{U,P} UPU^\dagger .
\end{equation}
With this notation, we obtain
\begin{align}
    \operatorname{Var}\!\bigl(\operatorname{Tr}(P \hat{\rho}_\sigma)\bigr)
    &=
    \frac{1}{N_U}
    \operatorname{Var}\!\bigl(
      \operatorname{Tr}\bigl(P(\hat{\rho}_U-\hat{\sigma}_U)\bigr)
    \bigr) \\
    &\le
    \frac{1}{N_U}
    \mathbb{E}\Bigl[
      \operatorname{Tr}\bigl(P(\hat{\rho}_U-\hat{\sigma}_U)\bigr)^2
    \Bigr] \\
    &=
    \frac{1}{N_U}
    \mathbb{E}\Bigl[
      \operatorname{Tr}\bigl(P\hat{\rho}_U\bigr)^2
      -2\,\operatorname{Tr}(P\hat{\rho}_U)\operatorname{Tr}(P\hat{\sigma}_U)
      +\operatorname{Tr}\bigl(P\hat{\sigma}_U\bigr)^2
    \Bigr] \\
    &=
    \frac{\widetilde m_P^{-2} m_P}{N_U N_s}
    +\frac{1}{N_U}
    \Bigl(
      \widetilde m_P^{-2}\mathbb{E}_U
      \bigl[
        \mathbf 1\{UPU^{\dagger}\in\pm\mathcal Z\}
        \lambda_{U,P}^2
      \bigr]
      \operatorname{Tr}(\rho P)^2
    \notag \\
    &\qquad\qquad
      -2 m_P^{-1}\operatorname{Tr}(\rho P)\operatorname{Tr}(\sigma P)
      + m_P^{-1}\operatorname{Tr}(\sigma P)^2
    \Bigr) \\
    &=
    \frac{\widetilde m_P^{-2} m_P}{N_U N_s}
    +\frac{m_P^{-1}}{N_U}
    \Biggl(
      \frac{
        \mathbb{E}_U\!\bigl[\mathbf 1\{UPU^{\dagger}\in\pm\mathcal Z\}\bigr]\,
        \mathbb{E}_U\!\bigl[\mathbf 1\{UPU^{\dagger}\in\pm\mathcal Z\}\lambda_{U,P}^2\bigr]
      }{
        \bigl(
          \mathbb{E}_U\!\bigl[\mathbf 1\{UPU^{\dagger}\in\pm\mathcal Z\}\lambda_{U,P}\bigr]
        \bigr)^2
      }
      \operatorname{Tr}(\rho P)^2
    \notag \\
    &\qquad\qquad
      -2\operatorname{Tr}(\rho P)\operatorname{Tr}(\sigma P)
      +\operatorname{Tr}(\sigma P)^2
    \Biggr)
\end{align}
In the presence of errors, the variance of the error mitigated estimator may not become zero even if $\sigma$ matches $\rho$. However, if the hardware error is sufficiently small so that $\lambda_{U,P}$ is close to 1, we can sufficiently reduce the second term with a good bias state $\sigma$. To completely resolve this,  $\hat \sigma _U^{{\rm{(new)}}} = {\sum _b}{\rm{Tr}}({\Lambda _U}(\sigma )|b\rangle \langle b|)\widetilde{\mathcal{M}}^{-1}({U^{\dagger}}|b\rangle \langle b|U)$ can be used as an estimator for the bias state instead of ${\hat \sigma _U}$. However, in this case, estimating  ${\rm{Tr}}(\hat \sigma _U^{{\rm{(new)}}}P)$ requires $\lambda_{U,P}$, which may not be practical.

\subsection{Entanglement-enhanced machine learning}
\label{appx:ee_ml}

So far, we have studied how classical shadows with shallow measurements improve the estimation of physical observables. 
Beyond observable estimation, recent theoretical~\cite{huang2022provably_55, lewis2024improved_56} and experimental~\cite{cho2024machine_57} work has used data obtained from randomized measurements (RM) as input to classical machine learning (ML) algorithms. 
Most prior studies used randomized measurements consisting only of single-qubit gates. 
Here, we extend this setting to shallow measurements with entangling gates and investigate whether the resulting data can improve classical ML performance.

\subsubsection{State preparation}
\begin{figure}[t]
  \centering
  \includegraphics[
    width=0.6\linewidth,
    trim={1cm 23cm 12cm 0cm},
    clip
    ]{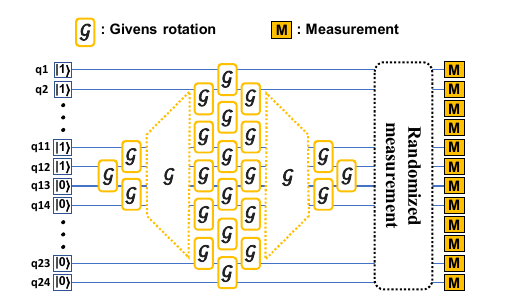}
    \caption{
    Quantum circuit for preparing the ground state of the Su--Schrieffer--Heeger (SSH) Hamiltonian.
    Each \(\mathcal{G}\) denotes a Givens rotation, which can be implemented with two controlled-Z (CZ) gates.
    The randomized measurement layer is applied after state preparation.
    }
  \label{fig:s7}
\end{figure}

In our experiments, we investigated at which values of w a phase transition occurs in the Su–Schrieffer–Heeger Hamiltonian $H_{\text{SSH}}(w) = {\sum _i}(a_{2i - 1}^\dagger {a_{2i}} + {\rm{h}}{\rm{.c}}{\rm{.}}) + w(a_{2i}^\dagger {a_{2i + 1}} + {\rm{h}}{\rm{.c}}{\rm{.}})$. To this end, we prepared the ground state $\rho_g(w)$ of $H_\text{SSH}(w)$ on a quantum computer and obtained its classical shadow. Because this state can be expressed as a Slater determinant, we employed a well-known circuit decomposition~\cite{jiang2018quantum} to implement it, as shown in Fig.~\ref{fig:s7}. 

We used $n$ = 24 qubits in our experiment, which required 46 layers of CZ gates for preparing the ground state of $H_{\text{SSH}}(w)$. After preparing the corresponding ground state $\rho_g(w)$, we obtained the classical shadow to use as training data by employing a block shadow with a block size $k=2$. Unlike in our previous experiments, here we employed a unitary ensemble $\mathbb{U} = \text{Haar}(2)^{\otimes n/2}$, where Haar(2) is the Haar measure on the group of two-qubit gates, instead of Cl(2). For our experiment on $\textit{ibm\_marrakesh}$, we used the following qubits: [41, 42, 43, 44, 45, 46, 47, 57, 67, 66, 65, 77, 85, 86, 87, 97, 107, 106, 105, 104, 103, 102, 101, 116]

\subsubsection{Error mitigation}
If $\mathbb{U} = \text{Cl}(k)$, then the inverse of the noise shadow channel $\widetilde{\mathcal{M}}$ can be expressed as
\begin{equation}
    {\widetilde {\cal M}^{ - 1}} = {\sum _{\lambda  \in R}}\widetilde f_\lambda ^{ - 1}{\Pi _\lambda } = {\widetilde f_0}^{ - 1}{\Pi _0} + {\widetilde f_1}^{ - 1}{\Pi _1} = {\Pi _0} + \alpha ({D_k} + 1){\Pi _1},
\end{equation}
where ${\Pi _0} = {\rm{ }}|I\rangle \!\rangle \langle \!\langle I|,{\rm{ }}{\Pi _1} = {\sum _{P \ne I}}|P\rangle \!\rangle \langle \!\langle P|$, and $\alpha \geq 1$. When expressed in the form of a standard quantum channel, it is written as
\begin{equation}
    {\widetilde {\cal M}^{ - 1}}(A) = \alpha ({D_k} + 1)A - \frac{{\alpha ({D_k} + 1) - 1}}{{{D_k}}}I.
\end{equation}
In the noiseless case, $\alpha = 1$, turning it into $\mathcal{M}^{-1}$. When computing the shadow kernel defined by 
\begin{equation}
    {k^{{\rm{(shadow)}}}}({S_T}(\rho ),{S_T}(\sigma )) = \exp \left( {\frac{\tau }{{{T^2}}}\sum _{t,t' = 1}^T\exp \left( {\frac{1}{{n/k}}\sum _{i = 1}^{n/k}{\gamma _i}{\rm{Tr}}(\hat \rho _i^{(t)}\hat \sigma _i^{(t')})} \right)} \right).
\end{equation}
we need to use the noise shadow channel $\cal \widetilde{M}$ for $\hat{\rho}_i^{(t)}$  and $\hat{\sigma}_i^{(t)}$, which requires additional calibration experiments. However, if we assume the error channel is translationally invariant and choose appropriate hyperparameters $\tau$ and $\{\gamma_i\}_i$, we can compute the error-mitigated (EM) shadow kernel without conducting further calibration experiments.

Given that
\begin{equation}
    {\rm{Tr}}({\widetilde {\cal M}^{ - 1}}({A_i}){\widetilde {\cal M}^{ - 1}}({B_i})) = \alpha _i^2{\rm{Tr}}({{\cal M}^{ - 1}}({A_i}){{\cal M}^{ - 1}}({B_i})) + {\beta _i},
\end{equation}
where ${\beta _i}: = ({D_k} + 2)\alpha _i^2 - \frac{{{\alpha _i}({D_k} + 1) - 1}}{{{D_k}}}$, the following holds accordingly
\begin{equation}
    {k^{{\rm{(shadow)}}}}({\rm{EM}};{\{ {\gamma _i}\} _i},\tau ) = {k^{{\rm{(shadow)}}}}({\rm{Raw}};{\{ {\gamma _i}\alpha _i^2\} _i},\tau {e^{\bar \beta }}),
\end{equation}
where $\bar \beta  = \frac{{{\sum _i}{\gamma _i}{\beta _i}}}{{n/k}}$. If the error channel is translationally invariant, then all $\{\alpha_i\}_i$ values are the same. Consequently, we can redefine the hyperparameters to obtain the error-mitigated shadow kernel
\begin{equation}
    {k^{{\rm{(shadow)}}}}({\rm{EM}};\gamma ,\tau ) = {k^{{\rm{(shadow)}}}}({\rm{Raw}};\gamma {\alpha ^2},\tau {e^{\bar \beta }})
\end{equation}
without conducting additional calibration experiments. To approximate a uniform error channel in experiments, one could randomly sample the qubit layouts. However, this introduces additional errors due to swap gate overhead in restricted qubit connectivity. Because we employed a one-dimensional open-chain qubit layout, we altered the direction of the chain during the experiment to avoid using additional gates. For the experiments presented in the main text of Fig.~\ref{fig5}, to ensure the numerical stability of the shadow kernel, we used the average of 80 independent measurements as follows
\begin{equation}
    {k^{{\rm{(shadow)}}}}({S_T}(\rho ),{S_T}(\sigma )) = \exp \left( {\frac{\tau }{{{T^2}}}\sum _{t,t' = 1}^{T( = 30)}\exp \left( {\frac{{\sum _{i = 1}^{n/k}}}{{n/k}}\gamma \frac{{\sum _{s = 1}^{S( = 80)}}}{S}{\rm{Tr}}(\hat \rho _i^{(t,s)}\hat \sigma _i^{(t',s)})} \right)} \right),
\end{equation}
where $\hat \rho _i^{(t,s)} = {{\cal M}^{-1}}(U_t^{\dagger}|{b_s}\rangle \langle {b_s}|{U_t})$, $(\tau, \gamma)=(1,1)$ when $k$ = 1 and $(\tau, \gamma)=(1,0.25)$ when $k$ = 2.

\begin{figure}[t]
  \centering
  \includegraphics[
    width=0.6\linewidth,
    trim={1cm 23cm 12cm 0cm},
    clip
    ]{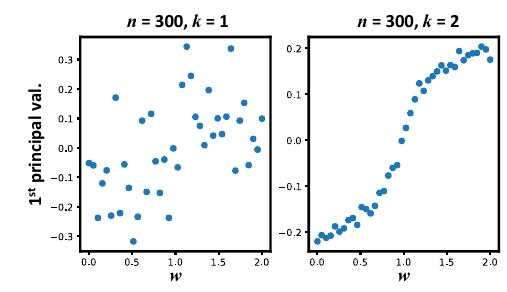}
    \caption{
    Noiseless numerical results for the Su--Schrieffer--Heeger (SSH) model with \(n=300\) and \(T=30\).
    The first principal component obtained from PCA is plotted as a function of \(w\).
    }
  \label{fig:s8}
\end{figure}
\subsubsection{ML performance with larger system size}
To demonstrate that the ML algorithm using shallow shadow also performs well for system sizes larger than the $n$ = 24 considered in the main text, we conducted a (noiseless) numerical experiment for the $n$ = 300 and $T$ = 30. The results are shown below.

For a larger system size, training with shallow shadows has been observed to enable more effective discrimination of quantum phases. Although quantum phases can also be distinguished using single-qubit measurement data by collecting classical shadows with a larger number of measurements $T$~\cite{huang2022provably_55}, this requires additional quantum experiments and extra post-processing time for classical machine learning.

\end{document}